\documentclass[12pt]{article}

\pdfoutput=1
\usepackage[latin9]{inputenc}
\usepackage[top=80pt,bottom=85pt,left=67pt,right=67pt]{geometry}
\usepackage{amssymb,amsmath}
\usepackage{xypic,subfigure}
\usepackage{graphicx,color}
\usepackage{bbm}
\usepackage{float}
\usepackage{cite}
\usepackage{tensor}
\usepackage{slashed}

\usepackage[hidelinks]{hyperref}

\definecolor{link}{rgb}{.8,.15,.1}
\hypersetup{colorlinks=true,linkcolor=link,citecolor=link,urlcolor=link,linktocpage}

\newcommand{\AdS}{\text{AdS}}
\newcommand{\Sp}[1]{\text{S}^{#1}}
\newcommand{\Mink}[1]{\text{Mink}_{#1}}

\makeatletter
\@addtoreset{equation}{section}
\makeatother

\newcommand{\beq}{\begin{equation}}
\newcommand{\eeq}{\end{equation}}
\newcommand{\bea}{\begin{eqnarray}}
\newcommand{\eea}{\end{eqnarray}}
\newcommand{\nn}{\nonumber}

\newcommand{\eq}{\begin{equation}}
\newcommand{\feq}{\end{equation}}
\newcommand{\eqn}{\begin{eqnarray}}
\newcommand{\feqn}{\end{eqnarray}}

\newcommand{\calA}{\mathcal{A}}

\newcommand{\calG}{\mathcal{G}}

\newcommand{\vol}{\text{vol}}

\begin{document}

\begin{titlepage}

\begin{center}

\vskip .5in 
\noindent

{\Large \bf{Type II embeddings for $d=6$ Einstein-Maxwell gauged supergravity}}

\bigskip\medskip

Niall T. Macpherson$^{a,b}$\footnote{macphersonniall@uniovi.es},  Ricardo Stuardo$^{a,b}$\footnote{ricardostuardotroncoso@gmail.com}\\

\bigskip\medskip
{\small 

a: Department of Physics, University of Oviedo,
Calle Leopoldo Calvo Sotelo 18, 33007 Oviedo}

\medskip
{\small and}

\medskip
{\small 

b: Instituto Universitario de Ciencias y Tecnolog\'ias Espaciales de Asturias (ICTEA),\\
Calle de la Independencia 13, 33004 Oviedo, Spain}\\

\vskip 2cm 

     	{\bf Abstract }\\[2mm]
        \end{center}
Bi-spinor and G-structure methods are used to classify the possible consistent truncations of type II supergravity to $d=6$ Einstein-Maxwell (gauged) supergravity, and its consistent sub-sectors. In the absence of R-symmetry gauging and a tensor multiplet we establish that every supersymmetric Mink$_6$ solution defines  an embedding of the $d=6$ theory. Adding a tensor multiplet places restrictions on these embeddings, but embeddings still exist. In the presence of R-symmetry gauging the internal spaces of the embeddings are neither related to Mink$_6$ or AdS$_6$. Under the assumption that the internal space contains a single U(1) isometry housing the $d=6$ gauge field we classify the possible embedding manifolds. We find two classes of embedding for the entire theory, one of which is governed by a Toda-like equation and contains at least one prototype bounded embedding, albeit containing likely un-physical singularities. In the absence of a tensor multiple the classes of embeddings become more permissive, though the PDEs governing them become more complicated in general. 
     	
     	\noindent

\noindent

\vfill
\eject

\end{titlepage}

\tableofcontents

\section{Introduction}
Whether ones interests tend towards the gravitational or the field theoretic, string theory has proven to be of much utility for theoretical physics. On the gravitational side, string theory is a leading contender for a theory of quantum gravity and the seminal work of Strominger and Vafa showed how it could provide a microscopic description of the entropy of black holes \cite{Strominger:1996sh}. Conversely many insights into the strong coupling limit of quantum field theories have been made since the advent of the AdS/CFT correspondence \cite{Maldacena:1997re} and its non-conformal extensions. However string theory is only well defined in 10 and 11 dimensions, while  one is often concerned with studying a theory that, at least at low energies and at least effectively, is $d$-dimensional with $d<10/11$. From a string phenomenological perspective one would of course like to have $d=4$ like the world around us. There are various reason that one might like to study field theories in diverse dimensions, but even in this context constraints on  dimensionality exist. For instance it is well known that superconformal field theories are only defined for $d\leq 6$. Thus the elephants in the room are the additional $10/11-d$ internal dimensions that need taking care of some how.

The lower energy weak curvature limits of string theory are supergravities in 10 and 11 dimensions. From this perspective the traditional way of dealing with  the extra dimensions is to demand that they should be compact, or at least bounded, such that integrals over the internal space yield a finite result\footnote{Or more properly that the effective $d$ dimensional Newtons constant is non-vanishing}. Then if the radius of the internal space is sufficiently small, the higher dimensional physics should decouple leaving an effective theory in $d$ dimensions.  This is all well and good as a philosophical principle, but at the end of the day one still needs to construct a 10 or 11 dimensional solution that has these properties. Constructing solutions of Einstein's equations is famously difficult and that difficulty scales with the total dimensions for the case at hand. Supersymmetry can of course help in this goal, but it is desirable to have some more explicit guiding principle.

A particularly useful method of constructing solutions  in 10 and 11 dimensional supergravities is to make use of consistent truncations to supergravity theories in  $d$ dimensions which may have R-symmetry gauging or not depending on the case at hand. The general idea is that it should be possible to embed the fields of the dimension $d$ theory into the higher dimensional theory in terms of some fixed embedding manifold such that the equation of motion in $d$ dimensions imply those in 10/11 dimensions. If a consistent truncation is constructed around a bounded internal space then the extra $10/11-d$ dimensions are automatically taken care of and one can construct a solution directly in $d$ dimensional supergravity before lifting it to higher dimensions. The issue is that to follow this path you need consistent a truncation in hand, and constructing one is a highly non-trivial task.

The majority of explicitly known consistent truncations fall into one of two categories. First there are consistent truncations to maximally supersymmetric gauged supergravities preserving 32 supercharges and a large gauge group. The construction of such truncation benefits from a large amount of symmetry which constrains the compact embedding manifolds to be spheres. Examples include the consistent truncations of $11$ dimensional supergravity on S$^4$ and S$^7$, type IIB on S$^5$ and massive IIA on S$^6$ \cite{deWit:1986oxb,Nastase:1999cb,Nastase:1999kf,Cvetic:2000nc,Guarino:2015vca,Varela:2015ywx,Baguet:2015sma}. It should be stressed though that the full non-linear embedding is still very challenging to construct and was often only found after the development of exceptional field theory techniques \cite{Coimbra:2011nw,Coimbra:2012af,Hohm:2013pua,Hohm:2013vpa,Hohm:2013uia,Hohm:2014fxa,Samtleben:2025fta}. For example the S$^5$  truncation of type IIB to $d=5$ maximally supersymmetric gauged supergravity was proposed when the lower dimensional theory was originally derived \cite{Gunaydin:1984qu,Pernici:1985ju}. However it took a further 30 years to construct the full embedding \cite{Baguet:2015sma}.

The second main category of known truncations are to gauged and un-gauged supergravities with only a gravity multiplet turned on and typically preserving minimal supersymmetry. Such supergravity theories typically have either SU(2) or U(1) R-symmetries groups that when gauged require only either a 2 or 1 sphere to embed into higher dimensions. Given the minimal bosonic field content, often only a metric and U(1) gauge field, it is relatively easy to find embeddings of these theories into 10 or 11 dimensions via brute force. In all cases we are aware of such minimal gauged supergravities are embedded into the internal space of AdS vacua (see for instance \cite{Gauntlett:2007ma,Gauntlett:2007sm, Liu:2010sa, Passias:2015gya,Larios:2019lxq,Couzens:2022aki}) while supergravities without R-symmetry gauging get embedded into those of Minkowski vacua (see for instance \cite{Bonetti:2011mw,Lima:2022hji,Lozano:2022ouq,Couzens:2022aki}).

G-structure and bi-spinor techniques, which geometresize the necessary conditions for supersymmetry,  have been very successfully applied to the construction and  classification of supersymmetric ``string vacua'' of $d=10/11$ supergravity - solutions containing AdS or Minkowski factors and preserving various amounts of supersymmetry (see for instance \cite{Grana:2004bg,Gauntlett:2004zh,Grana:2005sn,Gabella:2012rc,Apruzzi:2013yva,Dibitetto:2018ftj,Legramandi:2023fjr}). But beyond the realm of vacua, with some exceptions such as \cite{Giusto:2013rxa, Rosa:2013jja,Katmadas:2015ima,Couzens:2022aki,Macpherson:2024qfi}, they have been rather under utilized. Despite this, geometric conditions for totally generic backgrounds in 10 and 11 dimensions that preserve a single supercharge are known \cite{Gauntlett:2002fz,Gauntlett:2003wb,Tomasiello:2011eb}, so there is no particular barrier to using these methods to construct more general solutions. Generically a weakness of this approach for constructing solutions, in contrast to utilising consistent truncations,  is that one needs to ensure a compact internal space on a case by case basis. Furthermore, the approach is built around spinor bi-linears so typically are only used to construct supersymmetric solutions. However neither of these weaknesses are really relevant if one wants to use G-structure methods to construct an embedding of a lower dimensional supergravity into string theory. First off one typically wants a supersymmetric solution of the lower dimensional theory to be lifted to a supersymmetric solution in higher dimensions. Given that an embedding defines a fixed internal space, the only way for that to happen is if this internal space supports a Killing spinor, with which one can indeed define bi-linears. Also if one does not know an embedding, one needs some method of constructing it, and bi-linear and G-structure method give a systematic way to classify and construct the possibilities.

A seminal work for the application of G-structures to consistent truncations was \cite{Gauntlett:2007ma}. This work constructed consistent truncations to $d=4,5$ minimal gauged supergravites, containing a single U(1) gauge field ${\cal A}$, about certain general classes of supersymmetric AdS$_{4,5}$ vacua of $d=11$ and type IIB supergravities, with U(1) R-symmetries realised on their internal spaces. After making the, very reasonable, assumption that ${\cal A}$ was housed within the U(1) isometry of the internal space, without further modifications to the metric, the authors made an embedding ansatz for the fluxes and then derive constraints on the this ansatz with the Bianchi identities and equations of motion. Key to solving these constraints was knowledge of the G-structure classification of the internal spaces of these AdS vacua \cite{Gauntlett:2005ww,Gauntlett:2006ux}, indeed these were solved by fixing the embedding ansatz in terms of the forms spanning the G-structure on the internal space of the AdS vacua, given the differential constraints they were known to obey. At this point the authors were able to also confirm that supersymmetry of the gauged supergravity was lifted to higher dimensions via an explicit spinorial computation. Later in \cite{Couzens:2022aki}, which embedded (gauged) $d=5$ supergravity into type IIA, it was shown how to streamline this process by making use of bi-spinors\footnote{Tensor products of the spinors a supersymmetric solutions supports that map to poly-forms under the Clifford map.} defined in 10-dimensions and the conditions they should obey such that supersymmetry is preserved \cite{Tomasiello:2011eb}. Levering the construction of \cite{Tomasiello:2011eb} allowed the lifting of lower dimensional supersymmetry to higher dimensions to be  baked into the first step of the uplift process and,  as consistency with supersymmetry implies many details about the fluxes, this greatly reduced the guess work with the embedding ansatz - indeed in this case the fluxes were fixed uniquely by supersymmetry alone.

There are many more (gauged) supergravities than the maximally supersymmetric and minimal ones, indeed many physically interesting solutions exist in minimal supergravities coupled to additional matter multiplets - it would be useful to have consistent truncations to such theories, but how to construct them? One option would be to leverage the machinery of exceptional field theory which provides a framework to systematically deal with all consistent truncations, as spelt out in \cite{Cassani:2019vcl}.  Indeed such methods have been successfully employed to construct consistent truncations to half maximal gauged supergravities with and without matter multiplets \cite{Malek:2017njj,Cassani:2019vcl,Malek:2020jsa,Josse:2021put,Guarino:2024gke,Rovere:2025jks}, as  well as to minimally supersymmetric gauged supergravites by making use of ``exceptional generalized co-sets''  \cite{Hassler:2023axp,Hassler:2025qhh}. Here however we will follow a path more similar to \cite{Gauntlett:2007ma} and \cite{Couzens:2022aki}: G-structures and bi-spinor methods have already been shown to be well suited to construct consistent truncations to minimal supergravities, thus it is natural to explore their application to scenarios with additional matter multiplets turned on. It should be stressed that this is not an entirely new idea,  indeed  G-structures were employed to embed matter coupled gauged supergravities in \cite{Cassani:2011fu,Cassani:2012pj}, but it is an avenue that merits further exploration and development.

A main purpose of this work is to further develop the combined usage of G-structure and bi-spinor methods to construct consistent truncations. Specifically to embed minimal (gauged) supergravities with additional matter multiplets into string dimensions.  A particularly interesting theory to consider in this context is $d=6$ Einstein-Maxwell gauged supergravity \cite{Salam:1984cj}, which is also  referred to as the Salam-Sezgin model. This is minimal ${\cal N}=(1,0)$  supergravity in $d=6$ coupled to a vector and a tensor multiplet with U(1) R-symmetry gauging, and is famously consistent with a positive cosmological constant, which leads to it containing some interesting solutions: It provided an early example of a Mink$_4$ vacuum with chiral fermions though a consistent truncation on S$^2$. It also contains  AdS$_3$ solutions with squashed S$^3$ internal space whose entire spectra were recently shown to be consistent with scale separation, with and without supersymmetry \cite{Proust:2025vmv}. These solutions arise as the near horizon limits of dyonic string solutions found in \cite{Gueven:2003uw}. What the theory does not contain is an AdS$_6$ vacuum, so any uplift of the \textbf{gauged}\footnote{Which is to say, containing R-symmetry gauging, not merely containing a gauge field} version of the theory cannot be based around a known type II AdS$_6$ vacua, rather we must construct embedding manifolds from first principles.

There is only one uplift of $d=6$ Einstein-Maxwell gauged supergravity that exists in the literature \cite{Cvetic:2003xr}, but unfortunately the embedding manifold in this case is non-compact. There is also an F-theory embedding of 6d supergravity coupled to an arbitrary number of vectors, tensor and hyper multiplets in \cite{Bonetti:2011mw}, but this has no R-symmetry gauging so only includes, as a special case, an uplift of un-gauged Einstein-Maxwell supergravity. Additionally, being expressed in terms of the Kahler and complex moduli of an elliptically fibred CY$_3$, this embedding is necessarily some what implicit. Thus constructing consistent truncations of $d=6$ Einstein-Maxwell (gauged) supergravity about bounded embedding manifolds is an interesting and mostly unexplored avenue.

For the  above reasons we find $d=6$ Einstein-Maxwell gauged supergravity the perfect candidate for a G-structure and bi-linear based approach to constructing consistent truncations. In this work we will use these methods to classify the possible embeddings of this theory, its un-gauged limit and all its consistent sub-sectors (gravity multiplet only, gravity and vector multiplets,  gravity and tensor multiplets) into type II supergravity. We will make the assumption that when the gauge field ${\cal A}$ appears in the metric, as it must when we have R-symmetry gauging, it appears inside a single U(1) isometry direction in the internal space providing a circle fibration over the external $d=6$ directions. This differs from the embedding of \cite{Cvetic:2003xr} which contains a 2-torus fibation over the external direction  - as that only leave 2 undetermined directions, we believe such an ansatz is too constrained to yield an interesting embedding beyond \cite{Cvetic:2003xr}. \\
~\\
The layout of this work is as follows\\
~\\
We begin in section \ref{sec: the 6dtheory} by collecting the salient features of Einstein-Maxwell gauged supergravity. In section \ref{sec: the 6dtheoryrev} we review its matter content, symmetries and supersymmetry preservation. Next in section \ref{sec:6dsusy} we derive necessary and sufficient geometric conditions for the theory to preserve supersymmetry in terms of forms spanning an SU(2)$\ltimes \mathbb{R}^4$-structure. These conditions will be the foundation of our embeddings into ten dimensions. Finally in \ref{sec:6dsoltions} we give details of some interesting solutions in $d=6$ and test the results of the previous section by confirming that they do indeed solve our geometric constraints for supersymmetry.

The purpose of section \ref{sec:two} is to derive all the conditions that the $d=4$ embedding must obey for: 1) Supersymmetry in $d=6$ to imply supersymmetry in type II supergravity. 2) A solution to the $d=6$ equations of motion to imply a solution in $d=10$.  We present the general idea of how we derive these conditions in section \ref{sec:twop1} before presenting the necessary and sufficient conditions for 3 cases: Section \ref{sec:case1} deals with the strictly un-gauged limit of the theory where  ${\cal A}$ does not appear in the metric. Section \ref{sec:case2} presents the gauge compatible case where ${\cal A}$ does appear in the metric. Section \ref{sec:case2} presents uplift formula for an certain simple but inconsistent sub-sector of the 6d theory\footnote{What we mean here is that the embeddings do not support generic values of all the bosonic  6d fields required for the $d=6$ supersymmetry algebra to close. Not that the uplifts or 6d solutions themselves are sick in some way.}. Finally in section \ref{sec:internalbilinears} we give an explicit parametrisation of the bi-linears and G-structure the embedding manifolds support.

The next sections derive  explicit classes of embedding manifolds: In section \ref{eq:Mink6vac} we recover classes of supersymmetric Mink$_6$ vacua. In part this serves as a warm up, but it will also turn out that when there is no R-symmetry gauging it is the internal spaces of such solutions, possibly up to additional constraints, that provide the embedding manifolds for the $d=6$ theory. We consider uplifts of the various limits of the un-gauged $d=6$ theory, for which ${\cal A}$ does not appear in the internal metric, in section \ref{sec:ungagued}. We find that every supersymmetric Mink$_6$ vacua provides an embedding of minimal un-gauged supergravity coupled to a vector multiplet in section \ref{sec:universal}, and that while solutions with either a tensor multiplet or vector and tensor multiplet are more constrained they do still exist. We then turn our attention to explicit classes of embedding manifolds where ${\cal A}$ does appear in the internal metric (requiring the vector multiplet to be non-trivial), first without R-symmetry gauging in section \ref{eq:section 6}, then with it in section \ref{eq:section 7}. In all but one example, with R-symmetry gauging studied in section \ref{eq:todaclass}, we find that the embeddings are much more permissive in the absence of the tensor multiplet. We find two classes of embeddings that are compatible with with all the fields of Einstein-Maxwell gauged supergravity. The first, in section \ref{eq:todaclass}, is governed by a Toda-like equation and has the potential to contain well defined bounded embeddings. The second, presented in section \ref{sec:secondfullembedding}, gives rise to manifestly un-bounded embeddings, however it does contain the previously known embedding of \cite{Cvetic:2003xr}, modulo T-duality, in a certain limit that we show explicitly.

Next in section \ref{sec:towardscompacted} we explore the possibility of realising a concrete embedding of full Einstein-Maxwell gauged supergravity that has a bounded embedding. We first derive the effective $d=6$ Newtons constant, which for a bounded embedding should be non-vanishing. We then show that a least one bounded embedding does indeed exist, although it comes with singularities that we do not recognise as being associated to physical objects within a strings context, such as D-branes or O planes. Failure to recognise an objects physicality is not the same as it being un-physical, but it is likely that in this case the singularities are not physical. As such this embedding probably does not represent a true solution of the, matter coupled, type II equations of motion - the issue being the singular points, elsewhere the equations of motion are solved. However this prototype embedding follows from the most simple solution to the defining PDE of the class in section \ref{eq:todaclass}, there are likely many more solutions and some of these my not be plagued by likely un-physical singularities.

Finally we present our conclusions and discuss future directions in section \ref{eq:section 8}.

This work is supplemented by extensive technical appendices referred to throughout the main text.

\section{Einstein-Maxwell (gauged) supergravity in \texorpdfstring{$d=6$}{d=6} and G-structures}\label{sec: the 6dtheory}
In this section we review Einstein-Maxwell gauged supergravity \cite{Salam:1984cj}, also commonly referred to as the Salam-Sezgin model. We will also derive necessary and sufficient conditions for its solutions to preserve supersymmetry in terms of spinor bi-linears that give rise to differential conditions on the forms that span an  SU(2)$\ltimes \mathbb{R}^4$-structure. This has been done before in \cite{Cariglia:2004kk} but not in a manor that is particularly conducive to the procedure we will employ to embed this 6d theory into type II supergravity -  \cite{Cariglia:2004kk} also employs mostly negative signature conventions for the metric while we elect mostly positive conventions.

\subsection{Summary of the theory}\label{sec: the 6dtheoryrev}
Minimal $d=6$ supergravity consists of only the gravity multiplet whose bosonic part consists of the metric $g^{(6)}_{\mu\nu}$ and an, in our conventions for the Hodge dual (see appendix \ref{eq:convensions}), anti-self dual 3-form ${\cal G}^{-}$. It is possible to couple this theory to a tensor multiplet whose bosonic elements are a scalar $\varphi$ and a self-dual 3-form ${\cal G}^{+}$ and a vector multiplet containing the 1-form ${\cal A}$. The resulting model preserves ${\cal N}=(1,0)$ supersymmetry and has an SU(2) R-symmetry. It is then possible to gauge a U(1) subgroup of the R-symmetry which introduces a coupling $g$, which results in Einstein-Maxwell gauged supergravity \cite{Salam:1984cj}. In summary the Bosonic field content of the theory and what multiplet they belong to is
\beq
\textbf{Gravity}:~~~(g^{(6)}_{\mu\nu},~{\cal G}^-),~~~\textbf{Tensor}:~~~(\varphi,~{\cal G}^+),~~~\textbf{Vector}:~~~ {\cal A}.
\eeq
If we introduce a 2-form potential ${\cal B}$ we can now define a generic 3-form and 2-form field strength as
\beq
{\cal G}= {\cal G}^-+{\cal G}^+= d{\cal B}+{\cal A}\wedge {\cal F},~~~~{\cal F}=d{\cal A}.
\eeq
In terms of these the action of the bosonic part of the theory then takes the form (see appendix \ref{eq:convensions} for our conventions on form contractions)
\beq
S^{(6)}=\int d^6x \sqrt{-\det g^{(6)}}\bigg[R^{(6)}-(\partial\varphi)^2-2e^{2\varphi}{\cal G}^2-2e^{\varphi}{\cal F}^2-2 g^2e^{-\varphi}\bigg],\label{eq:action}
\eeq
where we have set the 6-dimensional Newtons constant to 1. Famously this action is compatible with a positive cosmological constant. This leads to equations of motion that can be written in the form
\begin{subequations}
\begin{align}
&d\star_6d\varphi+e^{\varphi}\star_6{\cal F}\wedge {\cal F}+2e^{2\varphi}\star_6{\cal G}\wedge{\cal G}=g^2e^{-\varphi}\text{vol}_6,\label{eq:6dEOM1}\\[2mm]
&R^{(6)}_{\mu\nu}-\nabla^{(6)}_{\mu}\varphi\nabla^{(6)}_{\nu}\varphi=2\left(e^{\varphi}{\cal F}^2_{\mu\nu}+e^{2\varphi}{\cal G}^2_{\mu\nu}\right)+\frac{1}{2}\left(e^{-\varphi}g^2-e^{\varphi}{\cal F}^2-2e^{2\varphi}{\cal G}^2\right)g_{\mu\nu}^{(6)},\label{eq:6dEOM2}\\[2mm]
&d(e^{2\varphi}\star_6 {\cal G})=0,~~~~ d(e^{\varphi}\star_6 {\cal F})=+2 e^{2\varphi}\star_6 {\cal G}\wedge {\cal F} ,\label{eq:6dEOM3}
\end{align}
\end{subequations}
where the  Bianchi identities are
\beq
d{\cal F}=0,~~~~d{\cal G}={\cal F}\wedge {\cal F}.\label{eq:6dEOM4}
\eeq
Supersymmetry is preserved in terms of a Wely spinor $\zeta_-$ with negative chirality  with respect to the chirality matrix  $\hat\gamma^{(6)}=(\gamma^{(6)})_{0...5}$. When the fermionic fields are set to zero a background preserves supersymmetry if a non-trivial $\zeta_-$ exists which obeys the conditions
\begin{subequations}
\begin{align}
\left({\cal F}-ig e^{-\varphi}\right)\zeta_-=0,\label{eq:BPS1}\\[2mm]
\left(d\varphi-e^{\varphi}{\cal G}\right)\zeta_-=0,\label{eq:BPS2}\\[2mm]
(\nabla_{\mu}-i g {\cal A}_{\mu})\zeta_-+\frac{1}{4}e^{\varphi}{\cal G}\gamma^{(6)}_{\mu}\zeta_-=0,\label{eq:BPS3}
\end{align}
\end{subequations}
where a $k$-form $C_k$ acts on a spinor as
\beq
C_k\zeta:=  \frac{1}{k!}(C_k)^{\alpha_1...\alpha_k}\gamma^{(6)}_{\alpha_1...\alpha_k}\zeta,
\eeq
\textit{i.e.} forms act on spinors and gamma matrices through the Clifford map. 

We conclude our summary with some general observations that will be useful later. First off we note that the action \eqref{eq:action} is invariant under the scaling symmetry
\beq
({\cal G},~e^{-\varphi},~{\cal F},~g)\to (\lambda{\cal G},~\lambda e^{-\varphi},~\lambda^{\frac{1}{2}}{\cal F},~ \lambda^{-\frac{1}{2}}g),\label{eq:invariances}
\eeq
for $\lambda$ a constant - notice that this also leaves the supersymmetry conditions \eqref{eq:BPS1}-\eqref{eq:BPS3} intact. Through this symmetry we have that a constant  dilaton $\varphi$ is equivalent to fixing $\varphi=0$. We also note that when $(g=0,{\cal F}=0)$ the action is symmetric under the following mapping of the fields
\beq
({\cal G},~e^{2\varphi}\star_6{\cal G},~\varphi) \to (-e^{2\varphi}\star_6{\cal G},~-{\cal G},~-\varphi),
\eeq  
realising an S-duality like symmetry, in that like S-duality of type IIB supergravity the dilaton is inverted. The specific signs in the flux terms are required so that the supersymmetry conditions that remain non-trivial when $(g=0,{\cal F}=0)$, \textit{i.e.} \eqref{eq:BPS2} and \eqref{eq:BPS3}, likewise respect this symmetry given that
\beq
\star_6{\cal G}= \hat\gamma^{(6)}{\cal G}=- {\cal G}\hat\gamma^{(6)},\label{eq:clifrelation}
\eeq
under the Clifford map.

Finally let us make some comments that apply to supersymmetric solutions specifically: It should be clear from \eqref{eq:BPS1} that a non-trivial gauge coupling requires ${\cal F}$ to also be non-trivial - this is a bit unusual, indeed many gauged supergravities admit AdS vacua which require ${\cal F}=0$ and $g\neq 0$. Next, as becomes clear by using \eqref{eq:clifrelation}, \eqref{eq:BPS2} and \eqref{eq:BPS3} only actually contain ${\cal G}^{+}$ and ${\cal G}^{-}$ respectively - so in particular \eqref{eq:BPS2} constrains only the tensor multiplet.\\
~\\
In the next section we will derive geometric conditions for solutions in this theory to preserve supersymmetry.

\subsection{Supersymmetry in terms of  \texorpdfstring{SU(2)$\ltimes \mathbb{R}^4$}{SU(2)|x R4}-structure forms}\label{sec:6dsusy}
In this section we derive geometric conditions for solutions of $d=6$ Einstein-Maxwell gauged supergravity to preserve supersymmetry. These will be important for our method of embedding this theory into type IIB.  

One can show a single Weyl spinor $\zeta_-$ in 6  Lorentzian dimensions supports an SU(2)$\ltimes \mathbb{R}^4$-structure. This consists of a null 1-form $k$, and real and holomorphic 2-forms $(J,\Omega)$ spanning an SU(2)-structure orthogonal to $k$. Specifically these are defined in terms of $\zeta_-$ as 
\begin{align}
&k=- \overline{\zeta_-}\gamma^{(6)}_{\alpha}\zeta_- dx^{\alpha},\nn\\[2mm]
&k\wedge J=\frac{i}{3!}\overline{\zeta_-}\gamma^{(6)}_{\alpha\beta\delta}\zeta_-dx^{\alpha\beta\delta},~~~k\wedge \Omega=-\frac{1}{3!}\overline{\zeta^c_-}\gamma^{(6)}_{\alpha\beta\delta}\zeta_-dx^{\alpha\beta\delta}.
\end{align}
where $\overline{\zeta}=(\gamma^{(6)}_0\zeta)^{\dag}=\zeta^{\dag}(\gamma^{(6)})^0$ and $\zeta^c=B^{(6)}\zeta^*$ for $(B^{(6)})^{-1}\gamma^{(6)}_{\mu}B^{(6)}=\gamma^{(6)*}_{\mu}$ and  $B^{(6)}B^{(6)*}=-\mathbb{I}$ (we also assume $B^{(6)\dag}=(B^{(6)})^{-1}=B^{(6)}$). To derive geometric conditions that are equivalent to \eqref{eq:BPS1}-\eqref{eq:BPS3} it is useful to introduce the following bi-spinors/poly-forms
\begin{align}
\slashed{\psi}^{(6)}_-&=\zeta_-\otimes \overline{\zeta_-}~~~~\Rightarrow~~~~\psi^{(6)}_-= -\frac{1}{8} k\wedge e^{-i J},\nn\\[2mm]
\tilde{\slashed{\psi}}^{(6)}_-&=\zeta_-\otimes \overline{\zeta^c_+}~~~~\Rightarrow~~~~\tilde{\psi}^{(6)}_-= \frac{1}{8} k\wedge \Omega\label{eq:6dpolyforms},
\end{align}
where we note that
\beq
\psi^{(6)}_5=\frac{1}{8}\iota_k \text{vol}_6.
\eeq

The first conditions we will deal with are \eqref{eq:BPS1}-\eqref{eq:BPS2}. Being independent of derivatives of the spinor, it is a  relatively simple matter to derive what conditions are equivalent to them by for instance working in the canonical frame of appendix \ref{canonicalframe}. We find that \eqref{eq:BPS1} is equivalent to
\begin{align}
&\iota_k{\cal F}=0,~~~~{\cal F}\wedge \psi^{(6)}_1= \frac{1}{8}\iota_k\star_6 {\cal F}+i g e^{-\varphi}\psi^{(6)}_3,~~~{\cal F}\wedge \psi^{(6)}_3=ig e^{-\varphi}\psi^{(6)}_5,\label{eq:geometric1}
\end{align}
while the condition \eqref{eq:BPS2} is equivalent to
\beq
{\cal L}_k \varphi=0,~~~~\iota_k({\cal G}+\star_6{\cal G})=-8 e^{-\varphi}d\varphi\wedge \psi^{(6)}_1.\label{eq:geometric2}
\eeq
Together these imply several conditions that are useful for the embedding into 10 dimensions, namely
\begin{align}
&{\cal F}\wedge \tilde{\psi}^{(6)}_-={\cal G}\wedge \tilde{\psi}^{(6)}_-=\star_6{\cal G}\wedge \tilde{\psi}^{(6)}_-=0,~~~k\wedge ({\cal G}+\star_6 {\cal G})= e^{-\varphi}\iota_k\star_6d\varphi\nn\\[2mm]
&\psi^{(6)}_3\wedge {\cal G}=\psi^{(6)}_3\wedge \star_6{\cal G}=0,~~~~d\varphi\wedge \psi^{(6)}_5=0\label{eq:wedgerules}.
\end{align}
We now turn our attention to \eqref{eq:BPS3},  this  does contain a derivative of the spinor so deriving conditions that imply it is more involved. First off it is not too hard to establish that \eqref{eq:BPS3} implies
\beq
\nabla_{(\mu}k_{\nu)}=0,~~~~e^{-\varphi}d\psi^{(6)}_-=\frac{1}{8}\iota_k({\cal G}-\star_6{\cal G}),~~~~d\tilde{\psi}^{(6)}_-=2g i {\cal A}\wedge \tilde{\psi}^{(6)}_-,\label{eq:geometric3}
\eeq
by making use of the identities in \eqref{eq:derivativeidentities}. The real issue is establishing whether \eqref{eq:geometric3} implies  \eqref{eq:BPS3}, it in fact does not on its own - we will return to this point momentarily.  For now we observe that  \eqref{eq:geometric2} and \eqref{eq:geometric3} imply that the null vector $k^{\mu}\partial_{\mu}$ is Killing with respect to the metric and dilaton $\varphi$. The conditions derived so far can be combined to give several other, a particularly useful one is
\beq
d(e^{-\varphi}k)=-2 \iota_k{\cal G}.\label{eq:usefulcond6d}
\eeq
From this and \eqref{eq:geometric1} it follows that if we assume the Bianchi identities of ${\cal G}$ and ${\cal F}$ it then follows that
\beq
{\cal L}_k{\cal F}=0,~~~~{\cal L}_k{\cal G}=0.
\eeq
Thus $k^{\mu}\partial_{\mu}$ is an isometry of an entire supersymmetric solution. Another useful piece of information going forward will be the charge of the spinor under this isometry, this can be established with the Lie derivative
\beq
{\cal L}_k\zeta_-=(k^{\mu}\nabla_{\mu}+\frac{1}{4}\nabla_{\mu}k_{\nu}(\gamma^{(6)})^{\mu\nu})\zeta_-
\eeq
Given what has been derived thus far and by making use of the canonical frame in appendix \ref{canonicalframe} we establish that
\beq
{\cal L}_k\zeta_-=(k^{\mu}\nabla_{\mu}+\frac{1}{4}\nabla_{\mu}k_{\nu}(\gamma^{(6)})^{\mu\nu})\zeta_-=i g\iota_{k}{\cal A}\zeta_-,
\eeq
so if we choose a gauge in which
\beq
\iota_k{\cal A}=0,
\eeq
then $\zeta_-$ is a singlet with respect to $k^{\mu}\partial_{\mu}$ - we will indeed elect such a gauge.

We now return to the issue of sufficient conditions to imply \eqref{eq:BPS3}. Indeed, as explained in appendix \ref{eq:6dpairing}, given that a chiral spinor in 5+1 dimensions supports an SU(2)$\ltimes \mathbb{R}^4$ structure \eqref{eq:BPS3} contains a total of 48 independent conditions, while \eqref{eq:geometric3} only yields 45 of these. To access the remaining 3 conditions it is necessary to introduce a second null 1-form $v$ such that
\beq
\iota_vk=1,~~~\iota_vJ=0,~~~\iota_v\Omega=0.
\eeq
We can then take $(k,v)$ to be vielbein directions such that the $d=6$ line element decomposes as
\beq
g^{(6)}_{\mu\nu}dx^{\mu}dx^{\nu}= 2k v+ ds^2(\text{M}_{\text{SU}(2)}),
\eeq
where $\text{M}_{\text{SU}(2)}$ is spanned by space-like vielbein directions with respect to which $(J,\Omega)$ are defined. 
We show through a long computation in appendix \ref{eq:6dpairing} that the remaining 3 constraints contained in \eqref{eq:BPS3}, that do not appear in \eqref{eq:geometric3}, can be expressed in terms of $v$ as
\begin{subequations}
\begin{align}
&v\wedge\Omega\wedge\bigg[  d(k\wedge v-i J)+ 2 e^{\varphi} {\cal G}\bigg]=0,\\[2mm]
&v\wedge\bigg[d \Omega\wedge \overline{\Omega}-(\nabla.v)k\wedge J\wedge J-2i k \wedge J\wedge dv-4i(g{\cal A}\wedge J\wedge J-e^{\varphi}{\cal G}\wedge J)\bigg]=0.
\end{align}
\end{subequations}
Note that $dv$ should be constrained as
\beq
\iota_{v}dv=0
\eeq
since $k^{\mu}\partial_{\mu}$ is an isometry of the metric.

In summary necessary and sufficient conditions for supersymmetry of Einstein-Maxwell gauged supergravity are
\begin{subequations}
\begin{align}
&\nabla_{(\mu}k_{\nu)}=0,~~~~{\cal L}_k \varphi=0,~~~~\iota_k{\cal F}=0,\label{eq:bpscond6d1}\\[2mm]
&e^{-\varphi}d\psi^{(6)}_-=\frac{1}{8}\iota_k({\cal G}-\star_6{\cal G}),~~~~d\tilde{\psi}^{(6)}_-=2g i {\cal A}\wedge \tilde{\psi}^{(6)}_-,\label{eq:bpscond6d2}\\[2mm]
&\iota_k({\cal G}+\star_6{\cal G})=-8 e^{-\varphi}d\varphi\wedge \psi^{(6)}_1,\label{eq:bpscond6d3}\\[2mm]
&{\cal F}\wedge \psi^{(6)}_1= \frac{1}{8}\iota_k\star_6 {\cal F}+i g e^{-\varphi}\psi^{(6)}_3,~~~{\cal F}\wedge \psi^{(6)}_3=ig e^{-\varphi}\psi^{(6)}_5,
\label{eq:bpscond6d4}\\[2mm]
&v\wedge\Omega\wedge\bigg[  d(k\wedge v-i J)+ 2 e^{\varphi} {\cal G}\bigg]=0,\label{eq:bpscond6d5}\\[2mm]
&v\wedge\bigg[d \Omega\wedge \overline{\Omega}-(\nabla.v)k\wedge J\wedge J-2i k \wedge J\wedge dv-4i(g{\cal A}\wedge J\wedge J-e^{\varphi}{\cal G}\wedge J)\bigg]=0,\label{eq:bpscond6d6}
\end{align}
\end{subequations}
we should stress that the real part of the final condition is redundant, but we keep it as it gives a definition for $(\nabla.v)$ which is useful for the $d=10$ pairing constraint computation in appendix \ref{sec:missing10dconds}. The above conditions are also valid in the un-gauged limit, one need only fix $g=0$, and also in the absence of the tensor or vector multiplets, one need only tune the 6d fields appropriately.\\
~\\
In the next section we present some notable supersymmetric solutions of $d=6$ gauged and un-gauged Einstein-Maxwell supergravity as well as how they solve the geometric constraints of this section.

\subsection{Some notable supersymmetric solutions}\label{sec:6dsoltions}
In this section we present some solutions that lie within the various sub-sectors of Einstein-Maxwell supergravity. This serves in part to provide examples of the sort of solutions that can be uplifted to type II supergravity using the results of the later sections of this work, but also as a text of our geometric conditions for supersymmetry \eqref{eq:bpscond6d1}-\eqref{eq:bpscond6d6}

\subsubsection{\texorpdfstring{$\AdS_{3} \times \Sp{3}$}{AdS3 x S3}}
The first solution we consider is the $d=6$ black-string near horizon, which is a solution of minimal $d=6$ supergravity with $g=0$. It has non-trivial fields  
    \begin{subequations}
    \begin{align}
        ds^{2} = \ell^{2} \left( ds^{2}(\AdS_{3}) + ds^{2}(\Sp{3}) \right),\\[2mm]
        \mathcal{G} =\ell^{2}\left( \vol(\AdS_{3}) +  \vol(\Sp{3})\right),
    \end{align}
    \end{subequations}
where in particular $\varphi = 0$ and ${\cal G}=-\star_6{\cal G}$ . Here the AdS and $\Sp{3}$ factors are of unit radius. This solution preserves 8 supercharges.

To show that this solution does indeed preserve supersymmetry we will show that it solves the geometric condtions of the previous section. To this end we use the following parametrization of AdS
    \begin{equation}\label{eq:ParamAdS3}
        ds^{2}(\AdS_{3}) = e^{2r}(-dt^2+dx^2) + dr^{2},
    \end{equation}
and take the 3-sphere to be spanned by a set of left invariant SU(2)-forms $L_i$ obeying on the 3-sphere satisfying
    \begin{equation}
        dL_{i} = \frac{1}{2}\epsilon_{ijk} L_{j}\wedge L_{k},
    \end{equation}
		such that
		\beq
		ds^2(\text{S}^3)=\frac{1}{4}(L_i)^2,~~~~\text{vol}(\text{S}^3)=\frac{1}{8}L_1\wedge L_2\wedge L_3
		\eeq
We find that the conditions for supersymmetry \eqref{eq:bpscond6d1}-\eqref{eq:bpscond6d6} are solved when
    \begin{subequations}
    \begin{align}
        k &= -\ell\,  e^{2r} \left(dt+dx^{1}\right),\quad
        v =  \frac{\ell}{2} \left( dt-dx^{1}\right),\\
        J &= -\frac{\ell^{2}}{2}\left( dr\wedge L_{3} + \frac{1}{2} L_{1}\wedge L_{2} \right)  ,\\
        \Omega &= \frac{i\, \ell^{2}}{2}\left( dr - \frac{i}{2}L_{3}\right)\wedge \left( L_{1}-i L_{2} \right).
    \end{align}
    \end{subequations}
At first sight this appears to only prove that this solution preserves a single supercharge. However notice that $(ds^2,{\cal G})$ are expressed in terms of SO(4) invariants while $(J,~\Omega)$ are only invariant under SU(2)$_L\subset$ SO(4), they are charged charged under SU(2)$_R$ and by acting with this symmetry one can generate a further $3$ independent versions of $(J,~\Omega)$ that also solve  \eqref{eq:bpscond6d1}-\eqref{eq:bpscond6d6} for the same $(k,v)$ taking us to $4$ supercharges. This is enhanced to $8$ because we have elected an SO(1,1) invariant parameterisation of $(k~,v~,J,~\Omega)$, there is a second choice of the forms on AdS$_3$ which obey the same constraints but are not SO(1,1) invariant. This is nothing more than the geometrisation of the Poinc\'are and conformal supercharges supported by an AdS$_3$ Killing spinor.  

\subsubsection{Salam-Sezgin (\texorpdfstring{Minkowski$_{4}\times \Sp{2}$}{Minkowski4 x S2})}

We now consider a solution with a vector multiplet. This correspond to the $\Mink{4}\times \Sp{2}$ solution of \cite{Salam:1984cj}. The configuration there is presented with a constant dilation. Here we use \eqref{eq:invariances} to set the dilaton to zero. With this consideration, the background configuration reads
    \begin{subequations}
    \begin{align}
        ds^{2} &= dx^{2}_{1,3} + \ell^{2} ds^{2}(\Sp{2}),\\
        \mathcal{A} &= - \frac{\ell}{\sqrt{2}}\left( \cos\theta \pm 1 \right)d\phi.
    \end{align}
    \end{subequations}
As before, the factor of $\Sp{2}$ is of unit radius. This configuration preserves four real supercharges. Using the flat Minkowski metric,  we can write 
    \begin{subequations}
    \begin{align}
        k &=  -dt+ dx^{1} , \quad
        v = \frac{1}{2}\left( dt + dx^{1} \right),\\
        J &=  dx^{2}\wedge dx^{3} + \ell^{2} \vol(\Sp{2}) ,\\
        \Omega &= -i \ell\, e^{\mp i \phi} \left(dx^{2}+i dx^{3}\right)\wedge \left(d\theta+i \sin\theta d\phi\right)
    \end{align}
    \end{subequations}
which solve the conditions supersymmetric conditions in  \eqref{eq:bpscond6d1}-\eqref{eq:bpscond6d6}.

\subsubsection{\texorpdfstring{$\AdS_{3}\times \text{squashed } \Sp{3}$ }{AdS3 x squashed S3}}\label{sec:SolutionHenning}

In \cite{Proust:2025vmv}, a solution containing both tensor and vector multiplets was reported. This solution is of the form $\AdS_{3}$ with a squashed-$\Sp{3}$. Using \eqref{eq:invariances} to set the constant dilaton to zero, we write this configuration as 
    \begin{subequations}
    \begin{align}
        ds^{2} &= \ell^{2}ds^{2}(\AdS_{3}) + \frac{1}{4}\cosh^{2}\beta\left( L^{2}_{1} + L^{2}_{2} \right)+ \frac{1}{4}L^{2}_{3},\\
        \mathcal{A} &= -\frac{\sinh\beta}{2\sqrt{2}}L_{3},\\
        \mathcal{G} &=  \ell^{2} \vol(\AdS_{3}) + \vol(\Sp{3}).
    \end{align} 
    \end{subequations}
Here the AdS factor is of unit radius, and $\vol(\Sp{3})$ is the volume form of the unit radius, un-squashed ($\beta=0$) 3-sphere. Also, the gauge coupling and the constant $\ell$ are fixed as
    \begin{equation}
        g = \sqrt{2} \frac{\tanh\alpha}{\cosh\beta},\quad
        \ell = \cosh\alpha\cosh\beta.
    \end{equation}
This solution is supersymmetric when $\alpha=\beta$, and it preserves four supercharges. Using the AdS parametrization in \eqref{eq:ParamAdS3} we write the solution to \eqref{eq:bpscond6d1}-\eqref{eq:bpscond6d6} as
    \begin{subequations}
    \begin{align}
        k &= \ell\, e^{2r} \left( dt + dx \right), \quad
        v = \frac{\ell}{2}\left( dt - dx \right),\\
        J &= -\frac{1}{2}\left( \ell\, dr \wedge L_{3} + \frac{1}{2}\cosh^{2}\beta L_{1}\wedge L_{2} \right),\\
        \Omega &= \frac{i}{4}\left( 2 \ell\, dr - i L_{3} \right)\wedge \left(L_{1} -i L_{2}\right).
    \end{align}
    \end{subequations}
were we have again chosen an SO(1,1) invariant parametrization. We count four real supercharges.

\subsubsection{Solution with non-constant dilaton}

Finally, we consider the following ansatz
    \begin{subequations}
    \begin{align}
        ds^{2} &= e^{2A}dx^{2}_{1,1} + e^{2k}dr^{2} + \frac{e^{2g}}{4}\left( 
        L^{2}_{1}+L^{2}_{2}\right) + \frac{e^{2h}}{4}L^{2}_{3},\\
        \mathcal{A} &= \lambda\, L_{3}
    \end{align} 
    \end{subequations}
where $A,k,g,h,\lambda$ and the dilaton $\varphi$ are functions of $r$ only. We keep $\mathcal{G}$ arbitrary as it can be fixed by requiring supersymmetry. To this aim, we also use an ansatz for the bi-linears 
    \begin{subequations}
    \begin{align}
        k &= \, e^{2A}\left( -dt + dx \right),\quad
        v = \frac{1}{2}\left( dt + dx \right),\\
        J&=  \frac{1}{2}e^{k+h} dr\wedge L_{3} + \frac{1}{4}e^{2g} L_{1}\wedge L_{2}, \\
        \Omega &= \frac{e^{g}}{2}\left( e^{k}dr + i \frac{e^{h}}{2}L_{3}  \right)\wedge \left( L_{1} + i L_{2}  \right).
    \end{align}
    \end{subequations}
Using the supersymmetry conditions in \eqref{eq:bpscond6d1}-\eqref{eq:bpscond6d6} we find an expression for $\calG$
    \begin{equation}
    \begin{aligned}
        \mathcal{G} &= k\wedge X^{(1,1)} - \frac{1}{2}e^{-A} \left(2e^{-\varphi}(e^{A})' + e^{A}(e^{-\varphi})' \right) k\wedge v \wedge dr 
        ,\\
        &\phantom{=} + \frac{1}{16}e^{2g+h-A-k}\left(-2e^{-\varphi}(e^{A})' + e^{A}(e^{-\varphi})' \right) L_{1}\wedge L_{2} \wedge L_{3}
    \end{aligned}
    \end{equation}
where $X^{(1,1)}$ is a primitive $(1,1)$ form given by
    \begin{equation}
    \begin{aligned}
        X^{(1,1)} &= e^{A} \left( g_{1}(r) \left(  -\frac{1}{2}e^{k+h} dr\wedge L_{3} + \frac{1}{4}e^{2g} L_{1}\wedge L_{2} \right) 
    + g_{2}(r) \left(  -\frac{1}{2}e^{k+g} dr\wedge L_{1} + \frac{1}{4}e^{g+h} L_{2}\wedge L_{3} \right) \right. \\
        &\phantom{=} \left. + g_{3}(r) \left(  -\frac{1}{2}e^{k+g} dr\wedge L_{2} - \frac{1}{4}e^{g+h} L_{1}\wedge L_{3} \right)\right),
    \end{aligned}
    \end{equation}
with $g_{1}$, $g_{2}$ and $g_{3}$ arbitrary. The supersymmetry conditions are solved provided that there exists a solution to the following BPS equations
    \begin{subequations}
    \begin{align}
        (e^{g})' &= - e^{h+k-g} - e^{g-A}(e^{A})',\\
        (e^{h})' &= e^{k}\left( -2 + e^{2h-2g} -4g \lambda \right) - e^{h-A}(e^{A})',\\
        \lambda' &= \frac{1}{2}e^{h+k}\left( 4\lambda e^{-2g} + g e^{-\varphi} \right).
    \end{align}
    \end{subequations}
To solve the equations of motion it is still necessary to solve the Bianchi identities for the fluxes. At this point, we note that the solution of section \ref{sec:SolutionHenning} is a particular case of the family of solutions above.

A solution to this system containing a non-trivial dilaton is the supersymmetry dyonic string found in \cite{Gueven:2003uw}, which preserves two supercharges.  The metric functions, $\lambda$ and dilaton are given by
    \begin{subequations}
    \begin{align}
        e^{2A} &= (H_{P} H_{Q})^{\frac{1}{2}}, \quad
        e^{2k} = \frac{\kappa^{2}P}{g^{2}r^{6}}H^{-\frac{5}{2}}_{P}H^{\frac{1}{2}}_{Q},\quad
        e^{2g} = \frac{2 \kappa}{g} (H^{-1}_{P} H_{Q})^{\frac{1}{2}},\\
        e^{2h} &= 4P (H^{-1}_{P} H_{Q})^{\frac{1}{2}},\quad
        e^{2\varphi} = \frac{H_{Q}}{H_{P}},\quad
        \lambda = \frac{P}{\kappa}-\frac{1}{2g}.
    \end{align}
    \end{subequations}
where 
    \begin{equation}
        H_{P} = P_{0} + \frac{P}{r^{2}},\quad
        H_{Q} = Q_{0} + \frac{Q}{r^{2}}.
    \end{equation}
Also, the primitive (1,1)-form in $\calG$ is set to zero, so that we have
    \begin{equation}
        \mathcal{G} = \frac{1}{2} dt\wedge dx\wedge d(H^{-1}_{Q}) + \frac{P}{2} L_{1}\wedge L_{2} \wedge L_{3}.
    \end{equation}
In order for this configuration to be a solution of the equations of motion, the following constraint needs to be satisfied 
    \begin{equation}
        g =  \frac{\kappa}{\kappa^{2}+2P}.
    \end{equation}

\section{Constraints on internal spaces for type II embeddings}\label{sec:two}
Our goal in this section is to derive constraints on internal manifolds that allow for embeddings of $d=6$ gauged and un-gauged Einstein-Maxwell supergravity, and their various sub-sectors, into type II supergravity. This will be achieved with spinor bi-linear techniques that follow the general pattern of methods that will be familiar  to those who have delved into the construction and classification of Minkowski and AdS string vacua. 

\subsection{General idea and preliminary details}\label{sec:twop1}
The way in which we will go about constructing uplifts of Einstein-Maxwell gauged supergravity is to use  bi-linear techniques to establish what conditions the internal manifold M$_4$ of such solutions must obey for supersymmetry to be preserved in type II supergravity if it is preserved in $d=6$. Our philosophy throughout will be that the bosonic fields of the $d=10$ background may only depend on the $d=6$ data through the bosonic fields of the $d=6$ theory
\beq
g,~~~{\cal A},~~~{\cal F},~~~ {\cal G},~~~\varphi,~~~g^{(6)}_{\mu\nu},~~~\text{vol}_6.
\eeq
In particular they should not depend on any of the G-structure forms in 6 dimensions
\beq
k,~v,~ J,~~\Omega
\eeq 
or the associated poly-forms $(\psi^{(6)},~\tilde{\psi}^{(6)})$. In this way it should follow that when the the equations of motion and Bianchi identities of type II supergravity are implied for a supersymmetic class of  $d=10$  solutions, they are actually closing on the conditions that imply internal supersymmetry only, with the necessary $d=6$ conditions being \eqref{eq:6dEOM1}-\eqref{eq:6dEOM4} which only requires a solution to hold. In this way our result should apply also to uplifts of non-supersymmetric $d=6$ solutions

We shall begin by assuming that the metric decomposes as
\beq
ds^2= e^{2A}g^{(6)}_{\mu\nu} dx^{\mu}dx^{\nu}+ ds^2(\text{M}_4), \label{eq: 10dmet}
\eeq
where $g^{(6)}_{\mu\nu}$ is the metric of $d=6$ supergravity, and M$_4$ is some internal space which $e^{2A}$ and the $d=10$ dilaton $\Phi$ have support on. There will be two case we need to consider: 1) The metric  \eqref{eq: 10dmet} is a warped product. 2) The metric is a fibre bundle with M$_4$ fibred over the $d=6$ directions in terms of ${\cal A}$. We shall address these in detail in the following sections. In either case the way that we will deal with the embedding of  $\varphi$ into $d=10$ is to essentially treat it as if it where an additional coordinate that $(e^{2A},\Phi)$ and (function but not component-wise) the metric on M$_4$, fibered or otherwise, can depend on. This will allow us to be agnostic about the scalar embedding and allow supersymmetry to decide for us, and it will as we shall see.

We now need to establish what form the NS, $H$, and RR, $F_{\pm}$ fluxes should take. For the RR fluxes we will work with the RR poly-form $F_{\pm}$, \textit{i.e.}
\beq
F_+=F_0+F_2+F_4+F_6+F_8+F_{10},~~~~F_-=F_1+F_3+F_5+F_7+F_9,
\eeq
in IIA and IIB respectively. Note that in objects like $F_{\pm}$ we employ notation such that the upper/lower sign is taken in type IIA/IIB. The poly forms $F_{\pm}$ in general contain twice the degrees of freedom that the type II RR sector should, this is remedied by imposing the self duality constraint
\beq
F_{\pm}=\star\lambda(F_{\pm}), \label{eq:selfduality}
\eeq 
where  $\lambda(C_k)= (-)^{[\frac{k}{2}]}C_k$ for $C_k$ a k-form.  The NS and RR fluxes should obey the Bianchi identities
\beq
d H=0,~~~~~d_H F_{\pm}=0,\label{eq:10dbinanchis}
\eeq
away from possible source terms - note that the EOM of the RR fluxes is implied through \eqref{eq:selfduality}. Of course we also need to solve the other EOMs of type II supergravity to actually have a solution, however we prove that these are implied by \eqref{eq:10dbinanchis} and the $d=6$ EOM when $d=10$ supersymmetry holds in appendix \ref{sec:Integrability}. We believe that  it should also follow that the $d=10$ EOM are implied also for uplifts of $d=6$ solutions that do not preserve supersymmetry that utilise the same internal spaces, but have not proved this. Our reason to believe this is that we impose that the  $d=10$ fields depend on $d=6$ data only through the $d=6$ bosonic fields. Generically in such a scenario the individual terms in the $d=6$ EOM will arise quite naturally from the $d=10$ EOM, what would not generically happen is that the internal $d=4$ data that also appears would be arranged such that the individual $d=6$ terms can close on \eqref{eq:6dEOM1}-\eqref{eq:6dEOM3}. But we already know that this does happen when $d=10$ supersymmetry holds, so should still hold when external (but not internal) supersymmetry is broken.

Given the $d=6$ fields available to us, the most general form that the NS 3-form can possibly take is
\beq
H= H_3+ H_0 {\cal G}+ \tilde{H}_0e^{2\varphi}\star_6{\cal G}+ H_1\wedge {\cal F}+H_2\wedge d\varphi
\eeq
where $(H_0,~\tilde{H}_0)$ will in general have to be constants for $dH=0$ to hold, and $(H_3,H_2,H_1)$ have support on  M$_4$ but can also depend on  $(\varphi,~{\cal A})$. One might think of including $d\varphi\wedge {\cal F}$, but there is no $d=6$ condition that the Hodge dual of this needs to close on for solutions in general, meaning that it gets ruled out by the EOM of $H_3$.

 Due to the self duality constraint, the most general decomposition that could plausibly close on the $d=6$ Bianchi identities and flux and scalar equations of motion of the $d=6$ theory is\footnote{Terms such as ${\cal F}\wedge {\cal F}$, or ${\cal F}\wedge {\cal G}$ can be excluded as their Hodge duals obey no special relation in general.}
\begin{align}
F_{\pm}&=f_{\pm} +e^{2A}{\cal F}\wedge g_{\pm}+ e^{3A}{\cal G}\wedge g_{\mp}+e^{6A}\text{vol}_6\wedge\star_4\lambda(f_{\pm}) -e^{4A}\star_6{\cal F}\wedge \star_4\lambda(g_{\pm})- e^{3A}\star_6{\cal G}\wedge \star_4\lambda(g_{\mp})\nn\\[2mm]
&+ e^{5A}\star_6 d\varphi\wedge h_{\mp}+ e^{A}d\varphi\wedge \star_4\lambda(h_{\mp}),
\end{align}
where $(f_{\pm},g_{\pm},g_{\mp},h_{\mp})$ have support on M$_4$ but we also allow to depend on $d=6$ data through $(\varphi,{\cal A})$.

The final general point about the embedding we need to address is how supersymmetry will be preserved in ten dimensions when it holds in six dimensions: To this end we will decompose the $d=10$ gamma matrices in terms of their analogues in 6 and 4 dimensions as
\begin{align}
\Gamma_{\mu}&=e^{A}\gamma^{(6)}_{\mu}\otimes\hat\gamma^{(4)},~~~~\Gamma_a=\mathbb{I}\otimes \gamma^{(4)}_a,\nn\\[2mm]
\hat\Gamma&=\hat\gamma^{(6)}\otimes \hat\gamma^{(4)},~~~~B=B^{(6)}\otimes B^{(4)},
\end{align}
where $(\hat\Gamma,B)$ are the $d=10$ chirality matrix and intertwiner for $(B^{(4)})^{-1}\gamma_a^{(4)}B^{(4)}=(\gamma^{(4)})^*_a$ and $\hat\gamma^{(4)}=-\gamma_{1234}$. We take the following spinor ansatz
\beq
\epsilon_1=\zeta_-\otimes \chi^1_-+\text{m.c},~~~~\epsilon_2=\zeta_-\otimes \chi^2_{\mp}+\text{m.c}\label{eq:spinoransatz10s}
\eeq
where m.c stands for Majorana conjugate, $\zeta_-$ is the spinor of $d=6$ Einstein-Maxwell supergravity obeying \eqref{eq:BPS1}-\eqref{eq:BPS2} and $(\chi^1_-,\chi^1_{\pm})$ are chiral (with respect to $\hat\gamma^{(4)}$) spinors on M$_4$, the upper/lower signs are again taken in type IIA/IIB.\\
~\\
Having established our embedding ansatz we can now make use of an existing set of geometric conditions that are necessary and sufficient for supersymmetry for general type II solutions \cite{Tomasiello:2011eb}. These are phrased in terms of the two 1-forms $(K,\tilde{K})$ and a $d=10$ poly-form $\Psi_{\pm}$ defined through
\beq
K=\frac{1}{2}(K_1+K_2),~~~\tilde{K}=\frac{1}{2}(K_1-K_2),~~~\slashed{\Psi}_{\pm}=\epsilon_1\otimes \overline{\epsilon}_2,~~~~K_{1,2}=\frac{1}{32}\bar{\epsilon}_{1,2}\Gamma_M\epsilon_{1,2}dX^M,\label{eq:10dformdefs}
\eeq
via the Clifford map. Supersymmetry requires that the following condition on these forms and the type II bosonic fields are obeyed
\begin{subequations}
\begin{align}
&d\tilde{K}=\iota_{K}H,\label{eq:10dBPS1}\\[2mm]
&\nabla^{(10)}_{(M}K_{N)}=0,~~~~{\cal L}_{K}\Phi=0,\label{eq:10dBP2}\\[2mm]
&d_H(e^{-\Phi}\Psi_{\pm})= - (\tilde{K}\wedge+ \iota_{K})F_{\pm}. \label{eq:10dBP3}
\end{align}
\end{subequations}
Note in particular that \eqref{eq:10dBP2} implies that $K^{M}\partial_M$ defines a Killing vector of the metric and $\Phi$, this can be either time-like or null. Further it is possible to show that 
\beq
{\cal L}_KH={\cal L}_KF_{\pm}=0,
\eeq
follows from \eqref{eq:10dBPS1} and \eqref{eq:10dBP3} when the $d=10$ Bianchi identities \eqref{eq:10dbinanchis} are assumed to hold, making $K^M\partial_M$ a symmetry of the entire background.  The conditions \eqref{eq:10dBPS1}-\eqref{eq:10dBP3} are necessary for supersymmetry but are not in general sufficient. In \cite{Tomasiello:2011eb} they are supplemented with an additional two so called ``pairing'' constraints which make the entire system sufficient for supersymmetry. Dealing with these is a rather messy computation which we sketch in appendix \ref{sec:missing10dconds}.

We find for our particular spinor ansatz of \eqref{eq:spinoransatz10s}  that the 1-forms forms in \eqref{eq:10dformdefs} decompose as
\beq
K=-\frac{e^A}{32}(|\chi^1_-|^2+|\chi^2_{\pm}|^2)k,~~~K=-\frac{e^A}{32}(|\chi^1_-|^2-|\chi^2_{\pm}|^2)k,
\eeq
where $k$ is the 1-form dual to the $d=6$ null Killing vector. The first supersymmetry condition we will deal with is that $K$ must be dual to a Killing vector, under our earlier gauge choice $\iota_k{\cal A}=0$, this implies that
\beq
{\cal L}_k A=0,~~~{\cal L}_k ds^2(\text{M}_4)=0~~~~ d(e^{-A}(|\chi^1_-|^2+|\chi^2_{\pm}|^2))=0,\label{eq:condssss1}
\eeq
whether ${\cal A}$ appears in the metric or not. The first of these tells us that $e^{A}$ must be independent of the isometry directions while $ds^2(\text{M}_4)$ must also respect this isometry - note that this does not exclude the possibility of either depending on $\varphi$ as ${\cal L}_k \varphi=0$ is necessary for external supersymmetry. We can solve the second of \eqref{eq:condssss1} by decomposing 
\beq
\chi^1_-=\cos\left(\frac{\beta}{2}\right)e^{\frac{A}{2}}\sqrt{2c}\eta^1_-,~~~\chi^2_{\pm}=\sin\left(\frac{\beta}{2}\right)e^{\frac{A}{2}}\sqrt{2c}\eta^2_{\pm}\label{eq:4dspinor}
\eeq 
where $c$ is a constant and $(\eta^1_{-},\eta^2_{\pm})$ are unit norm, we now have
\begin{align}
K&=-\frac{e^{2A}c}{16}k,~~~~\tilde{K}=-\frac{e^{2A}c}{16}\cos\beta k.
\end{align}
Now, as we already assume that $\Phi$ depends only on the $d=6$ coordinates through $\varphi$, we have solved \eqref{eq:10dBP2}. 
We will next deal with \eqref{eq:10dBP2}: As external supersymmetry demands $(\iota_{k}{\cal F}=0,~\iota_{k}d\varphi=0)$ and we choose a gauge in which $\iota_k {\cal A}=0$ the only terms in $H$ that can contribute are 
\beq
H=H_0{\cal G}+e^{2\varphi}\tilde{H}_0 \star_6{\cal G}+...
\eeq
where $...$ gives zero when acted on by $\iota_k$. We then find through the conditions \eqref{eq:usefulcond6d} and \eqref{eq:bpscond6d3} that \eqref{eq:10dBP2} gives rise to
\beq
d(e^{2A+\varphi}\cos\beta)=e^{2\varphi}\tilde{H}_0 d\varphi,~~~\left(e^{2\varphi}\tilde{H_0}-(H_0+ 2 e^{2A+\varphi}\cos\beta)\right)\iota_k{\cal G}=0, \label{eq:consff}
\eeq
the first of these in general gives a constraint on the internal fields but the second only gives a $d=4$ constraint when $\iota_k{\cal G}\neq 0$. Notice that $\iota_k{\cal G}=0$ is not a necessary condition of either Einstein-Maxwell gauged supergravity or any consistent sub-sector of it (for instance with the tensor or vector multiplet turned off). Instead this is an additional condition one can impose on supersymmetric solutions only, as $k$ requires a Killing spinor to define. As such the second of  \eqref{eq:consff} does not conform to our general uplift philosophy,  we will thus instead imposes the stronger constraint
\beq
\left(e^{2\varphi}\tilde{H_0}-(H_0+ 2 e^{2A+\varphi}\cos\beta)\right) {\cal G}=0,\label{eq:strongercodn}
\eeq 
which while not a general condition for any sub-sector of the $d=6$ theory at least makes sense in the absence of external supersymmetry.

The last condition we must deal with is \eqref{eq:10dBP3}, which is by a considerable margin the most involved, indeed to really make progress with it  we will need to get specific about the precise form of M$_4$, as we will in the following subsections. However we will push a bit further in this section before doing this. The first thing we need to do is compute $\Psi_{\pm}$ which requires us to  introduce some $d=4$ poly-forms $(\psi_{\mp},~\tilde{\psi}_{\mp})$ defined through
\beq
\slashed{\psi}_{\mp}=\chi^1_{-}\otimes \chi^{2\dag}_{\pm},~~~~\tilde{\slashed{\psi}}_{\mp}=\chi^1_{-}\otimes \chi^{2c\dag}_{\pm},\label{eq:4dbispinors}
\eeq
which we will make more explicit in section \ref{sec:internalbilinears}, as the conditions we derive in the  sections that proceed this constrain them somewhat. We find that the $d=10$ poly-form decomposes in terms of $(\psi_{\mp},\tilde{\psi}_{\mp})$ and the $d=6$ bi-linears $(\psi^{(6)}_{-},~\tilde{\psi}^{(6)}_{-})$ as
\beq
\Psi_{\pm}=\mp 2\bigg(e^{A}\psi^{(6)}_1\wedge \text{Re}\psi_{\mp}+ e^{3A}i\psi^{(6)}_{3}\wedge \text{Im}\psi_{\mp}+e^{3A}\text{Re}\big(\tilde{\psi}^{(6)}_-\wedge\tilde{\psi}_{\mp}\big)+e^{5A}\psi^{(6)}_5\wedge \text{Re}\psi_{\mp}\bigg),\label{eq:10dbispinors}
\eeq
which despite initial appearances is actually real, as is clear from \eqref{eq:6dpolyforms}. As we insist that  $(H,F)$ do not depend on $(\psi^{(6)}_{-},~\tilde{\psi}^{(6)}_{-})$, and there is no condition in \eqref{eq:bpscond6d1}-\eqref{eq:bpscond6d6} that would convert the term involving $\tilde{\psi}^{(6)}_{-}$  into something related to the $d=6$ bosonic fields when substituted into \eqref{eq:10dBP3}, we must have that this decouples from the rest, \textit{i.e.}
\beq
d_H\left(e^{3A-\Phi}\tilde{\psi}^{(6)}_-\wedge\tilde{\psi}_{\mp}\right)=0.
\eeq
This is a term that it is possible to make some general statements about: First off one should appreciate that the only terms in $H$ that enter this expression are
\beq
H= H_3+d\varphi\wedge H_2+...
\eeq
as $...$ only contains terms proportional to $({\cal G},\star_6{\cal G},{\cal F})$ which drop out of the above expression through \eqref{eq:wedgerules}, we find that 
\beq
\left(d_{H_3}(e^{3A-\Phi}\tilde{\psi}_{\mp})-e^{3A-\Phi}d\varphi\wedge H_2\wedge \tilde{\psi}_{\mp}+2i g  e^{3A-\Phi} {\cal A}\wedge \tilde{\psi}_{\mp}\right)\wedge \tilde{\psi}^{(6)}_-=0.\label{eq:firstdeq10cond}
\eeq
When $g=0$ there is nothing particularly interesting about this condition, but when $g\neq 0$ (and likewise ${\cal A}$) it cannot be solved without assuming that  M$_4$ contains at least one U(1) isometry direction $\partial_{\phi}$ such that we can decompose
\beq
\psi_{\mp}= (\psi^{(3)}_{\mp}+ e^{C}D\phi\wedge \psi^{(3)}_{\mp}),~~~~\tilde{\psi}_{\mp}= e^{i n \phi}(\tilde{\psi}^{(3)}_{\mp}+ e^{C}D\phi\wedge \tilde{\psi}^{(3)}_{\pm}),~~~~D\phi= d\phi+ p {\cal A}+V,
\eeq
where $(\psi^{(3)}_{\pm},~\psi^{(3)}_{\mp},~\tilde{\psi}^{(3)}_{\mp},~\tilde{\psi}^{(3)}_{\pm},~V,~e^{C})$ are independent of $\phi$. Assuming that there is exactly one U(1) isometry in which ${\cal A}$ is housed, \eqref{eq:firstdeq10cond} is implied by
\beq
d_{H_3}(e^{3A-\Phi}\tilde{\psi}_{\mp})-e^{3A-\Phi}d\varphi\wedge H_2\wedge \tilde{\psi}_{\mp}\bigg\lvert_{{\cal A}\to 0}=0,~~~~2g=n p\label{eq:gencond1},
\eeq
\textit{i.e.} \eqref{eq:firstdeq10cond} decomposes as $\tilde{\psi}^{(6)}_-\wedge $ 4 distinct terms of which 3 are parallel to one of $(D\phi,{\cal A},{\cal F})$ and one is orthogonal to all of these. The ${\cal F}$ term vanishes as ${\cal F}\wedge \tilde{\psi}^{(6)}_-=0$, the ${\cal A}$ term yields $2g=np$ and what remains are two conditions on M$_4$ alone that are equivalent to the first expression in  \eqref{eq:gencond1}.   It follows from this that $(\chi^1_{\mp},\chi^2_{\mp})$ are charged under the U(1) isometry of $\partial_{\phi}$ when $g\neq 0$.

The remaining terms in the decomposition of $\Psi_{\pm}$ are more complicated as they can mix in \eqref{eq:10dBP3} through the conditions the $d=6$ bi-linears and bosonic fields must obey when external supersymmetry holds. Our ansatz for $F_{\pm}$ leads to the terms appearing in the right hand side of \eqref{eq:10dBP3} decomposing as 
\begin{align}
\iota_KF_{\pm}&=-\frac{c}{2}\bigg[\frac{e^{3A}}{2}\left(-e^{-\varphi} d\varphi \wedge \psi^{(6)}_1\wedge( g_{\mp}-\star_4\lambda(g_{\mp}))+e^{-\varphi}d\psi^{(6)}_1\wedge( g_{\mp}+\star_4\lambda(g_{\mp}))\right)\\[2mm]
&+ e^{6A}\psi^{(6)}_5\wedge\star_4\lambda(f_{\pm}) -e^{4A}\left({\cal F}\wedge \psi^{(6)}_1-i g e^{-\varphi}\psi^{(6)}_3\right)\wedge \star_4\lambda(g_{\pm})- e^{5A+\varphi}\psi^{(6)}_1\wedge({\cal G}+\star_6{\cal G})\wedge h_{\mp}\bigg],\nn\\[2mm]
\tilde{K}\wedge F_{\pm}&=\frac{c}{2}e^{2A}\cos\beta\psi^{(6)}_1\wedge \bigg[ f_{\pm} +e^{2A}{\cal F}\wedge g_{\pm}+ e^{3A}{\cal G}\wedge g_{\mp}- e^{3A}\star_6{\cal G}\wedge \star_4\lambda(g_{\mp})+e^{A}d\varphi\wedge \star_4\lambda(h_{\mp})\bigg]\label{eq:fluxbit}.
\end{align}
Using this and \eqref{eq:bpscond6d1}-\eqref{eq:bpscond6d6} one must then expand  \eqref{eq:10dBP3} in a basis of $d=6$ forms that are generically independent from each other, wedged with expressions involving the 4d bi-linears, 4d fields and $\varphi$. We will solve \eqref{eq:10dBP3} by setting these $d=4$ conditions to zero yielding constraints on our $d=10$ embedding that lift $d=6$ supersymmetry to type II supergravity. Generically such a basis of $d=6$ forms is given by
\begin{align}
\tilde{\psi}^{(6)}_-,~~~{\cal A}\wedge \tilde {\psi}^{(6)}_-,~~~\psi^{(6)}_1,~~~d\psi^{(6)}_1,~~~\psi^{(6)}_1\wedge {\cal F},~~~~\psi^{(6)}_3~~~~~{\cal G}\wedge {\tilde \psi}^{(6)}_1,~~~\star_6{\cal G}\wedge {\tilde \psi}^{(6)}_1,~~~{\tilde \psi}^{(6)}_5.
\end{align}
However some terms, such as $\psi^{(6)}_1\wedge {\cal F}$, only appear when certain multiplets are turned on, yet others such as $d\psi^{(6)}_1$ may be zero on specific solutions. We will keep track of what $d=6$ multiplets are turned on but ignore possibilities like $d\psi^{(6)}_1=0$ which don't make sense as constraints on non-supersymmetric $d=6$ solutions. Specifically what this means is that for certain classes of \textbf{supersymmetric} $d=6$ solutions we may be imposing non-necessary constraints on M$_4$. The constraints we derive will be necessary for consistent truncations to Einstein-Maxwell gauged supergravity, its consistent sub-sectors (\textit{i.e.} without tenor or vector multiplets or both) and their respective limits with $g=0$.

While the precise details depend on what multiplets are non-trivial and whether $g=0$ or not, it turns out that the main distinction comes from whether or not ${\cal A}$ appears in the metric when the vector muliplet is turned on.
We begin our detailed analysis in section \ref{sec:case1} and \ref{sec:case2} by constructing condition for internal supersymmetry that will sever as constraints on internal manifolds that provide consistent truncations to either $d=6$ Einstein-Maxwell supergravity or one of its well defined sub-sectors. In particular that means that we cannot assume that ${\cal G}=0$ and must solve \eqref{eq:strongercodn} as
\beq
e^{2\varphi}\tilde{H}_0-H_0-2e^{2A+\varphi}\cos\beta=0,
\eeq 
we need to consider both the case where ${\cal A}$ does not (section \ref{sec:case1}) and does (section \ref{sec:case2}) appear in the internal 4-manifold. In section \ref{sec:case3}  we will make the assumption that ${\cal G}=0$. The conditions we derive here will not provide internal manifolds for consistent truncations, but will provide uplift formulae for restricted $d=6$ solutions obeying
\beq
{\cal G}=0,~~~~\varphi=0,~~~~{\cal F}\wedge {\cal F}=0,~~~~{\cal F}\wedge \star_6{\cal F}=g^2\text{vol}_6. \label{eq:bigconstriant}
\eeq
The reasons for doing so is two-fold: First this is necessary for recovering general conditions for Mink$_6$ vacua, second there are actually many important solutions which obey the constraint \eqref{eq:bigconstriant}, an important one being the Mink$_4\times$S$^2$ Salam-Sezgin solution \cite{Salam:1984cj}.

\subsection{\texorpdfstring{${\cal A}$}{A} not in metric: The strictly un-gauged case}\label{sec:case1}
When ${\cal A}$ does not appear as part of the metric on M$_4$ the condition \eqref{eq:firstdeq10cond} leads to 
\beq
g=0
\eeq
so we are strictly considering uplifts of un-gauged $d=6$ Einstein-Maxwell supergravity.  We take the following ansatz for the fluxes
\begin{align}
H&=H_3+H_1\wedge {\cal F}+H_0 {\cal G}+e^{2\varphi}\tilde{H}_0 \star_6{\cal G}+d\varphi\wedge H_2,\nn\\[2mm]
F_{\pm}&=\left(1+\star\lambda\right)\left(f_{\pm} +e^{2A}{\cal F}\wedge g_{\pm}+ e^{3A}{\cal G}\wedge g_{\mp}+ e^{5A}\star_6 d\varphi\wedge h_{\mp}\right),\label{eq:fluxansatz}
\end{align}
where $(H_0,\tilde{H}_0)$ must be constant for $dH=0$ to hold, $(e^{A},~H_3,~H_2,~H_1,~f_{\pm},~g_{\pm},~g_{\mp},~h_{\mp})$ have support on $\text{M}_4$ and like wise the $d=10$ dilaton $\Phi$. We allow all the internal fields to also depend on the $d=6$ dilaton $\varphi$, however the only dependence on $d\varphi$ is written explicitly. 

We find that necessary and sufficient conditions for internal supersymmetry in the presence of non-trivial gravity, tensor and vector multiplets are given by the following general constraints
\begin{subequations}
\begin{align}
&e^{2\varphi}\tilde{H}_0-H_0-2e^{2A+\varphi}\cos\beta=0,\label{eq:sugbps1}\\[2mm]
&d_{H_3}(e^{3A-\Phi}\text{Im}\psi_{\mp})-d\varphi\wedge H_2\wedge \text{Im}\psi_{\mp}=0,\label{eq:sugbps2}\\[2mm]
&d_{H_3}(e^{3A-\Phi}\tilde{\psi}_{\mp})- e^{3A-\Phi}H_2\wedge d\varphi \wedge \tilde{\psi}_{\mp}=0\label{eq:sugbps3},\\[2mm]
&\frac{c}{8}e^{3A-\varphi}(1+\star_4\lambda)g_{\mp}=\mp e^{A-\Phi}\text{Re}\psi_{\mp},\label{eq:sugbps4}\\[2mm]
&d_{H_3}(e^{5A-\Phi}\text{Re}\psi_{\mp})\mp \frac{c}{4}e^{6A}\star_4\lambda(f_{\pm})\bigg\lvert_{d\varphi\rightarrow 0}=0,\label{eq:sugbps5}\\[2mm]
&d_{H_3}(e^{A-\Phi}\text{Re}\psi_{\mp})-d\varphi\wedge \bigg[\pm \frac{c}{8}e^{3A-\varphi}(1-\star_4\lambda)g_{\mp}\nn\\[2mm]
&\mp \frac{c}{4}e^{3A}\cos\beta \star_4\lambda(h_{\mp})+ e^{A-\Phi} H_2 \wedge \text{Re}\psi_{\pm}\bigg]=\mp \frac{c}{4} e^{2A}\cos\beta f_{\pm},\label{eq:sugbps6}
\end{align}
\end{subequations}
where one needs to fix $d\varphi\to 0$ in \eqref{eq:sugbps5} because this term arises from \eqref{eq:10dBP3} in the form \eqref{eq:sugbps5}$\wedge \psi^{(6)}_5$ and  $d\varphi\wedge \psi^{(6)}_5=0$ is a consequence of external supersymmetry. Note that these conditions are independent of $(H_1,~g_{\pm})$ which couples to the vector multiplet through ${\cal F}$ in \eqref{eq:fluxansatz}, so the same conditions hold in the absence of the vector multiplet. The effect of turning off the tensor multiplet amounts to tuning the 4d fields and $\varphi$ in the above expressions as
\begin{align}
\textbf{No tensor:}&~~~\Rightarrow~~~~ \tilde{H}_0=\varphi= H_2=h_{\mp}=(1-\star_4\lambda)g_{\mp}=0 \label{eq:novecten}.
\end{align}
In addition to the general constraints we also find an addition 3 constraints that should only be applied when one or both of the tensor and vector multiplets are non-trivial, namely .
\begin{subequations}
\begin{align}
\textbf{Tensor}:&~~~\frac{c}{4}e^{5A-\varphi}\left(h_{\mp}- e^{-\varphi}\cos\beta \star_4\lambda(g_{\mp})\right)=\pm e^{A-\Phi}\tilde{H}_0 \text{Re}\psi_{\mp}\label{eq:sugbps7},\\[2mm]
&~~~\partial_{\varphi}(e^{2A}\sin\beta)=0,~~~~ \partial_{\varphi}(e^{4A-2\Phi}\sqrt{\det g^{(4)}})=0, \label{eq:sugbpsparing}\\[2mm]
\textbf{Vector}:&~~~\frac{c}{4}e^{4A}(\cos\beta+\star_4\lambda)g_{\pm}=\pm e^{A-\Phi}H_1\wedge \text{Re}\psi_{\mp}\label{eq:sugbps8},
\end{align}
\end{subequations}
where only \eqref{eq:sugbps8} contains $(g_{\pm}, H_1)$ and in \eqref{eq:sugbpsparing} $g^{(4)}$ is the metric on M$_4$. Note that \eqref{eq:sugbps7} and \eqref{eq:sugbps8} follow from the parts of \eqref{eq:10dBP3} that appear wedged with $k\wedge({\cal G}+\star_6{\cal G})$ and ${\cal F}$ respectively, while \eqref{eq:sugbpsparing} implies the pairing constraints. This means that strictly speaking, due to \eqref{eq:wedgerules}, the tensor multiplet conditions \eqref{eq:sugbps7} only need be imposed when $d\varphi \neq 0$ - however while one can derive embeddings for solutions with $d\varphi = 0$ without imposing \eqref{eq:sugbps7}, they do not define consistent truncations\footnote{\textit{i.e.} a consistent truncation should be a truncation to the bosonic part of a self consistent 6d theory, which   setting to zero one part of the tensor multiplet without the other is not.} unless we also fix ${\cal G}+\star_6{\cal G}=0$. \\

With necessary and sufficient conditions for supersymmetry in hand we can now study the Bianchi identities of the fluxes. We assume that the $d=6$ Bianchi identities and equations of motion hold and derive what $d=4$ conditions imply
\beq
dH=0,~~~d_HF_{\pm}=0,\label{eq:10dBIS}
\eeq
which should hold away from sources, given this assumption. First off for $dH=0$  to hold we require (in addition to $(H_0,\tilde{H}_0)$ being constant) that
\beq
dH_3=d\varphi\wedge dH_2,~~~dH_1=0,~~~~ H_0 {\cal F}\wedge {\cal F}=0,\label{eq:NSbisnotmet}
\eeq
which in particular means that we must have either  $H_0=0$ or ${\cal F}\wedge {\cal F}=0$,  the latter of which obviously holds when there is no vector multiplet but also for on shell solutions of the 6d theory without a tensor multiplet\footnote{This follow from the consistency of the $d=6$ conditions $d(e^{2\varphi}\star_6{\cal G})=0$ and $d{\cal G}={\cal F}\wedge {\cal F}$ with $\varphi=0$ and $\star_6{\cal G}=-{\cal G}$.}. We thus have that
\beq
\textbf{Tensor + Vector}~~~~\Rightarrow~~~~ H_0=0.
\eeq
We should also have that $d_H F_{\pm}=0$ which branches into many distinct conditions on the 4 dimensional fields in general. However, though a long computation, it is possible to show that the vast majority of these are implied by the geometric conditions for supersymmetry and \eqref{eq:NSbisnotmet}  when a small sub-set of these $d=4$ conditions are assumed to hold. In general we find that away from sources it is necessary to impose
\beq
d_{H_3}f_{\pm}-d\varphi\wedge (H_2\wedge f_{\pm}\pm d_{H_3}\star_4\lambda(h_{\pm}))=0,\label{eq:fpmbi1}
\eeq
which follows from the part of the Bianchi identity along vol$_6$, 
while we get additional conditions that depend on exactly what multiplets are turned on in addition to the gravitational one, namely
\begin{subequations}
\begin{align}
\textbf{Tensor}:~~& d_{H_3}(e^{3A}g_{\mp})+H_0 f_{\pm}+ d\varphi\wedge(e^{A}H_0\star_4\lambda(h_{\mp})-e^{3A}H_2\wedge g_{\mp})=0,\label{bissngtensor}\\[2mm]
\textbf{Vector}:~~& d_{H_3}(e^{2A}g_{\pm})-H_1\wedge f_{\pm}-d\varphi\wedge\left(e^{A}H_1\wedge \star_4\lambda(h_{\mp})+e^{2A} H_2\wedge g_{\pm}\right)=0,\label{bissngvector}
\end{align}
\end{subequations}  
which respectively follow from the Bianchi identity along ${\cal G}$ and ${\cal F}$. These conditions must be imposed whenever the respective multiplet is turned on - \textit{i.e.} when the tensor multiplet is turned off \eqref{bissngtensor} combines with a condition along $\star_6 {\cal G}$ with the result being implied due to \eqref{eq:sugbps4}. When both muliplets are turned on \eqref{bissngtensor}-\eqref{bissngvector} still hold, but one gets an additional constraint from the ${\cal F}\wedge {\cal F}$ term in $d_{H}F_{\pm}$, namely we must have
\beq
\textbf{Tensor + Vector}:~~ e^{A}g_{\mp}= H_1\wedge g_{\pm}.
\eeq
One can show that this conditions actually implies \eqref{bissngtensor}, but not \eqref{bissngvector}. Finally it also possible to show that $\cos\beta$\eqref{eq:fpmbi1} is implied in general, but this only implies  \eqref{eq:fpmbi1} when $\cos\beta \neq 0$.

\subsection{\texorpdfstring{${\cal A}$}{A} in metric: The gauge compatible case}\label{sec:case2}
We now consider the case where  ${\cal A}$ does appear in the metric, which is compatible with 
\beq
g\neq 0
\eeq
although does not required it and indeed, as far as the conditions for supersymmetry that we will present are concerned, the $g\to 0$ limit is not problematic. As it is a U(1) gauge field, embedding ${\cal A}$ inside the internal metric  requires us to assume that it contains at least 1 U(1) isometry direction $\partial_{\phi}$. As such  the internal spaces  decomposes as U(1) $\hookrightarrow \text{M}_4\rightarrow \text{M}_3$ and the vector field should appear as a connection term which fibers M$_4$ over the $d=6$ directions as
\beq
ds^2(\text{M}_4)=ds^2(\text{M}_3)+e^{2C}D\phi^2,~~~~ D\phi=d\phi+p {\cal A}+ V
\eeq 
where we will need to take $p$ to be a constant and $(e^{A},~e^C,~V)$ and the dilaton now have support on M$_3$, but can  additionally have dependence on $\varphi$. One could of course assume that M$_4$ contains additional U(1) directions that also house ${\cal A}$ in this fashion, indeed \cite{Cvetic:2003xr} contains an uplift with ${\cal A}$ appearing in two distinct U(1) directions, albeit with a non-compact internal space. We will not consider this possibility here however, primarily because it more greatly constrains the space of possible internal manifolds. 

We will again formally decompose the NS and RR fluxes as in \eqref{eq:fluxansatz}, though one must appreciate that every internal flux term, and like wise the internal bi-linears $(\psi_{\mp},~\tilde{\psi}_{\mp})$ can have a portion along $D\phi$ and a portion orthogonal to it, \textit{i.e.} if in the previous section we had a $k$-form $C_k$ this now takes the form
\beq
C_k= C^{(3)}_k+  e^C D\phi\wedge C^{(3)}_{k-1},\label{eq:gendecomp}
\eeq
where generically $(C^{(3)}_k,~C^{(3)}_{k-1})$ are independent of $\phi$, the point being that now ${\cal F}$ can be generated from $dC_k$. The one exception to the decomposition of \eqref{eq:gendecomp} is $\tilde{\psi}_{\mp}$ as we need to allow this to be charged under $\partial_{\phi}$ to end up with $g\neq 0$. Specifically we will take
\beq
\tilde{\psi}_{\mp}= e^{i n \phi}(\tilde{\psi}^{{(3)}}_{\mp}+ e^{C}D\phi\wedge \tilde{\psi}^{(3)}_{\pm}),
\eeq 
for $n$ some constant. Now one needs to perform a long and tedious computation to extract the conditions that imply supersymmetry from \eqref{eq:10dBP3}, we will omit the details.

We find that necessary and sufficient conditions for internal supersymmetry in the presence of non-trivial gravity, tensor and vector multiplets are the following general conditions
\begin{subequations} 
\begin{align}
&2g=n p,\label{eq:gbps0}\\[2mm]
&e^{2\varphi}\tilde{H}_0-H_0-2e^{2A+\varphi}\cos\beta=0,\label{eq:gbps1}\\[2mm]
&d_{H_3}(e^{3A-\Phi}\text{Im}\psi_{\mp})-d\varphi\wedge H_2\wedge \text{Im}\psi_{\mp}\mp \frac{c}{4}g e^{4A-\varphi}\star_4\lambda(g_{\pm})\bigg\lvert_{{\cal A}\to 0}=0 ,\label{eq:gbps2}\\[2mm]
&d_{H_3}(e^{3A-\Phi}\tilde{\psi}_{\mp})- e^{3A-\Phi}H_2\wedge d\varphi \wedge \tilde{\psi}_{\mp}\bigg\lvert_{{\cal A} \to 0}=0\label{eq:gbps3},\\[2mm]
&\frac{c}{8}e^{3A-\varphi}(1+\star_4\lambda)g_{\mp}=\mp e^{A-\Phi}\text{Re}\psi_{\mp},\label{eq:gbps4}\\[2mm]
&d_{H_3}(e^{5A-\Phi}\text{Re}\psi_{\mp})\mp \frac{c}{4}e^{6A}\star_4\lambda(f_{\pm})+ ge^{3A-\Phi-\varphi}\left(H_1\wedge \text{Im}\psi_{\pm}-p\iota_{\partial_{\phi}}\text{Im}\psi_{\pm}\right)\bigg\lvert_{(d\varphi,~{\calA})\rightarrow 0}=0,\label{eq:gbps5}\\[2mm]
&\bigg[d_{H_3}(e^{A-\Phi}\text{Re}\psi_{\mp})-d\varphi\wedge \bigg(\pm \frac{c}{8}e^{3A-\varphi}(1-\star_4\lambda)g_{\mp}\nn\\[2mm]
&\mp \frac{c}{4}e^{3A}\cos\beta \star_4\lambda(h_{\mp})+ e^{A-\Phi} H_2 \wedge \text{Re}\psi_{\pm}\bigg)\pm \frac{c}{4} e^{2A}\cos\beta f_{\pm}\bigg]\bigg\lvert_{{\cal A}\to 0}=0,\label{eq:gbps6}
\end{align}
\end{subequations}
and the following that should only be imposed when the tensor and/or vector multiplets are non-trivial
\begin{subequations}
\begin{align}
\textbf{Tensor}:~~~&\frac{c}{4}e^{5A-\varphi}\left(h_{\mp}- e^{-\varphi}\cos\beta \star_4\lambda(g_{\mp})\right)=\pm e^{A-\Phi}\tilde{H}_0 \text{Re}\psi_{\mp}\label{eq:gbps7}\\[2mm]
~~~&\partial_{\varphi}(e^{2A}\sin\beta)=0,~~~~\partial_{\varphi}(e^{4A+C-2\Phi}\sqrt{\det g^{(3)}})=0,\label{eq:gbpspairing}\\[2mm]
\textbf{Vector}:~~~&\frac{c}{4}e^{4A}(\cos\beta+\star\lambda)g_{\pm}=\pm e^{A-\Phi}\left(H_1\wedge \text{Re}\psi_{\mp}-p \iota_{\partial_{\phi}}\text{Re}\psi_{\mp}\right)\label{eq:gbps8},
\end{align}
\end{subequations}
where $g^{(3)}$ is the metric on M$_3$. Note that as in the previous section, strictly speaking, one only needs to impose \eqref{eq:gbps7} when $d\varphi \neq 0$, but if we do not impose \eqref{eq:gbps7} for solutions with $d\varphi=0$ the resulting embeddings would not define consistent truncations unless ${\cal G}=-\star_6{\cal G}$ is also imposed.   Apart from the need to send ${\cal A}\to 0$ in conditions containing an exterior derivative\footnote{This is purely a presentational device: As every form on M$_4$ decomposes as in \eqref{eq:gendecomp} we have that each condition on M$_4$ that \eqref{eq:10dBP3} implies has a component parallel and orthogonal to $D\phi$ that define conditions on M$_3$ alone. These conditions on M$_3$ are implied by what we write explicitly if ${\cal A}$ is sent to zero at the end of each computation.}, the above conditions are modified with respect to those of the previous section with additional  $g$ dependent terms and interior products with the Killing vector $\partial_{\phi}$ which follow from the ${\cal F}$ terms generated from $dD\phi$ when expanding out \eqref{eq:10dBP3} and collecting the terms that appear wedged with common 6d forms. It no longer makes much sense to turn off the vector multiplet, as this would lead to $g=0$ and a restricted form of the conditions of the previous section, however the effect of turning off the tensor multiplet is to again tune the fields in the above expression as in \eqref{eq:novecten}. Finally we note that \eqref{eq:gbps0} implies that $\tilde{\psi}_{\mp}$ must indeed be charged under $\partial_{\phi}$ to have $g\neq 0$ as claimed earlier, but we also note that a perfectly well defined $g\to 0$ limit exists in \eqref{eq:gbps0}-\eqref{eq:gbps8}, it just demands that $n=0$, so that in this case  $\tilde{\psi}_{\mp}$ is not charged under $\partial_{\phi}$. In the following sections we will fix the constant $p$ as follows
\begin{align}
(g=0,~p=1)&,~~~~\Rightarrow~~~n=0,\nn\\[2mm]
(g\neq 0,~ p=2g)&~~~~\Rightarrow ~~~n=1
\end{align}
without loss of generality, however it will be convenient to keep it arbitrary for now.

We now once more turn our attention to the Bianchi identities of the 10 dimensional fluxes in \eqref{eq:10dBIS}. First off $dH=0$ now demands that we fix 
\begin{align}
&dH_0=0,~~~d\tilde{H}_0=0,~~~(H_0+p\iota_{\partial_{\phi}}H_1){\cal F}\wedge {\cal F}=0,\nn\\[2mm]
&dH_1+p\iota_{\partial_{\phi}}(H_3+d\varphi\wedge H_2)\bigg\lvert_{{\cal A}\to 0}=0,\nn\\[2mm]
&dH_3-d\varphi\wedge dH_2\bigg\lvert_{{\cal A}\to 0}=0 ,\label{eq:NSbismet} 
\end{align}
where we note that the last of these conditions can only be non-trivial when $d\varphi \neq 0$ and we again have a condition that only holds when only both the tensor and vector multiplets are non-trivial, this time 
\beq
\textbf{Tensor + Vector}~~~~\Rightarrow~~~~ H_0=-p\iota_{\partial_{\phi}}H_1.
\eeq
Again we should also impose $d_H F_{\pm}=0$ away from sources and as before when supersymmetry and \eqref{eq:NSbismet} hold most of the 4d conditions that follow from this are implied by a small sub-set. We find again that one must always impose
\beq
d_{H_3}f_{\pm}-d\varphi\wedge (H_2\wedge f_{\pm}\pm d_{H_3}\star_4\lambda(h_{\pm}))\bigg\lvert_{{\cal A}\to 0}=0,
\eeq
while we again get additional conditions that depend on exactly which non-minimal multiplets are turned on, namely we have when either tensor or vector multiplets are non-trivial that
\begin{subequations}
\begin{align}
\textbf{Tensor}:~~& d_{H_3}(e^{3A}g_{\mp})+H_0 f_{\pm}+ d\varphi\wedge(e^{A}H_0\star_4\lambda(h_{\mp})-e^{3A}H\wedge g_{\mp})\bigg\lvert_{{\cal A}\to 0}=0,\label{bissngtensorcalAinmet}\\[2mm]
\textbf{Vector}:~~& d_{H_3}(e^{2A}g_{\pm})-(H_1\wedge -p\iota_{\partial_{\phi}}) f_{\pm}-d\varphi\wedge\left(e^{A}(H_1\wedge -p\iota_{\partial_{\phi}}) \star_4\lambda(h_{\mp})+e^{2A} H_2\wedge g_{\pm}\right)\bigg\lvert_{{\cal A}\to 0}=0.\label{bissngvectorcalAinmet}
\end{align}
\end{subequations}  
But when both are turned on simultaneously there is an additional term following from the ${\cal F}\wedge {\cal F}$ term in $d_{H}F_{\pm}$, namely we have
\beq
\textbf{Tensor + Vector}:~~ e^{3A}g_{\mp}= e^{2A}(H_1\wedge-p\iota_{\partial_{\phi}}) g_{\pm}.
\eeq
As before this implies \eqref{bissngtensorcalAinmet}, but not \eqref{bissngvectorcalAinmet}.

\subsection{Uplift formulae with \texorpdfstring{${\cal G}=\varphi=0$}{G=varphi=0} }\label{sec:case3}
In this section we consider the special case of solutions in which we fix
\beq
{\cal G}=\varphi=0,\label{eq:commentthing}
\eeq
this  means we are talking about a restricted class of solutions within the minimal theory coupled to a vector multiplet only. 
We will simply present conditions for internal supersymmetry when ${\cal A}$ appears in the metric explicitly, the case without ${\cal A}$ in the metric can be extracted from these by setting $g=0$ and the terms with $\iota_{\partial_{\phi}}$ acting on them to zero in what we do present -  ${\cal A}\to 0$ when it appears no longer does anything. This time we will take the ansatz
\begin{align}
H&=H_3+H_1\wedge {\cal F},\nn\\[2mm]
F_{\pm}&=\left(1+\star\lambda\right)\left(f_{\pm} +e^{2A}{\cal F}\wedge g_{\pm}\right),\label{eq:fluxansatz3}
\end{align}
for the fluxes. From which it follows that necessary and sufficient conditions for supersymmetry are the following
\begin{subequations}
\begin{align}
&2g=n p,\label{eq:gbps0s}\\[2mm]
&d(e^{2A}\cos\beta)=0\label{eq:gbps1s}\\[2mm]
&d_{H_3}(e^{3A-\Phi}\text{Im}\psi_{\mp})-\mp \frac{c}{4}g e^{4A}\star_4\lambda(g_{\pm})\bigg\lvert_{{\cal A}\to 0}=0 ,\label{eq:gbps2s}\\[2mm]
&d_{H_3}(e^{3A-\Phi}\tilde{\psi}_{\mp})\bigg\lvert_{{\cal A}\to 0}=0\label{eq:gbps3s},\\[2mm]
&d_{H_3}(e^{5A-\Phi}\text{Re}\psi_{\mp})\mp \frac{c}{4}e^{6A}\star_4\lambda(f_{\pm})+ ge^{3A-\Phi-\varphi}(H_1\wedge -p\iota_{\partial_{\phi}})\text{Im}\psi_{\pm}\bigg\lvert_{{\cal A}\to 0}=0,\label{eq:gbps4s}\\[2mm]
&d_{H_3}(e^{A-\Phi}\text{Re}\psi_{\mp})\pm \frac{c}{4} e^{2A}\cos\beta f_{\pm}\bigg\lvert_{{\cal A}\to 0}=0,\label{eq:gbps5s}\\[2mm]
&\frac{c}{4}e^{4A}(\cos\beta+\star\lambda)g_{\pm}=\pm e^{A-\Phi}(H_1\wedge-p \iota_{\partial_{\phi}})\text{Re}\psi_{\mp},\label{eq:gbps6s}.
\end{align}
\end{subequations}
This time it is possible to show that imposing $dH=0$ amounts to imposing
\beq
dH_1+p\iota_{\partial_{\phi}}(H_3+d\varphi\wedge H_2)\bigg\lvert_{{\cal A}\to 0}=0,~~~~dH_3-d\varphi\wedge \tilde{d}H_2\bigg\lvert_{{\cal A}\to 0}=0 ,\label{eq:NSbismetg}
\eeq
and when these and \eqref{eq:gbps0s}-\eqref{eq:gbps6s} are assumed to hold then $d_HF_{\pm}=0$ is implied by 
\beq
d_{H_3}f_{\pm}\bigg\lvert_{{\cal A}\to 0}=0,~~~~ d_{H_3}(e^{2A}g_{\pm})-e^{A-\Phi}(H_1\wedge-p \iota_{\partial_{\phi}}) f_{\pm}\bigg\lvert_{{\cal A}\to 0}=0.
\eeq
Note that it is also possible to show that
\beq
\cos\beta d_{H_3}f_{\pm}\bigg\lvert_{{\cal A}\to 0}=0,
\eeq
but $\cos\beta =0$ is possible meaning that $d_{H_3}f_{\pm}\bigg\lvert_{{\cal A}\to 0}=0$ is not in general implied. 

Ultimately the only difference between the internal spaces defined by this class which do not define a consistent truncations to a $d=6$ supergravity, and the result of turning off the tensor multiplet in those of the previous section which do define consistent truncations, is that fixing the NS flux as in \eqref{eq:fluxansatz3} here does not require $\cos\beta=0$.

\subsection{Parametrising the internal bi-linears}\label{sec:internalbilinears}
In this section we will present a parametrisation of the internal bi-linears $(\psi_{\pm},~\tilde{\psi}_{\mp})$ which appear in the conditions for internal supersymmetry.\\
~\\
The bi-linears $(\psi_{\pm},~\tilde{\psi}_{\mp})$ are defined in terms of a pair of chiral spinors in $d=4$ $(\chi^1_-,~\chi^2_{\mp})$ as in \eqref{eq:4dbispinors}. They must in general obey a constraint which allows them to be decomposed in terms of unit norm 4d spinors $(\eta^1_{-},\eta^2_{\pm})$ as in \eqref{eq:4dspinor}. 

In type IIA the spinors have opposite chirality and, as explained at length in section 3 of \cite{Apruzzi:2014qva}, define an identity-structure spanned by two complex vielbein components $(U,W)$ that span M$_4$ with orientation such that 
\beq
\text{vol}(\text{M}_4)= \text{Re}U\wedge \text{Im}U\wedge \text{Re}W\wedge \text{Im}W. 
\eeq
Following  \cite{Apruzzi:2014qva} we have that 
\begin{align}
\slashed{\psi}^0_-&=\eta_-\otimes \eta^{\dag}_+~~~\Rightarrow~~~\psi^0_-= \frac{1}{4}\overline{U}\wedge e^{\frac{1}{2}W\wedge \overline{W}},\nn\\[2mm]
\tilde{\slashed{\psi}}^0_-&=\eta_-\otimes \eta^{\dag c}_+~~~\Rightarrow~~~\tilde{\psi}^0_-= \frac{1}{4}W\wedge e^{-\frac{1}{2}U\wedge \overline{U}}
\end{align}
where $(\psi_-,~\tilde\psi_-)= e^{A}c \sin\beta( \psi^0_-,~\tilde \psi^0_-)$.

In type IIB the spinors have the same chirality and define an SU(2)-structure on the internal space. One can decompose $(\eta_-^1,~\eta^2_-)$ in a basis of one unit norm spinor $\eta_-$ as
\beq
\eta^1_-=\eta_-,~~~~\eta^2_-=a \eta_-+\frac{b}{2}\overline{W}\eta_-,~~~|a|^2+b^2=1.
\eeq
Again following \cite{Apruzzi:2014qva} we see that we can work in conventions such that
\begin{align}
\slashed{\psi}^0_+&=\eta_-\otimes \eta^{\dag}_-~~~\Rightarrow~~~\psi^0_-= \frac{1}{4}e^{-i j},\nn\\[2mm]
\tilde{\slashed{\psi}}^0_+&=\eta_-\otimes \eta^{\dag c}_+~~~\Rightarrow~~~\tilde{\psi}^0_-= -\frac{1}{4}\omega
\end{align}
where $(j,\omega)$ are SU(2) structure forms which decompose in terms on $(U,~W)$ as
\beq
j=\frac{i}{2}\bigg(-U\wedge\overline{U}+W\wedge \overline{W}\bigg),~~~~\omega=\overline{U}\wedge W.\label{eq:j2omega2decomp}
\eeq
In isolation this would imply that our 4d bi-linears in type IIB take the form
\beq
\psi_+=\frac{c}{4}e^{A}\sin\beta\left(\overline{a}e^{-i j}+b \omega\right),~~~\tilde{\psi}_+=\frac{c}{4}e^{A}\sin\beta\left(b e^{-i j}-a \omega\right),
\eeq
however it is possible to establish that one can fix
\beq
b=0,
\eeq
without loss of generality for the classes of solution that we consider. The reason for this is actually different depending on whether $g=0$ or $g \neq 0$. When $g\neq 0$ both \eqref{eq:gbps3} and \eqref{eq:gbps3s} contain the term
\beq
d(e^{4A-\Phi}\sin\beta e^{i n\psi} b)=0,
\eeq
which, given that $g\neq 0$ implies $n\neq 0$, and that we can not set $\sin\beta=0$ without turning off the RR flux means we must fix $b=0$. Conversely if $g=0$ we  have $n=0$, but then \eqref{eq:sugbps2}-\eqref{eq:sugbps3} contain the terms
\beq
d(e^{4A-\Phi}\sin\beta  b)=0,~~~~d(e^{4A-\Phi}\sin\beta  a_2)=0,
\eeq
which if we parametrise $(b=\rho \sin\theta,~a_2=\rho \cos\theta)$ fixes $d\theta=0$.  One can then effectively fix $\theta=0$ within the $d=10$ bi-linear $\Psi_{\pm}$ with separate frame rotations on the  external and internal vielbein directions. Thus we can without loss of generality fix $b=0$ which makes $a$ simply a phase, we can  also then simply send  $a\omega_2\to \omega_2$ in \eqref{bispinors4D IIB}  which further simplifies our IIB bi-linears.

In summary, for the classes of solution we consider in this work, the internal bi-linears can be parameterised as
\begin{subequations}
\begin{align}
\psi_-&=\frac{c}{4}e^{A}\sin\beta \overline{U}\wedge e^{\frac{1}{2}W\wedge \overline{W}},~~~\tilde{\psi}_-=\frac{c}{4}e^{A}\sin\beta W\wedge e^{-\frac{1}{2}U\wedge \overline{U}},\label{bispinors4D IIA}\\[2mm]
\psi_+&=\frac{c}{4}e^{A}\sin\beta\overline{a}e^{-i j},~~~\tilde{\psi}_+=-\frac{c}{4}e^{A}\sin\beta \omega\label{bispinors4D IIB},
\end{align}
\end{subequations}
where $a=a_1+i a_2$ for $(a_1,a_2)$ real and constrained such that
\beq
a_1^2+a_2^2=1.
\eeq

\section{\texorpdfstring{Minkowski$_6$}{Minkowski6} vacua: A warm up}\label{eq:Mink6vac}
In this section we will extract necessary and sufficient conditions for supersymmetric Mink$_6$ vacua of type II supergravity and review some explicit classes of solutions. This serves in part as a warm up for the more demanding derivation of internal spaces that allow embeddings of Einstein-Maxwell supergravity into type II. However it will also turn out that the explicit classes we present can be used to uplift more general solutions of Einstein-Maxwell supergravity with $g=0$.\\
~\\
The bosonic fields of Mink$_6$ solutions decompose as
\begin{align}
ds^2&=e^{2A}ds^2(\text{Mink}_6)+ds^2(\text{M}_4),\nn\\[2mm]
H&= H_3,~~~~F_{\pm}=f_{\pm} +e^{6A}\text{vol}(\text{Mink}_6)\wedge  \star_4\lambda(f_+).
\end{align}
Since Mink$_6$ solutions obviously have ${\cal G}=\varphi=0$ by definition, the appropriate formulae for their supersymmetry preservation are contained in section \ref{sec:case3}, subject to the comment below  \eqref{eq:commentthing} and additionally fixing ${\cal F}=0$. Specifically the conditions are
\begin{subequations}
\begin{align}
&d(e^{2A}\cos\beta)=0,\label{eq:Mink6bps0},\\[2mm]
&d_{H_3}(e^{3A-\Phi}\tilde{\psi}_{\mp})=0,\label{eq:Mink6bps1}\\[2mm]
&d_{H_3}(e^{3A-\Phi}\text{Im}\psi_{\mp})=0,\label{eq:Mink6bps2}\\[2mm]
&d_{H_3}(e^{A-\Phi}\text{Re}\psi_{\mp})=\mp\frac{c}{4}e^{2A}\cos\beta f_{\pm},\label{eq:Mink6bps3}\\[2mm]
&d_{H_3}(e^{5A-\Phi}\text{Re}\psi_{\mp})= \pm\frac{c}{4}e^{6A}\star_6\lambda(f_{\pm})\label{eq:Mink6bps4}.
\end{align}
\end{subequations}
Whenever the above conditions hold it is only necessary to impose the NS and RR Bianchi identities \eqref{eq:10dbinanchis} to be guaranteed to have a solution. 
When $dH_3=0$ the above conditions imply that 
\beq
d_{H_3}(e^{6A}\star_6\lambda(f_{\pm}))=0,~~~~e^{2A}\cos\beta d_{H_3}f_{\pm}=0.
\eeq
As such the electric part of the RR flux is implied in general by supersymmetry, but also the magnetic part when $\cos\beta\neq 0$. One might imagine that fixing $\cos\beta \neq 0$ such that one only needs to solve \eqref{eq:Mink6bps0}-\eqref{eq:Mink6bps4} and $dH_3=0$ would be the best strategy to find solutions, but this does not turn out to be the case. A main issue is that $\cos\beta =0$ is a necessary condition for orbifold planes and Dp brane sources, as shown in \cite{Martucci:2005ht,Koerber:2007hd}, the former of which provides a source of negative tension which can circumvent the no-go theorem for Minkowski solutions with compact internal spaces \cite{Maldacena:2000mw}. It should also be clear that \eqref{eq:Mink6bps3} forces the Romans mass to be zero in type IIA, while in general a source for the RR sector requires source corrections to $dH_3=0$ when $\cos\beta\neq 0$. Finally we note that in type IIB one can generate solutions with $\cos\beta \neq 0$  (and even $\sin\beta=0$) via the SL(2,$\mathbb{R}$) duality the theory enjoys. As such we will focus on solutions with
\beq
\cos\beta=0.
\eeq
In the next section we will derive the unique class in IIA which is D8-D6-NS5 system that first appeared in \cite{Legramandi:2019ulq} and yields the, in hindsight, obvious generalisation of \cite{Imamura:2001cr} to branes without SO(3) rotational invariance in their co-dimension. In section \ref{eq:IIBMink} we will recover two classes, one with D5 branes back-reacted on a CY$_2$ manifold, one with an internal space that is the base of an elliptically fibred CY$_3$, with obvious F-theory significance.

\subsection{The type IIA class}\label{eq:MinkIIA}
In this section we derive the unique class of solutions in type IIA compatible with $\beta=\frac{\pi}{2}$.\\
~\\
In this case \eqref{eq:Mink6bps0} is trivial and $f_{\pm}$ drops out of \eqref{eq:Mink6bps3}. This allows us to use the 2 form parts of \eqref{eq:Mink6bps1}-\eqref{eq:Mink6bps3} to define a vielbein - specifically these yield
\beq
d(e^{2A-\Phi} \text{Re}U)=d(e^{4A-\Phi} \text{Im}U)=d(e^{4A-\Phi} W)=0.
\eeq
These conditions can be solved in terms of local coordinates $(\rho,y_i)$ for $i=1,2,3$ as
\beq
\text{Re}U= e^{-2A+\Phi}d\rho,~~~~\text{Im}U=e^{-4A+\Phi}dy_1,~~~W=e^{-4A+\Phi}d(y_2+i y_3).
\eeq
The 4-form part of \eqref{eq:Mink6bps1}-\eqref{eq:Mink6bps3} then define $H_3$, which takes a simpler form in terms of arbitrary functions $h=h(\rho,y_i)$ and $u=u(\rho,y_i)$ defined though
\beq
e^{2A}=\frac{1}{\sqrt{h}},~~~~e^{-\Phi}= \frac{h^{\frac{3}{4}}}{\sqrt{u}}.
\eeq
With these redefinitions of the fields we find that the NS flux takes the form
\beq
H_3= \partial_{\rho}(hu)dy_{123}-\frac{1}{2}\epsilon_{ijk}\partial_{y_i}u d\rho\wedge dy_j\wedge dy_k.
\eeq
Given that we have a definition of the vielbein it is a simple matter to extract $f_+$ from \eqref{eq:Mink6bps4}, we find that its non-trivial parts are
\beq
f_0=\frac{\partial_{\rho}h}{u},~~~f_2=-\frac{1}{2}\epsilon_{ijk}\partial_{y_i}h dy_j\wedge dy_k.
\eeq
This solves all of all of \eqref{eq:Mink6bps3}-\eqref{eq:Mink6bps4}\\
~\\
In summary we have recovered the  D8-D6-NS5 class of \cite{Legramandi:2019ulq}, which generalises \cite{Imamura:2001cr}. The $d=10$ fields given by
\begin{align}
ds^2&= \frac{1}{\sqrt{h}}ds^2(\text{Mink}_6)+\sqrt{h}u(dy_i)^2+\frac{u}{\sqrt{h}}d\rho^2,~~~e^{-\Phi}=\frac{h^{\frac{3}{4}}}{\sqrt{u}},\nn\\[2mm]
H_3&= \partial_{\rho}(hu)dy_{123}-\frac{1}{2}\epsilon_{ijk}\partial_{y_i}u d\rho\wedge dy_j\wedge dy_k,\nn\\[2mm]
F_0&=\frac{\partial_{\rho}h}{u},~~~F_2=-\frac{1}{2}\epsilon_{ijk}\partial_{y_i}h\wedge dy_j\wedge dy_k
\end{align}
The Bianchi identities away from sources impose that 
\beq
dF_0=0,~~~\partial^2_{y_i}u+\partial_{\rho}^2(uh)=0,~~~~\partial^2_{y_i}h+F_0\partial_{\rho}(uh)=0.
\eeq
Note however that when $F_0 \neq 0$ one can fix $u= F_0 (\partial_{\rho}h)^{-1}$ which leads to the above reducing to a single PDE
\beq
\partial_{y_i}^2h+ \frac{1}{2}\partial_{\rho}^2(h^2)=0.
\eeq
Notable solutions in this class include all supersymmetric AdS$_7$ solutions of type II supergravity \cite{Macpherson:2016xwk}, however it also contains compact Mink$_6$ solution such as that in \cite{Legramandi:2019ulq}. 

\subsection{The type IIB classes}\label{eq:IIBMink}
In this section we derive the classes of solution in type IIB with $\beta=\frac{\pi}{2}$.\\
~\\
The case of type IIB is a little more complicated than IIA because it contains to classes of solution determined by whether or not $a_2=0$ for
\beq
a=a_1+i a_2,
\eeq
With a little work it is possible to establish that \eqref{eq:Mink6bps0}-\eqref{eq:Mink6bps4} in general contain the conditions
\begin{subequations}
\begin{align}
&d(e^{2A-\Phi}a_1)=0,\label{eq:bpsiibnobetamink61}\\[2mm]
&d(e^{4A-\Phi}a_2)=0,\label{eq:bpsiibnobetamink62}\\[2mm]
&d(e^{4A-\Phi}\omega)=0,\label{eq:bpsiibnobetamink63}\\[2mm]
&d(e^{4A-\Phi}a_1 j)-e^{4A-\Phi}a_2 H_3=0,\label{eq:bpsiibnobetamink64}\\[2mm]
&d(e^{2A-\Phi}a_2 j)+ e^{2A-\Phi}a_1 H_3=0,\label{eq:bpsiibnobetamink65}\\[2mm]
&d(e^{6A-\Phi}a_1)-d(e^{6A-\Phi}a_2 j)-e^{6A-\Phi}a_1 H_3+e^{6A}\star_4\lambda(f_-)=0\label{eq:bpsiibnobetamink66}.
\end{align}
\end{subequations}
We will now set about deriving the two distinct classes these contain.\\
~\\
\textbf{D5 branes back-reated on CY$_2$}\\
~\\
In this section we recover a class of solutions with D5 branes back-reacted on a general CY$_2$.\\
~\\
The class of this section follows from fixing
\beq
(a_1,~a_2)=(1,~0)
\eeq
This means that we can solve \eqref{eq:bpsiibnobetamink61} by introducing a constant $c_0$ such that
\beq
e^{2A-\Phi}=c_0,
\eeq
however we can fix $c_0=1$ without loss of generality by rescaling Mink$_6$. Now
\eqref{eq:bpsiibnobetamink62} is implied, 
while \eqref{eq:bpsiibnobetamink63}-\eqref{eq:bpsiibnobetamink65} become
\beq
d(e^{2A}j)=0,~~~~d(e^{2A}\omega)=0,~~~~H_3=0.
\eeq
The first two of these tell us that our internal space is conformally a Calabi-Yau 2-fold, \textit{i.e}
\beq
ds^2(\text{M}_4)= e^{-2A} ds^2(\text{CY}_2),
\eeq
while the last tells us that the NS flux is trivial. All that remains to solve is \eqref{eq:bpsiibnobetamink66}, which by introducing an arbitrary function $h$ with support on $\text{CY}_2$ becomes
\beq
e^{-4A}=h,~~~~\Rightarrow ~~~~\star_4\lambda(f_-)= \frac{dh}{\sqrt{h}},
\eeq 
which is easily inverted to give
\beq
f_-=  \star_4\left(\frac{dh}{\sqrt{h}}\right)=   \hat{\star}_4dh
\eeq
where $\hat{\star}_4$ is the Hodge dual on the unwarped CY$_2$.\\
~\\
In summary we have recovered the class of formal D5 branes backreated on CY$_2$ whose $d=10$ fields take the form
\begin{align}
ds^2&=\frac{1}{\sqrt{h}}ds^2(\text{Mink}_6)+\sqrt{h}ds^2(\text{CY}_2),~~~~e^{-\Phi}=  \sqrt{h},~~~~F_3=  \hat \star_4 dh.
\end{align}
One has a solution when the Bianchi identity of $F_3$ is imposed which, away from sources, requires
\beq
\hat{\nabla}^2h=0,
\eeq
where $\hat{\nabla}^2$ is the Lapacian on the unwarped CY$_2$.\\
~\\
\textbf{F-theory class}\\
~\\
The second class follows from assuming $a_2 \neq 0$ which means we can solve \eqref{eq:bpsiibnobetamink61}-\eqref{eq:bpsiibnobetamink62} as
\beq
e^{4A-\Phi}a_2=1,~~~e^{2A-\Phi}a_1= b_0,
\eeq
for $b_0$ a constant, the former of which we assume does not vanish. We then again introduce an arbitrary function of the internal space $h$ such that
\beq
e^{-4A}=h-b_0^2,~~~~e^{-\Phi}=\sqrt{1-\frac{b_0}{h}}h,~~~~a_1=\frac{b_0}{\sqrt{h}}.
\eeq
Given this it is possible to manipulate \eqref{eq:bpsiibnobetamink63}-\eqref{eq:bpsiibnobetamink66} to the form
\begin{subequations}
\begin{align}
&d\hat j=0,~~~~ d\hat \omega=\frac{1}{2}d\log h\wedge \hat\omega_2,\label{eq:ftheorybps1}\\[2mm]
&H_3= d(b_0 h^{-1} \hat j),~~~~\star_4\lambda (f_-)=\left(-b_0 \sqrt{\frac{h}{1- \frac{b_0^2}{h }}} d\log h+\sqrt{\frac{1-\frac{b_0^2}{h}}{h}}dh\wedge \hat j\label{eq:ftheorybps2}\right)
\end{align}
\end{subequations}
where we have defined 
\beq
\sqrt{\frac{h}{1-\frac{b_0^2}{h}}}( j,~\omega)=(\hat j,~\hat \omega).
\eeq
The conditions \eqref{eq:ftheorybps1} imply that the internal space decomposes as
\beq
ds^2(\text{M}_4)= \sqrt{\frac{1-\frac{b_0^2}{h}}{h}} ds^2(\text{B}_4)
\eeq
where B$_4$ is a Kahler manifold that defines the base of an elliptically fibered Calabi-Yau 3-fold as in \cite{Couzens:2017way}. Specifically one has
\beq
d\hat \omega=i \hat P\wedge \hat\omega,~~~~\hat P=-\frac{1}{2}d^c\log h
\eeq
where $d^c$ is defined such that $dh+i d^c h$ is holomorphic- The Ricci form on $B_4$ is then defined as $d\hat P= \hat {\cal R}$.
 On the other hand \eqref{eq:ftheorybps2} implies that
\beq
f_-= d^c h\wedge(1+ B_2 ),~~~~B_2=\frac{b_0}{h}  \hat j,~~~~dB_2=H_3.
\eeq

In summary we find a class with $d=10$ fields of the form
\begin{align}
ds^2&=\frac{1}{\sqrt{h-b_0^2}}ds^2(\text{Mink}_6)+ \sqrt{\frac{1-\frac{b_0^2}{h}}{h}}ds^2(\text{B}_4),~~~~e^{-\Phi}=\sqrt{1-\frac{b_0}{h}}h,\nn\\[2mm]
H&= dB_2,~~~~ B_2=\frac{b_0}{h}  \hat j,\nn\\[2mm]
F_1&=  d^c h,~~~~~F_3= B_2\wedge F_1.
\end{align}
Supersymmetry demands that $B_4$ is a Kahler manifold, with $\hat j$ its associated Kahler form, defined such that
\beq
d^c\log h=-2\hat{ \cal R}.
\eeq
When this holds one again has a solution when the Bianchi identities of the fluxes hold, away from sources this amounts to imposing the existence of a potential $C_0$ such that 
\beq
d C_0= F_1,
\eeq
which holds when $ -C_0+ i  h$ is holomorphic - which when $b_0=0$ means that $\tau=C_0+ i e^{-\Phi}$ is anti-holomorphic. A similar solution with Mink$_6\to$ AdS$_3\times$S$^3$ was found in \cite{Lozano:2019emq}, generalising a solution of \cite{Couzens:2017way} by turning on non-trivial 3-form fluxes. In \cite{Couzens:2017way} an example of a compact internal manifold that is compatible with this class of Mink$_6$ solutions (at least for $b_0=0$) is also presented.

\section{Internal spaces for the strictly un-gauged case}\label{sec:ungagued}
In this section we will derive internal spaces that permit Einstein-Maxwell supergravity with $g=0$ to be embedded into type II supergravity without ${\cal A}$ appearing in the $d=10$ metric. This consists of two steps, solving the necessary conditions for supersymmetry of section \ref{sec:case1} and then solving the Bianchi identities of the fluxes. In the first step we will need to distinguish between the case where the tensor multiplet is turned on or not, and in the second whether, if the tensor multiplet is turned on, the vector is also turned on. 

We begin by considering the case without a tensor multiplet where we will be able to give a universal uplift valid for generic $\beta$ in section \ref{sec:universal}. Later in section \eqref{eq:sticktnongaguedwithtensor} we will consider uplifts of solutions that include a tensor multiplet with or without the vector present.

\subsection{A universal uplift for solutions with gravity and vector multiplets}\label{sec:universal}
In this section we will consider uplift of minimal $d=6$ un-gauged supergravity coupled to a vector multiplet only. In particular this means that we consider $d=6$ solutions obeying the following constrains
\beq
{\cal G}=-\star_6{\cal G},~~~~\varphi=0,~~~~g=0,~~~~{\cal F}\wedge {\cal F}={\cal F}\wedge \star_6{\cal F}=0.
\eeq
where the first 3 of these define the truncation to the sector of current interest and the final 2 are necessary to have a solution given this.
We take the following ansatz for the $d=10$ bosonic fields
\begin{align}
ds^2&=e^{2A}g^{(6)}_{\mu\nu}dx^{\mu}dx^{\nu}+ ds^2(\text{M}_4),~~~~H=H_3+ H_0 {\cal G}+ H_1\wedge {\cal F},\\[2mm]
F_{\pm}&=f_{\pm}+ e^{2A}{\cal F}\wedge g_{\pm}- e^{4A}\star_6{\cal F}\wedge \star_4\lambda(g_{\pm}) + e^{3A}{\cal G}\wedge (g_{\mp}+ \star_4\lambda(g_{\mp}))+e^{6A}\text{vol}_6\wedge \star_4\lambda(f_{\pm}).\nn
\end{align}
The sufficient conditions for supersymmetry then reduce to
\begin{subequations}
\begin{align}
&H_0= -2 e^{2A}\cos\beta,\label{eq:univeralbps1}\\[2mm]
&e^{3A} (1+ \star_4\lambda)g_{\mp}=\mp \frac{8}{c}e^{A-\Phi}\text{Re}\psi_{\mp},\label{eq:bpss1}\\[2mm]
&\frac{c}{4}e^{4A}(\cos\beta g_{\pm}+ \star_4\lambda(g_{\pm}))= \pm e^{A-\Phi}H_1\wedge \text{Re}\psi_{\mp},\label{eq:bpss2}
\end{align}
\end{subequations}
and precisely the conditions required to have a supersymmetric Mink$_6$ vacua, namely \eqref{eq:Mink6bps0}-\eqref{eq:Mink6bps4}.
We note that the new conditions \eqref{eq:univeralbps1},\eqref{eq:bpss1} and \eqref{eq:bpss2} simply serve to define the components of $(H,~F_{\pm})$ that couple to $({\cal G},~{\cal F})$, with the notable exception of $H_1$ which is merely constrained.  From this it is clear that as far as supersymmetry is concerned the internal manifolds of Mink$_6$ solutions also serve as internal manifolds for the more general $d=6$ solutions currently under consideration. Of course supersymmetry is not enough to have a solution: The additional constraints that we must impose so that the above implies the Bianchi identities in $d=10$  are
\begin{align}
&dH_3=0,~~~~ d_{H_3} f_{\pm}=0,\nn\\[2mm]
&dH_1=0,~~~~d_{H_3}(e^{2A}g_{\pm})=H_1\wedge f_{\pm}\label{eq:notbiimplied}.
\end{align}
The first line here implies that, if an uplift exists, then M$_4$ must be the internal space of a supersymmetric Mink$_6$ solution, the caveat on existence is down to the second line: One obvious solution to this is to simply fix
\beq
g_{\pm}=H_1=0,
\eeq 
which truncates the $d=6$ theory to just the gravity multiplet. It was already shown in \cite{Lozano:2022ouq} that the class of solutions in \eqref{eq:MinkIIA} provides an uplift of minimal $d=6$ supergravity, though its status regarding supersymmetry was not checked. We now see that there are as many uplifts of the minimal theory as there are supersymmetric Mink$_6$ solutions, one for each.

We would now like to establish if the second constraint in \eqref{eq:notbiimplied} can be solved in a non-trivial fashion allowing us to uplift solutions in $d=6$ with gravity and vector multiplets turned on. To show that this is indeed possible we find it useful to introduce a function $B_0$ and poly-form $\tilde{g}_{\pm}$ such that 
\beq
H_1=dB_0,~~~~ e^{2A}g_{\pm}= B_0 f_{\pm}+ \tilde{g}_{\pm}.
\eeq
Then under the assumption that the Bianchi identity of $f_{\pm}$ holds, that of $e^{2A}g_{\pm}$ becomes
\beq
d_{H_3}\tilde{g}_{\pm}=0.
\eeq
By making use of the  conditions for supersymmetry we can then bring \eqref{eq:bpss2} to the form
\beq
\frac{c}{4}e^{2A}(\cos\beta \tilde{g}_{\pm}+\star_4\lambda(\tilde{g}_{\pm}))=\pm d(e^{-4A} B_0)\wedge e^{5A-\Phi}\text{Re}\psi_{\mp}.
\eeq
It is possible that more general solutions exist for specific classes of internal space, but the simplest way to solve this is
\beq
\tilde{g}_{\pm}=0,~~~~ B_0=e^{4A},
\eeq
where we have set a possible integration constant to one through the scaling symmetry of the $d=6$ theory. Thus we find that each supersymmetric Mink$_6$ solution also defines an uplift for the minimal theory coupled to a vector multiplet.\\
~~\\
In summary we have found a universal uplift to type II supergravity for solutions with $d=6$ minimal supergravity (with $g=0$) coupled to a vector multiplet. The uplift takes the form
\begin{align}
ds^2&= e^{2A}g^{(6)}_{\mu\nu}dx^{\mu}dx^{\nu}+ds^2(\text{M}_4),~~~ H=H_3-2 e^{2A}\cos\beta {\cal G}+ d(e^{4A})\wedge {\cal F},\nn\\[2mm]
F_{\pm}&=\left(1+e^{4A}{\cal F}\right)\wedge f_{\pm} \mp \frac{8}{c}e^{A-\Phi}{\cal G}\wedge \text{Re}\psi_{\mp}+e^{6A}\left(\text{vol}_6-\star_6{\cal F}\right) \wedge \star_4\lambda(f_{\pm}),
\end{align}
where $(e^{A},~e^{-\Phi},~\beta,~ds^2(\text{M}_4),~f_{\pm},~H_3,~H_0,~\text{Re}\psi_{\mp})$ can be the the internal fields and bi-linears of any supersymmetric Mink$_6$ solution. The values of the internal fields for 3 classes of solution with $\beta=\frac{\pi}{2}$ can be extracted from sections \ref{eq:MinkIIA} and \ref{eq:IIBMink}.\\

\subsection{Uplifts with a tensor multiplet}\label{eq:sticktnongaguedwithtensor}
In this section we derive uplifts of minimal supergravity that couple to a tensor multiplet and possibly also a vector multiplet. This means that we must solve the general supersymmetry conditions \eqref{eq:sugbps1}-\eqref{eq:sugbps6}, those that permit a tensor multiplet  \eqref{eq:sugbps7}-\eqref{eq:sugbpsparing} and \eqref{eq:sugbps8} and also the Bianchi identities of the flux. As every condition that holds when only the tensor multiplet is present must also hold when the vector is also turned on, we find it easiest to approach the problem of adding the tensor multiplet first. As we need to ascertain exactly how $\varphi$ is embedded in the internal space, we will not be able to make the sort of universal statement we did in the previous section. We will instead have to consider the classes of uplift on a case by case basis, for this reason we will fix
\beq
\cos\beta=0, \label{eq:constaintbeta}
\eeq
specifically $\beta=\frac{\pi}{2}$ as a simplifying assumption that makes the task more tractable. However as discussed in section \ref{eq:Mink6vac}, we have reason to believe that the physically interesting classes of uplifts will be captured by this assumption. Fixing $\cos\beta=0$ has secondary consequences: First off \eqref{eq:sugbps1} becomes incompatible with $(H_0,~\tilde{H}_0)$ being any constant value other than $(0,0)$ when $d\varphi \neq 0$. Then, given this, \eqref{eq:sugbps7} is uniquely solved by taking $h_{\mp}=0$, so we must also tune
\beq
H_0=\tilde{H}_0=h_{\mp}=0.
\eeq
That means we are considering backgrounds of the form
\begin{align}
ds^2&= e^{2A}ds_6^2+ ds^2(\text{M}_4),\nn \\[2mm]
F_{\pm}&=(1+\star\lambda)(f_{\pm}+e^{2A} {\cal F}\wedge g_{\pm}+e^{3A}{\cal G}\wedge g_{\mp}),~~~H=H_3+ H_1\wedge {\cal F}+d\varphi\wedge H_2,\label{eq:genericform}
\end{align}
which are in general subject to the supersymmetry constraints
\begin{subequations}
\begin{align}
&\partial_{\varphi}(e^{A})=0,~~~~ \partial_{\varphi}(e^{-2\Phi}\sqrt{\det g_4})=0,\label{eq:sugbpsrest0}\\[2mm]
&d_{H_3}(e^{3A-\Phi}\text{Im}\psi_{\mp})-d\varphi\wedge H_2\wedge \text{Im}\psi_{\mp}=0,\label{eq:sugbpsrest1}\\[2mm]
&d_{H_3}(e^{3A-\Phi}\tilde{\psi}_{\mp})- e^{3A-\Phi}d\varphi \wedge H_2\wedge  \tilde{\psi}_{\mp}=0\label{eq:sugbpsrest2},\\[2mm]
&\frac{c}{8}e^{3A-\varphi}(1+\star_4\lambda)g_{\mp}=\mp e^{A-\Phi}\text{Re}\psi_{\mp},\label{eq:sugbpsrest3}\\[2mm]
&d_{H_3}(e^{5A-\Phi}\text{Re}\psi_{\mp})\mp \frac{c}{4}e^{6A}\star_4\lambda(f_{\pm})\bigg\lvert_{d\varphi\rightarrow 0}=0,\label{eq:sugbpsrest4}\\[2mm]
&d_{H_3}(e^{A-\Phi}\text{Re}\psi_{\mp})-d\varphi\wedge \bigg[\pm \frac{c}{8}e^{3A-\varphi}(1-\star_4\lambda)g_{\mp}+ e^{A-\Phi} H_2 \wedge \text{Re}\psi_{\pm}\bigg]=0,\label{eq:sugbpsrest5}
\end{align}
\end{subequations}
and if and only if the vector multiplet is non-trivial, also
\beq
\frac{c}{4}e^{4A}\star_4\lambda(g_{\pm})=\pm e^{A-\Phi}H_1\wedge \text{Re}\psi_{\mp}\label{eq:sugbpsrest6}.
\eeq

Before moving onto the cases let us make one observation about \eqref{eq:sugbpsrest5} which will be a useful going forward: Note that this contains the combination $(1-\star_4\lambda)g_{\mp}$, which is anti self-dual under $\star_4\lambda$, this restricts it possible form to
\beq
(1-\star_4 \lambda)g_{\mp}=\left\{\begin{array}{l} X_1-\star_4X_1~~~\text{in IIA}\\[2mm]
p(1-\text{vol}(\text{M}_4))+ X^{(1,1)}~~~\text{in IIB}\end{array}\right.
\eeq
where $p$ is a function on M$_4$, $X_1$ a 1-form and $X^{(1,1)}$ a real primitive (1,1)-form, which is to say it obeys
\beq
X^{(1,1)}\wedge j= X^{(1,1)}\wedge\omega=0,~~~~\star_4 X^{(1,1)}= X^{(1,1)}.
\eeq
Each of these could in principle depend function-wise on $\varphi$ also. 

\subsubsection{D8-D6-NS5 embedding in IIA} \label{IIAGTM}
We begin by considering embeddings into type IIA.\\
~\\
In this case the 2-form parts of \eqref{eq:sugbpsrest1}-\eqref{eq:sugbpsrest2} give rise to
\beq
d(e^{4A-\Phi}W)=0,~~~d(e^{4A-\Phi}\text{Im}U)=0,
\eeq
which we can solve in terms of local coordinates $(y_1,y_2,y_3)$ as
\beq
e^{4A-\Phi}\text{Im}U=dy_1,~~~e^{4A-\Phi}W=dy_2+i dy_3,
\eeq
just like we did when considering Mink$_6$ vacua. If we had $ d\varphi=0$ the 2 form part of \eqref{eq:sugbpsrest5} would be $d(e^{2A-\Phi}\text{Re}U)=0$, which we could likewise solve in terms of a local coordinate. With $d\varphi\neq 0$  we have that the 2-form parts of \eqref{eq:sugbpsrest5}  becomes
\beq
d(e^{2A-\Phi}\text{Re}U)-\frac{c}{2}e^{3A-\varphi}d\varphi\wedge X_1.\label{eq:inter}
\eeq
We can proceed by locally taking 
\beq
e^{2A-\Phi}\text{Re}U=f(d\rho+ V_idy_i),
\eeq
where $(f,V_i)$ depend on $(\rho,y_i,\varphi)$, which contains no assumption. One can then use \eqref{eq:sugbpsrest1}, \eqref{eq:sugbpsrest2}, \eqref{eq:sugbpsrest5} and  \eqref{eq:sugbpsrest3} to fix $(g_-,H_2,H_3)$ and extract some PDEs relating $(e^{A},e^{-\Phi},V_i,f)$. However if one then tries to impose the Bianchi identities of the fluxes one finds that one can locally fix $f=f(\varphi)$ and $V_i=0$ without loss of generality. The derivation of this is long and tedious so let us continue our derivations from
\beq 
e^{2A-\Phi}\text{Re}U=fd\rho,~~~~f=f(\varphi).
\eeq  
We then find that the 4-form parts of \eqref{eq:sugbpsrest1}, \eqref{eq:sugbpsrest2},\eqref{eq:sugbpsrest5} with legs in $d\varphi$ fix
\beq
H_2=0,~~~~\partial_{\varphi}(e^{-\Phi}f^{-\frac{1}{2}})=0,
\eeq
given \eqref{eq:sugbpsrest0}, while the parts with no legs in $d\varphi$ fix 3 of the 4 components of $H_3$. Given this one then finds from  \eqref{eq:sugbpsrest5} that
\beq
e^{3A}(1-\star_4\lambda)g_-=2 f'(1-\star_4\lambda)d\rho,~~~~e^{-\Phi}=\sqrt{f}e^{-\hat{\Phi}}
\eeq
where $e^{-\hat{\Phi}}$ is independent of $\varphi$, and one also received the final component of $H_3$. We now are free to decompose
\beq
e^{2A}=\frac{1}{\sqrt{h}},~~~~ e^{-\hat{\Phi}}=\frac{h^{\frac{3}{4}}}{\sqrt{u}},
\eeq
similarly to how we did in the Mink$_6$ case with $(h,u)$ independent of $\varphi$ but otherwise free. We now have that
\beq
H_3= \frac{1}{f^2}\partial_{\rho}(hu)dy_{123}-\frac{1}{2}\epsilon_{ijk}\partial_{y_i}u d\rho\wedge dy_j\wedge dy_k,
\eeq
which indicates that we must have $\partial_{\rho}(hu)=0$ to be able to solve the Bianchi identity of $H_3$ when $d\varphi\neq 0$. At this point we can use \eqref{eq:sugbpsrest5} to extract
\beq
e^{3A}g_-=- e^{\varphi}(f-f')d\rho+ e^{\varphi}f^{-2}hu dy_{123}(f+f'),
\eeq
This should be closed if the Bianchi for the RR flux Bianchi identity to hold, given that we already established that non-constant $f$ demands $\partial_{\rho}(hg)=0$, this amounts to imposing
\beq
\partial_{\varphi}f=\pm f~~~~\Rightarrow~~~ f=   e^{\pm \varphi},\label{eq:fsign1}
\eeq
where we have used the invariance under \eqref{eq:invariances} to fix a possible constant factor in $f$ to 1. Notice that either choice of sign leads to
\beq
\partial_{\varphi}(e^{-2\Phi}\sqrt{g_4})=0,
\eeq
so \eqref{eq:sugbpsrest0} is now solved. Let us proceed with the minus sign and comment on the other choice at the end, we then find that
\beq
e^{3A}g_-=-2 d\rho,~~~~ e^{3A}\star_4\lambda(g_-)= 2e^{2\varphi}hu dy_{123}
\eeq
with the second term having the correct $e^{2\varphi}$ dependence to close on $d(e^{2\varphi}\star_6{\cal G})=0$. It is then a simple matter to extract $f_+$ from \eqref{eq:sugbpsrest4} which again yields the Mink$_6$ result
\beq
f_+= \frac{\partial_{\rho}h}{u}-\frac{1}{2}\epsilon_{ijk}\partial_{y_i}h d\rho\wedge dy_j\wedge dy_k,
\eeq
at which point the conditions for supersymmetry with a tensor multiplet turned on are solved.\\
~\\
In summary we find a class that, under the assumption that the tensor multiplet is non-trivial, has $d=10$ fields given by
\begin{align}
ds^2&= \frac{1}{\sqrt{h}}g^{(6)}_{\mu\nu}dx^{\mu}dx^{\nu}+\sqrt{h}e^{\varphi}u(dy_i)^2+e^{-\varphi}\frac{u}{\sqrt{h}}d\rho^2,~~~e^{-\Phi}=e^{-\frac{1}{2}\varphi}\frac{h^{\frac{3}{4}}}{\sqrt{u}},\nn\\[2mm]
H_3&= -\frac{1}{2}\epsilon_{ijk}\partial_{y_i}u d\rho\wedge dy_j\wedge dy_k,\nn\\[2mm]
F_0&=\frac{\partial_{\rho}h}{u},~~~F_2=-\frac{1}{2}\epsilon_{ijk}\partial_{y_i}h dy_j\wedge dy_k,~~~~F_4=-2d\rho\wedge {\cal G},\label{eq:IIAtensor}
\end{align}
which is subject to the constraint
\beq
\partial_{\rho}(hu)=0.\label{eq:constraintfjfj}
\eeq
The Bianchi identities of the fluxes demand that $F_0$ is constant and that
\beq
\partial_{y_i}^2u=0,~~~\partial_{y_i}^2h=0.
\eeq
There is a second embedding that follows from choosing the positive sign in \eqref{eq:fsign1}, but this take the form of \eqref{eq:IIAtensor} with the metric and dilaton modified by 
\beq
\varphi\to -\varphi,~~~~{\cal G}\to-e^{2\varphi}\star_6{\cal G}.
\eeq
This is nothing more than the S-duality like symmetry of the $d=6$ theory when $g={\cal A}=0$. 
\\
~\\
\textbf{Adding a vector multiplet}\\
~\\
We will now consider the compatibility of of the uplift we have derived with the addition of a tensor multiplet. This means that we must also include $(H_1,~g_{+})$ such that \eqref{eq:sugbpsrest6} is obeyed, which leads to
\beq
e^{2A}g_{+}=- (H_1)_{\rho}\frac{h}{g}+\frac{h}{2}\epsilon_{ijk} (H_1)_i dy_j\wedge dy_k ,
\eeq
for $(H_{\rho},H_i)$ the components of $H_1$ - this is all we need to be compatible with supersymmetry. Moving on to the Bianchi identities: As we are now in the presence of both a tensor and vector multiplet we must impose $e^{A}g_-=H_1\wedge g_+$, which implies one of
\begin{align}
f&=e^{-\varphi},~~~~ (H_1)_i=0,~~~~~2u= (H_{1})_{\rho}^2 h,\nn\\[2mm]
f&=e^{\varphi},~~~~ (H_1)_{\rho}=0,~~~~2u=(H_{1})_{i}^2.
\end{align}
However we also need to impose the Bianchi identity of $e^{2A}g_+$: For the first case above we find this forces $H_1=0$ leading to no vector multiplet. For the second case there is a non-trivial embedding but we must impose $\partial_{\rho}h=0$ which combined with \eqref{eq:constraintfjfj} makes $\partial_{\rho}$ an isometry. This makes this embedding,  modulo T-duality,  contained in the more general embedding of the next section, so we will not present it explicitly here.

\subsubsection{\texorpdfstring{CY$_2$}{CY2} embedding in IIB}
In this section we derive an embedding into type IIB. We will focus on generalising the Mink$_6$ class with a CY$_2$ manifold as this case permits the $d=6$ dilaton to simply appear as an overall warp factor in the internal metric. We have found that embeddings that generalise the IIB class with a Kahler manifold are quite restricted - indeed one can show that solving part of \eqref{eq:sugbpsrest5} along $\varphi$ requires that the two terms in $j$, as expressed in \eqref{eq:j2omega2decomp} have opposite powers of $e^{\varphi}$. It then also follows from \eqref{eq:sugbpsrest0} that $\partial_{\varphi}\Phi=0$. Given that $j$ also needs to be conformally closed with respect to the exterior derivative on M$_4$, this vastly constrains the form the manifold can take. We have doubts that, at least for $\cos\beta=0$, any such embedding with bounded internal space exists beyond $\mathbb{T}^2\times \mathbb{T}^2$ dressed by $\varphi$. This we know exists because we can take CY$_2=\mathbb{T}^4$ in embedding of this section then T-dualise twice to generate it.\\
~\\
We find that the conditions for supersymmetry in the case that $(a_1=1,a_2=0)$ reduce to \eqref{eq:sugbpsrest0} and
\begin{subequations}
\begin{align}
&d(e^{4A-\Phi}j)=0,~~~~d(e^{4A-\Phi}\omega)=0,\label{eq:cytensorbps1}\\[2mm]
&H_3=0,~~~H_2\wedge j=H_2\wedge \omega=0\label{eq:cytensorbps1b}\\[2mm]
&d\varphi\wedge (e^{3A}(1-\star_4\lambda)g_+)=-2e^{\varphi}d(e^{2A-\Phi}(1-\frac{1}{2}j\wedge j))-2e^{2A+\varphi-\Phi}d\varphi\wedge H_2,\label{eq:cytensorbps2}\\[2mm]
&e^{3A}(1+\star_4\lambda)g_+= 2 e^{2A+\varphi-\Phi}(1-\frac{1}{2}j\wedge j)\label{eq:cytensorbps3},\\[2mm]
&e^{6A}\star_4\lambda(f_-)=- d(e^{6A-\Phi}(1-\frac{1}{2}j\wedge j))\bigg\lvert_{d\varphi\to 0}\label{eq:cytensorbps4}.
\end{align}
\end{subequations}
Clearly \eqref{eq:cytensorbps1} of these implies that M$_4$ is conformally CY$_2$ and the 1-form part of \eqref{eq:cytensorbps2} that 
\beq
e^{2A-\Phi}= f
\eeq
where $f=f(\varphi)$. If we then define
\beq
e^{-4A}=h
\eeq
for $h$ with support on M$_4$ only as \eqref{eq:sugbpsrest0} demands We solve \eqref{eq:cytensorbps1} as 
\beq
ds^2(\text{M}_4)= \sqrt{h} f^{-1} ds^2(\text{CY}_2),
\eeq
which fixes an arbitrary multiplicative constant - note that we already have that  \eqref{eq:sugbpsrest0} is solved. Then \eqref{eq:cytensorbps1b} informs us that $H_2$ yields the only non-trivial part of the NS flux and that it must be a primitive (1,1)-form. The condtion $dH=0$ then demands
\beq
H_2= p(\varphi)X^{(1,1)},~~~~ dX^{(1,1)}=0,
\eeq
for $X^{(1,1)}$ a primitive (1,1)-form on CY$_2$ - note that
\beq
X^{(1,1)}\wedge X^{(1,1)}= \frac{1}{2}(X^{(1,1)})^2\text{vol}(\text{CY}_2),\label{eq:thsr}
\eeq
where form contraction here is performed on the unwarped CY$_2$ directions. Then from \eqref{eq:cytensorbps2}-\eqref{eq:cytensorbps4} we easily extract 
\beq
f_-=\hat{\star}_4 dh,~~~~e^{3A}g_+=e^{\varphi}\left(f-f'+ f H_2+(f+f')\frac{h}{f^2}\text{vol}(\text{CY}_2)\right).
\eeq
For the Bianchi identity of  $e^{3A}g_+$ to be satisfied it is necessary that
\beq
f''= f,~~~~(f^2p)'=0,~~~~ (f')^2+\frac{ (X^{(1,1)})^2}{4 h}f^4p^4-f^2=0.
\eeq
These are solved in general by
\beq
p=\frac{1}{f^2},~~~~f= c_+e^{\varphi}+c_- e^{-\varphi},~~~~(X^{(1,1)})^2=16 c_+c_- h,
\eeq 
for $c_{\pm}$ constant, though we should stress that it remains to be seen that the final constraint can actually be solved on a given CY$_2$ for $X^{(1,1)}\neq 0$ and $h$ non-constant. At this point all the necessary conditions for supersymmetry with a tensor multiplet turned on are solved and the Bianchi identities dealt with.\\
~\\
In summary taking the minus sign leads to the the class
\begin{align}
ds^2&=\frac{1}{\sqrt{h}}g^{(6)}_{\mu\nu}dx^{\mu}dx^{\nu}+\frac{\sqrt{h}}{f}ds^2(\text{CY}_2),~~~~e^{-\Phi}=  \sqrt{h}f,~~~f=c_+e^{\varphi}+c_- e^{-\varphi},\\[2mm]
F_3&=  \left(\hat \star_4 dh+2 (c_-{\cal G}-c_+ e^{2\varphi}\star_6{\cal G})\right),~~~~F_5=  \frac{e^{  \varphi}}{f}X^{(1,1)}\wedge({\cal G}+\star_6{\cal G}),~~~~H=\frac{1}{f^2} d\varphi\wedge X^{(1,1)}\nn
\end{align}
for $X^{(1,1)}$ a primitive (1,1) form on CY$_2$. The Bianchi identities of the fluxes demand that
\beq
\hat{\nabla}^2h=0,~~~~dX^{(1,1)}=0,~~~~(X^{(1,1)})^2=16 c_+c_-h
\eeq
away from possible sources. Notice that the embedding is invariant under
\beq
\varphi\to-\varphi,~~~({\cal G},~e^{2\varphi}\star_6{\cal G})\to -(e^{2\varphi}\star_6{\cal G},~{\cal G}),~~~c_{\pm}\to c_{\mp},~~~X^{(1,1)}\to - X^{(1,1)},
\eeq
reflecting the S-like duality of the $d=6$ theory in the limit currently under consideration.
\\
~\\
\textbf{Adding a vector multiplet}\\
~\\
We will now address whether a vector multiplet can also be added to the above background. Supersymmetry demands that we solve \eqref{eq:sugbpsrest6} which in this case means
\beq
e^{2A}g_-=- h \star_{\text{CY}_2}H_1.
\eeq
$H_1$ should also be closed, we thus introduce a function on CY$_2$ $w$ such that
\beq
H_1=d\left(\frac{w}{h}\right),
\eeq
with the factor of $h$ appearing to simplify later expressions.  As we now have both a tensor and vector multiplet turned the Bianchi identities require that we solve $e^{3A}g_+= H_1\wedge g_-$, which forces us to tune
\beq
X^{(1,1)}=0,~~~~c_-=0,~~~~ c_+ \left(d \left(\frac{w}{h}\right)\right)^2=2,
\eeq
from which it follows that $(\hat\nabla(\frac{w}{h}))^2$ should be constant. Finally we need to ensure that the Bianchi identity of $e^{2A}g_-$ is satisfied which, given that $h$ is a harmonic function leads to 
\beq
\hat\nabla^2w=0,
\eeq
away from sources, making $w$ another harmonic function on CY$_2$. At this point we have derived what is required to add the vector multiplet - though we are not totally clear  on whether  $(\nabla(\frac{w}{h}))^2$ constant and non-zero can be achieved for some CY$_2$ without fixing $h$ constant.\\
~\\
In summary a $d=10$ uplift for $d=6$ solutions containing both a vector and tensor multiplet is given by
\begin{align}
ds^2&=\frac{1}{\sqrt{h}}g^{(6)}_{\mu\nu}dx^{\mu}dx^{\nu}+e^{-\varphi}\sqrt{h}ds^2(\text{CY}_2),~~~~e^{-\Phi}=  e^{\varphi}\sqrt{h},\\[2mm]
F_3&=  \left(\hat \star_4 dh-2  e^{2\varphi}\star_6{\cal G}\right),~~~~F_5= (1+\star \lambda)e^{-\varphi} d\left(\frac{w}{h}\right)\wedge\star_6{\cal F},~~~~H= d\left(\frac{w}{h}\right)\wedge {\cal F}\nn,
\end{align}
where we have set $c_+=1$ without loss of generality. The Bianchi identities of the $d=10$ fluxes require that away from sources
\beq
\hat\nabla^2h=0,~~~~\hat\nabla^2w=0,~~~~ (\hat\nabla \left(\frac{w}{h}\right))^2=2.
\eeq
Let us stress though that it is not clear to us whether  it is possible to have $h$ non-constant and solve the last expression for some CY$_2$. With $h$ constant a bounded internal space would require that CY$_2$ is K3 or $\mathbb{T}^4$,  for the later case at least  $w=c_i y_i$ for $y_i$ coordinates on the torus can solve the required constraint. If its possible to have $h$ non-constant then the class is less restrictive as one can still construct a bounded internal space from a non-compact CY$_2$ when it has sources back-reacted on it as in \cite{Lima:2022hji}.  

\section{Internal spaces for the \texorpdfstring{$g=0$}{g=0} limit of the gauge compatible case}\label{eq:section 6}

In this section we consider the $g=n=0$ limit of the gauge compatible case in section \ref{sec:case2}. We will again assume that 
\beq
\cos\beta= H_0=\tilde{H}_0=h_{\pm}=h_{\mp}=0, \label{eq:tuninggeq0}
\eeq
for simplicity. This means we are again considering an embeddings into type II of the form in \eqref{eq:genericform} but where now
\beq
ds^2=e^{2C}D\phi^2+ds^2(\text{M}_3),~~~~D\phi= d\phi+ V+ {\cal A},
\eeq
where every internal $d=4$ field and bi-linear contains a part that is parallel and orthogonal to  $D\phi$.

To start with we note that after fixing $g=0$, the conditions \eqref{eq:gbps1}- \eqref{eq:gbps6},  essentially reproduce the earlier conditions  \eqref{eq:sugbps1}-\eqref{eq:sugbps6}, where ${\cal A}$ is assumed not to appear in the metric. The only condition that is not of this form is \eqref{eq:gbps8}, which fixes $g_{\pm}$ in terms of $H_1$. Thus upon tuning the fields as \eqref{eq:tuninggeq0},  solving the supersymmetry constraints for the classes of this section was basically already done in section \ref{sec:ungagued}, we need only impose a U(1) isometry on them that the bi-linears  $(\psi_{\mp},~\tilde{\psi}_{\mp})$ are singlets under. We will thus skip to this point in this section and proceed to impose the Bianchi identities, which are different to section \ref{sec:ungagued}. The only additional thing we need to decide on is where $D\phi$ will lie within bi-linears: We will assume that, with respect to \eqref{eq:j2omega2decomp} and \eqref{bispinors4D IIA}-\eqref{bispinors4D IIB}
\beq
W= (\tilde{W}+ i e^{C} D\phi),
\eeq  
with $(\text{Re}U,~\text{Im}U,~\tilde{W})$ defining a vielbein on M$_3$. This choice can be made without loss of generality in IIB but in IIA one could consider other possibilities. We will not do this here, primarily because when $g\neq 0$ this choice (without loss of generality) will become forced on us and this section in large part serves as a stepping stone towards, and comparison to, the gauged embeddings.

We will begin our analysis in type IIA where there is a single class of embeddings before moving onto type IIB where their are two.

\subsection{Type IIA embeddings}
It is possible to show that when \eqref{eq:tuninggeq0} is imposed the conditions \eqref{eq:gbps0}-\eqref{eq:gbps6} and \eqref{eq:gbpspairing} are solved as 
\begin{align}
U&=\sqrt{u }\left(\frac{\sqrt{f}}{h^{\frac{1}{4}}}d\rho+i \frac{h^{\frac{1}{4}}}{\sqrt{f}} dy_2\right) ,~~~ W= \frac{\sqrt{u} h^{\frac{1}{4}}}{f}(dy_1+i D\phi),~~~D\phi= d\phi+{\cal A},\nn\\[2mm]
e^{A}&=\frac{1}{h^{\frac{1}{4}}},~~~~e^C=\sqrt{u}h^{\frac{1}{4}},~~~~~e^{-\Phi}=\frac{\sqrt{f} h^{\frac{1}{4}}}{\sqrt{u}},\nn\\[2mm]
f_+&=\left(\frac{\partial_{\rho}h}{g}-\epsilon_{ij}\partial_{y_i}hD\phi\wedge  dy_j\right),\nn\\[2mm]
e^{3A}g_-&=-e^{\varphi}\left( (f-f')d\rho+ \frac{(f+f')}{f^2}uh D\phi\wedge dy_2\wedge dy_2\right),\nn\\[2mm]
H_3&= D\phi\wedge\left(\epsilon_{ij}\partial_{y_i}ud\rho\wedge dy_j-\frac{1}{f^2}\partial_{\rho}(hu)dy_1\wedge dy_2 \right),~~~~H_2=0
\end{align}
where ($h,u)$ have support on $(\rho,~y_1,~y_2)$ and $f=f(\varphi)$, which can be set to 1 without loss of generality when the tensor multiplet is not turned on. What remains is to solve \eqref{eq:gbps8} to ensure consistency with supersymmetry. For this we need to decompose $H_1$ as
\beq
H_1= h_0 D\phi+h_i dy_i+ h_{\rho}d\rho,~~~~dh_0=0,
\eeq
which leads to
\beq
e^{2A}g_+=- h\left(\frac{h_{\rho}}{g}-D\phi\wedge\left(\epsilon_{ij}h_idy_j- u d\rho\right)+h_0 dy_1\wedge dy_2\right)
\eeq
At this point the conditions for supersymmetry are solved, but we still need to solve the Bianchi identities of the NS and RR fluxes. In the case at hand this amounts to imposing in general  that
\begin{align}
&dH_3=0,~~~dH_1+\iota_{\partial_{\phi}}H_3=0,\nn\\[2mm]
&d_{H_3}f_{+}\bigg\lvert_{{\cal A}\to 0}=0,\nn\\[2mm]
&d_{H_3}(e^{2A}g_{+})-(H_1\wedge -\iota_{\partial_{\phi}}) f_{+}\bigg\lvert_{{\cal A}\to 0}=0,
\end{align}
and when we also have a tensor multiplet turned on that
\beq
h_0=0,~~~~e^{3A}g_{-}= e^{2A}(H_1\wedge-\iota_{\partial_{\phi}}) g_{+}.
\eeq
This leads to 4 classes of embeddings, 2 with and 2 without a tensor multiplet turned on, we will skip the details of their derivation and simply present the classes
\subsubsection{Massless embedding with gravity and vector multiplets}
The first class of embedding is in massless IIA and accommodates a non-trivial gravity and vector multiplet only. Locally the NS sector takes the form
\begin{align}
ds^2&= \frac{1}{\sqrt{h}} g^{(6)}_{\mu\nu}dx^{\mu}dx^{\mu}+ u \bigg(\frac{1}{\sqrt{h}}d\rho^2+ \sqrt{h}\left((dy_i)^2+ D\phi^2\right)\bigg),~~~e^{-\Phi}= \frac{h^{\frac{3}{2}}}{\sqrt{u}},\nn\\[2mm]
H&=d\left(D\phi\wedge \left(h_i dy_i- \frac{u G}{h}d\rho\right)\right)+ h_0 D\phi\wedge {\cal F},
\end{align}
where $y_i=(y_1,~y_2)^i$ and $h=h(y_i),~G=G(y_i)~u=u(\rho,~y_i),~ h_i=h_i(\rho,~y_i)$. The RR sector contains the following non-trivial fluxes
\begin{align}
F_2&=\epsilon_{ij}\partial_{y_i}h  dy_j\wedge D\phi+G{\cal F},\nn\\[2mm]
F_4&= -h\left(h_0 dy_1\wedge dy_2+\left(u d\rho+\epsilon_{ij}h_i dy_j\right)D\phi\right)\wedge {\cal F}+2  d\rho\wedge{\cal G}.
\end{align}
The $d=10$ Bianchi identities demand first that $h_i$ are constrained such that
\begin{align}
\epsilon_{ij}\partial_{y_i}h_j&= h \partial_{\rho}u,~~~~\partial_{y_i}(h^2 h_i)= h^2 G \partial_{\rho}u,\nn\\[2mm]
\partial_{\rho}h_i&=-\frac{G}{h}\partial_{y_i}u+\epsilon_{ij}\left(\frac{1}{2}\frac{u}{h^2}\partial_{y_j}\left(h^2+G^2\right)+\partial_{y_j}u\right),
\end{align}
that $h+i G$ is holomorphic on $(y_1,y_2)$, \textit{i.e}
\beq
d(h+i G)\wedge (dy_1+i dy_2)=0,
\eeq
and further that the following PDEs are satisfied away from the loci of sources
\begin{align}
\partial_{y_i}^2h=0,~~~\partial_{y_i}^2u+h \partial_{\rho}^2 u=0,
\end{align}
which are those of a flat space D6-NS5 system with U(1) rotational symmetry in its codimensions. Whenever these conditions are solved we have a consistent truncation to minimal $d=6$ un-gauged supergravity coupled to a vector multiplet.

\subsubsection{Massive embedding with gravity and vector multiplets}
The second class of embedding also permits the gravity and vector multiplet to be turned on but this time has non-trivial Romans mass $F_0$, which is constant (locally). Locally it is characterised by a NS sector of the form
\begin{align}
ds^2&= \frac{1}{\sqrt{h}} g^{(6)}_{\mu\nu}dx^{\mu}dx^{\mu}+\frac{\partial_{\rho}h}{F_0}\bigg(\frac{1}{\sqrt{h}}d\rho^2+\sqrt{h}((dy_i)^2+D\phi^2)\bigg),~~~e^{-\Phi}= \sqrt{\frac{F_0}{\partial_{\rho}h}}h^{\frac{3}{4}},\nn\\[2mm]
H&=\frac{1}{F_0}d\left(\left(\epsilon_{ij}\partial_{y_i}dy_j+d\left(\frac{G}{h}\right)\right)\wedge D\phi\right)
\end{align}
where $h$ is a function of $(\rho,~y_i)$, $G$ a function of $y_i$. In addition to $F_0$ the RR sector supports non-trivial 2 and 4 -forms 
\begin{align}
F_2&=\epsilon_{ij}\partial_{y_i}h  dy_j\wedge D\phi-\frac{G}{h}{\cal F},\nn\\[2mm]
F_4&=2d\rho\wedge {\cal G}+\frac{1}{2F_0}D\phi\wedge\left(d(h^2)-2 h \epsilon_{ij}\partial_{y_i}\left(\frac{G}{h}\right)dy_j\right).
\end{align}
Embeddings are then defined by the solutions of the following two PDEs,
\beq
\partial_{y_i}^2G=0,~~~~~ \partial_{y_i}^2h+ \frac{1}{2}\partial_{\rho}^2(h^2)=0,
\eeq
which imply the Bianchi identities of the RR and NS fluxes away from sources.\\
~\\
\textbf{Massless embedding with gravity, vector and tensor multiplets}\\
~\\
The third class of embedding, like the first, is again in massless IIA but this time accommodates all the fields of 6d Einstein-Maxwell supergravity with $g=0$. Its NS sector locally takes the form
\begin{align}
ds^2&=\frac{1}{\sqrt{h}} g^{(6)}_{\mu\nu}dx^{\mu}dx^{\mu}+\Delta\bigg(\frac{e^{-\varphi}}{\sqrt{h}}d\rho^2+e^{\varphi}\sqrt{h}((dy_i)^2+D\phi^2)\bigg),~~~~e^{-\Phi}= \frac{e^{-\frac{1}{2}\varphi}h^{\frac{3}{4}}}{\sqrt{\Delta}},\nn\\[2mm]
H&=4 d\left(\frac{G}{\Delta} d\rho\wedge D\phi\right),~~~~~\Delta= \frac{2 h}{h^2+G^2}
\end{align}
where $(h,~G)$ have support on $y_i$ only and $\partial_{\rho}$ is an isometry. The RR sector on the other hand takes the form
\beq
F_2=\epsilon_{ij}\partial_{y_i}h  dy_j\wedge D\phi+G{\cal F},~~~F_4= \frac{4h^2}{\Delta}D\phi\wedge d\rho \wedge {\cal F}+2d\rho \wedge {\cal G}.
\eeq
The $d=10$ Bianchi identities are solved whenever 
\beq
d(h+ i G)\wedge (dy_1+i dy_2)=0,
\eeq
\textit{i.e.} whenever $h+ i G$ is a holomorphic function on $(y_1,~y_2)$. When this is true one has a consistent truncation to the un-gauged limit of the full $d=6$ theory.
\subsubsection{Massive embedding with gravity, vector and tensor multiplets}

The forth class of embedding is in massive IIA and permits gravity, vector and tensor multiplets. Its NS sector can be locally expressed as
\begin{align}
ds^2&= \frac{1}{\sqrt{h}}g^{(6)}_{\mu\nu}dx^{\mu}dx^{\mu}+\frac{2e^{\varphi}}{\sqrt{h}}\left((dy_i)^2+ D\phi^2\right)+e^{-\varphi}\frac{\sqrt{h}}{2} d\rho^2,~~~e^{-\Phi}= \frac{1}{\sqrt{2}}e^{-\frac{1}{2}\varphi} h^{\frac{5}{4}},~~~H=0
\end{align}
where $h=h(\rho)$ and $y_i$ are isometry directions one can take to span a $\mathbb{T}^2$. The background supports the following $d=10$ fluxes
\beq
F_0=\partial_{\rho}h,~~~~F_4= h D\phi\wedge d\rho\wedge {\cal F}+ h d\rho\wedge {\cal G},
\eeq
and embeddings are defined in terms of solutions to the ODE
\beq
\partial_{\rho}^2h=0,
\eeq
which imposes that $F_0$ is constant and makes $h$ a linear function. This should hold away from possible sources, \textit{i.e.} $h$ is the warp factor of a D8 brane locally. Globally $F_0$ need only be piecewise constant with the discontinuities giving rise to D8 brane sources along the interior of the interval spanned by $\rho$. This fact can be used to glue local solutions to $\partial_{\rho}^2h=0$ together which allows one to construct global embeddings bounded between D8-O8 sources with D8 sources placed along the interval - see section 4.1 of \cite{Macpherson:2018mif}.

\subsection{Type IIB  embeddings}
In type IIB we still need to solve \eqref{eq:bpsiibnobetamink66}-\eqref{eq:bpsiibnobetamink61}, at least at constant values of $\varphi$ and with ${\cal A}\to 0$. This means there are two types of class, for which the internal space is conformally either a CY$_2$ or the base of an elliptically fibered CY$_3$. Each of these further splits into classes of embedding that are or are not compatible with a tensor multiplet, leading to 4 classes in total. The derivation of these classes mirrors what we have previously presented, so we will only present the results.

\subsubsection{\texorpdfstring{CY$_2$}{CY2} embedding with gravity and vector multiplets}
The first IIB class of embeddings is compatible with a gravity and tensor multiplet and has an internal space that is a warped CY$_2$ containing a U(1) fibre. Such CY$_2$ manifolds have been classified \cite{Bakas:1996gf}, and depending on whether their holomorphic 2-form is charged under the  U(1) or not, they are defined in terms of a  Toda or flat space Laplace  equation in $d=3$ respectively. Solutions with D5 branes back-reacted on such CY$_2$'s have also been considered \cite{Lima:2022hji}. As we are in the $g=0$ limit, we expect a connection to the Laplace type CY manifolds, this indeed turns out to be the case.

We find a class of embeddings whose NS sector takes the form
\begin{align}
ds^2&=\frac{1}{\sqrt{h}} g^{(6)}_{\mu\nu}dx^{\mu}dx^{\mu}+\sqrt{h}\bigg(h(dy_i)^2+ \frac{1}{h}D\phi^2\bigg),~~~~e^{-\Phi}= \sqrt{h},\nn\\[2mm]
H&= H_1\wedge {\cal F},~~~H_1=d\left(\frac{u}{h}\right)+h_0(D\phi-V),~~~D\phi= d\phi+V+ {\cal A},
\end{align}
where $y_i=(y_1,y_2,y_3)$,  $(h,~u)$ have support on $y_i$ and it is the part of $ds^2$ in parentheses that spans the CY$_2$, modulo the external gauge field ${\cal A}$.  Additionally the internal connection $V$ is such that
\beq
dV= -\frac{1}{2}\epsilon_{ijk}\partial_{y_i}h dy_j\wedge dy_k.\label{eq:g0intcon}
\eeq
The non-trivial RR fluxes on the other hand are given by
\beq
F_3=D\phi\wedge (dV-{\cal F})+2 {\cal G},~~~~F_5= (1+ \star) H_1\wedge\star_{6}{\cal F}.
\eeq
The $d=10$ Bianchi identities demand that
\beq
\partial_{y_i}^2h=0,~~~~h\partial_{y_i}^2u= \partial_{y_i}(h^2V_i)
\eeq
away from the loci of possible sources. One then has an embedding whenever these are satisfied.

\subsubsection{\texorpdfstring{CY$_2$}{CY2} embedding with gravity, vector and tensor multiplets}
It is also possible to embed solutions with gravity, vector and tensor multiplets turned on into internal spaces of CY$_2$ type. This time the class of embeddings takes the form 
\begin{align}
ds^2&=\frac{1}{\sqrt{h}} g^{(6)}_{\mu\nu}dx^{\mu}dx^{\mu}+\sqrt{h}e^{\frac{1}{2}\varphi}\bigg(h(dy_i)^2+ \frac{1}{h}D\phi^2\bigg),~~~~e^{-\Phi}= e^{-\varphi}\sqrt{h}\nn,\\[2mm]
H&=0,~~~~F_3=D\phi\wedge (dV-{\cal F})+ {\cal G},~~~~D\phi=d\phi+V+{\cal A}
\end{align}
where again $h=h(y_i)$ and the internal connection obeys \eqref{eq:g0intcon}. The only PDE one must solve this time is 
\beq
\partial_{y_i}^2h=0,
\eeq
and one has a consistent truncation to the $g=0$ limit of Einstein-Maxwell supergravity whenever this holds.

\subsubsection{F-theory type embedding with gravity and vector multiplets}
Solutions in $d=6$ with  gravity and vector multiplets non-trivial can also be embedded into IIB in terms of an internal space that is the base of an elliptically fibered CY$_3$, with a U(1) isometry imposed on it. Such embeddings have an NS sector of the form
\begin{align}
ds^2&= \frac{1}{\sqrt{h} \sqrt{\Delta}} g^{(6)}_{\mu\nu} dx^{\mu}dx^{\nu}+   \sqrt{\Delta}\bigg(u  \left(\sqrt{h}(dy_i)^2+\frac{1}{\sqrt{h}}d\rho^2\right)+ \frac{1}{u\sqrt{h}}D\phi^2\bigg),~~~e^{-\Phi}=\sqrt{\Delta}h,\nn\\[2mm]
H&=d\left(\frac{1}{h}\right)\wedge d\rho\wedge D\phi+H_1\wedge{\cal F},~~~~H_1=h_0 D\phi+ \frac{1}{h \Delta^2}h_i dy_i-\frac{b_0+\frac{u G}{\Delta}}{h}d\rho\nn\\[2mm]
\Delta&=  1- \frac{b_0^2}{h},~~~D\phi=d\phi+V+{\cal A}
\end{align}  
where $y_i=(y_1,~y_2)$,  $(h,~G)$ have support on $y_i$,  $(u,~h_i)$ on $(\rho,~y_i)$, $(b_0,~h_0)$ are constants and the internal connection is such that
\beq
dV= d\rho\wedge(\epsilon_{ij}\partial_{y_i}u dy_j)- h \partial_{\rho}u dy_1\wedge dy_2.
\eeq
The non-trivial RR fluxes are given by
\begin{align}
F_1&=- \epsilon_{ij}\partial_{y_i} h dy_j,\nn\\[2mm]
F_3&= b_0 \left(\epsilon_{ij}\partial_{y_i}\log h dy_j\wedge d\rho\wedge D\phi+ 2{\cal G}\right)+ \left(G D\phi+ h_0 \Delta u h d\rho+ \frac{1}{h\Delta} \epsilon_{ij}h_i dy_j\right)\wedge{\cal F},\nn\\[2mm]
F_5&= 2 \Delta(D\phi\wedge d\rho+ u h dy_1\wedge dy_2)\wedge {\cal G}+ (1+ \star)(b_0H_1-\Delta d\rho)\wedge \star_6 {\cal F}.
\end{align}
Embeddings are defined by first the branching rule
\beq
b_0 dh=0,
\eeq
and the following PDEs  
\beq
\partial_{y_i}h=0,~~~~\partial_{y_i}^2u+ h\partial_{\rho}^2 u=0,~~~~d( h_0 h+i G)\wedge d(y_1+i y_2)=0.
\eeq
Given a solution to the above $h_i$ must obey
\begin{align}
&\partial_{y_i}h_i= G h^2 \Delta\partial_{\rho}u,\nn\\[2mm]
&\epsilon_{ij}\partial_{y_i}\left(\frac{h_j}{h^2 \Delta^2}\right)=h_0 h \partial_{\rho}u,\nn\\[2mm]
&\partial_{\rho}h_i= u^2 \partial_{y_i}\left(\frac{G h \Delta}{u}\right)+h_0\epsilon_{ij}\partial_{y_j}\left(h^2 \Delta^2 u\right).
\end{align}
When these conditions are solved one has a consistent truncation.

\subsubsection{F-theory type embedding with gravity, vector and tensor multiplets}
Finally we find an F-theory like embedding with all $d=6$ multiplets turned on, its NS sector is locally of the form
\begin{align}
ds^2&= \frac{1}{\sqrt{h}}g_{\mu\nu}^{(6)}dx^{\mu}dx^{\nu} +G\bigg(e^{-\varphi}\sqrt{h}(dy_i)^2+\frac{e^{\varphi}}{\sqrt{h}\Delta}d\rho^2\bigg)+\frac{e^{\varphi}}{\sqrt{h}\sqrt{\Delta} G}D\phi^2 ,~~~e^{-\Phi} =\sqrt{\Delta } h,\\[2mm]
H&=d\left(\frac{w}{h}\right)\wedge {\cal F}+dB_2,~~~B_2=\frac{b_0e^{2\varphi}}{h \Delta}d\rho \wedge D\phi,~~~D\phi=d\phi+ {\cal A}+ \rho \epsilon_{ij}\partial_{y_i}Gdy_j,~~~~\Delta=1+\frac{b_0^2e^{2\varphi}}{h}\nn
\end{align}
where $y_i=(y_1,y_2)_i$, $(h,G,w)$ have support on $y_i$, $b_0$ is a constant and $d^2D\phi=0$ requires
\beq
\partial_{y_i}^2G=0.\label{eq:Gcond}
\eeq
The non-trivial $d=10$ RR fluxes are the following
\begin{align}
F_1&= \epsilon_{ij}\partial_{y_i}hdy_j,\nn\\[2mm]
F_3&= B_2\wedge F_1-2b_0 e^{2\phi}\star_6{\cal G}+{\cal F}\wedge h \epsilon_{ij}\partial_{y_i}\left(\frac{w}{h}\right)dy_j,\nn\\[2mm]
F_5&=(1+ \star)\left(2 G h {\cal G}+ e^{\varphi}\left(b_0 d\left(\frac{w}{h}\right)-d\rho\right)\wedge \star_6{\cal F}\right).
\end{align}
One has a consistent truncation whenever \eqref{eq:Gcond} and
\begin{align}
&b_0dh=0,~~~G=\frac{1}{2}\left(1+\left(\partial_{y_i}\left(\frac{w}{h}\right)\right)^2\right),\nn\\[2mm]
&\partial_{y_i}^2h=0,~~~~\partial_{y_i}^2w=0,
\end{align}
hold, which implying the Bianchi identities of the fluxes away from sources. However it remains to be seen that the above definition of $G$ can be made consistent with $G$ needing to be harmonic on $y_i$, beyond the case of $w\propto h$ and $2G=1$.

\section{Internal spaces for Einstein-Maxwell gauged supergravity}\label{eq:section 7}
In this section we turn our attention to embeddings of the $d=6$ theory into type II supergravity in the presence of non-trivial R-symmetry gauging. Contrary to the previous two section, where we had $g=0$ so no R-symmetry gauging, the internal manifolds of these embeddings will not be dressed versions of Mink$_6$ solutions. When $g\neq 0$ it follows that ${\cal F} = 0$ is inconsistent with external supersymmetry so we will always  assume that at least a gravity and tensor multiplet is turned on. We will however distinguish between case that do or do not have a non-trivial tensor multiplet in addition to this, as the former are rather more constrained generically.\\
~\\
We will again focus on the cases for which
\beq
\beta=\frac{\pi}{2},~~~~ H_0= \tilde{H}_0= h_{\pm}=0,\label{eq:constraint}
\eeq
though we should stress that our previous appeals to the existence of compact Mink$_6$ vacua are no longer valid so we do not claim that this is anything more than a simplifying assumption. As such we are considering uplifts of the form
\begin{align}
ds^2&= e^{2A} g^{(6)}_{\mu\nu} dx^{\mu}dx^{\nu}+ds^2(\text{M}_4),~~~~H= H_3+ H_1\wedge {\cal F}+ d\varphi\wedge H_2\nn\\[2mm]
F_{\pm}&= (1+\star \lambda)\left(f_{\pm}+ {\cal G}\wedge g_{\mp}+{\cal F}\wedge g_{\pm}\right),
\end{align}
where all internal forms and the bi-linears $(\psi_{\mp},~\tilde{\psi}_{\mp})$ have parts parallel and orthogonal to $D\phi$ such that
\beq
ds^2(\text{M}_4)=ds^2(\text{M}_3)+e^{2C}D\phi,~~~~D\phi= d\phi+V+2g {\cal A},
\eeq
where $V$ is a 1-form on M$_3$ and $\partial_{\phi}$ is an isometry of the bosonic fields and $\psi_{\mp}$ under which $\tilde{\psi}_{\mp}$ has charge 1. It is possible to establish that, with respect to \eqref{eq:j2omega2decomp} and \eqref{bispinors4D IIA}-\eqref{bispinors4D IIB}, we can choose to align the isometry direction $D\phi$ purely along $W$ as
\beq
W= e^{i\phi}(\tilde{W}+ i e^{C} D\phi),
\eeq  
with $(\text{Re}U,~\text{Im}U,~\tilde{W})$ defining a vielbein on M$_3$: With $g\neq 0$ this choice can be made without loss of generality. As before finding an embedding then amounts to a two step process: First one solves the relevant supersymmetry conditions, in this case \eqref{eq:gbps0}-\eqref{eq:gbps8} subject to \eqref{eq:constraint}, which actually makes \eqref{eq:gbps7} implied. Second one solves the Bianchi identities of the NS and RR fluxes, which for the case at hand amounts to imposing in general that
\begin{align}
&dH_1+2g\iota_{\partial_{\phi}}(H_3+d\varphi\wedge H_2)\bigg\lvert_{{\cal A}\to 0}=0,~~~~~ dH_3-d\varphi\wedge dH_2\bigg\lvert_{{\cal A}\to 0}=0,\nn\\[2mm]
&d_{H_3}f_{\pm}-d\varphi\wedge H_2\wedge f_{\pm}\bigg\lvert_{{\cal A}\to 0}=0,\nn\\[2mm]
&d_{H_3}(e^{2A}g_{\pm})-(H_1\wedge -2g\iota_{\partial_{\phi}}) f_{\pm}-e^{2A} d\varphi\wedge H_2\wedge g_{\pm}\bigg\lvert_{{\cal A}\to 0}=0,\label{eq:gneq0bianchis}
\end{align}
and only when the tensor multiplet is also non-trivial
\beq
\iota_{\partial_{\phi}}H_1=0,~~~e^{3A}g_{\mp}= e^{2A}(H_1\wedge-2g\iota_{\partial_{\phi}}) g_{\pm}.
\eeq
Our aim is to reduce the above to as few necessary conditions as possible that one must solve. As we shall see, there are several classes of embedding and how far we can progress that aim will depend on the class at hand.\\
~\\ 
We begin our analysis in type IIA in section \ref{sec:gaugedIIA} before moving onto type IIB in \ref{sec:gaugedIIB}.

\subsection{Type IIA embeddings for \texorpdfstring{$g\neq 0$}{g=/=0}} \label{sec:gaugedIIA}
Our focus in type IIA will be on embeddings that do not include a tensor multiplet for which we should fix
\beq
\varphi=0,~~~~\star_6{\cal G}=-{\cal G},
\eeq
which makes $H_2$ drop out of the ansatz. The reason to constrain things thus is that while IIA embeddings with a tensor multiplet turned on do exist, we found that they contain at least 1 U(1) isometry that the bi-linears are not charged under. As such, modulo T-duality, these are special cases of the embeddings we will derive for type IIB. 

Without a tensor multiplet turned on it is possible to extract the following conditions on M$_3$ from \eqref{eq:gbps0}-\eqref{eq:gbps8} that do not involve the RR fluxes
\begin{subequations}
\begin{align}
&V=0,~~~ d(e^{4A+C-\Phi})=e^{4A-\Phi} \tilde{W},~~~~~d(e^{2A-\Phi}\text{Re}U)=0,\label{eq:IIAgaguedbps1}\\[2mm]
&d(e^{4A-\Phi}\text{Im}U)+ g e^{2A-\Phi}(H_1\wedge \text{Re}U+ 2g e^C  \text{Re}U\wedge \tilde{W})=0,\label{eq:IIAgaguedbps2}\\[2mm]
&d(e^{4A+C-\Phi}\text{Re}U\wedge \text{Im}U)- e^{4A-\Phi}(\iota_{\partial_{\phi}}H_3+ \text{Re}U\wedge \text{Im}U)\wedge \tilde{W}=0,\label{eq:IIAgaguedbps3}\\[2mm]
&H_3\wedge D\phi=0,~~~\iota_{\partial_{\phi}}H_1=0,~~~d(e^{2A+C-\Phi}\text{Im}U\wedge \tilde{W})= e^{2A-\Phi}\iota_{\partial_{\phi}}H_3\wedge\text{Re}U,\label{eq:IIAgaguedbps4}\\[2mm]
&d(e^{4A+C-\Phi}\text{Re}U\wedge \tilde{W})+e^{2A-\Phi}(e^{2A}\iota_{\partial_{\phi}}H_3+g e^{C} H_1\wedge \tilde{W})\wedge\text{Im}U=0.\label{eq:IIAgaguedbps5}
\end{align}
\end{subequations}
The conditions in \eqref{eq:IIAgaguedbps1} tell us that $D\phi$ is not fibered over M$_3$, and that we can choose local coordinates $(\rho,x)$ such that
\beq
e^{4A+C-\Phi}=q,~~~~ \tilde{W}= e^{C}\frac{q'}{q}dx,~~~~~ \text{Re}U=e^{-2A+\Phi} d\rho
\eeq
where $q$ is an arbitrary functions of $x$ that can be fix with a coordinate transformation. We do not however have a condition allowing us to fix $\text{Im}U$ such that M$_3$ is in general diagonal, the best we can do is introduce a final local coordinate $y$ such that
\beq
\text{Im}U= e^{D}Dy,~~~~ Dy=(dy+ \tilde{\lambda} dx)
\eeq
where $(e^D,~\tilde{\lambda})$ have support on $(\rho,x,y)$. The remaining conditions \eqref{eq:IIAgaguedbps2}-\eqref{eq:IIAgaguedbps5} then constrain the components of the NS 3-form and give us a single PDE. To present these we find it helpful to introduce functions $(h,u,G)$ with support on $(\rho,x,y)$ that are related to the functions already appearing in the local ansatz as
\beq
e^A=\left(\frac{G}{h}\right)^{\frac{1}{4}},~~~e^C= \frac{h^{\frac{1}{4}}q\sqrt{u}}{G^{\frac{3}{4}}},~~~e^D=  (h G)^{\frac{1}{4}}\sqrt{u},~~~\tilde{\lambda}=\frac{\lambda}{G}.
\eeq 
In terms of these we find that \eqref{eq:IIAgaguedbps2}-\eqref{eq:IIAgaguedbps5} fix the non-trivial parts of the NS flux as
\begin{align}
H_1&=b d\rho+\frac{1}{g}(\partial_{\rho}G dy+ \partial_{\rho}\lambda dx),\\[2mm]
H_3&=D\phi\wedge \bigg[\frac{q}{q'}\left(u \partial_y\left(\frac{\lambda}{G}\right)-\tilde{\partial}_{x} u\right)Dy\wedge d\rho -q q'\partial_{\rho}\left(\frac{h u}{G}\right)dx\wedge dy\nn\\[2mm]
&+ \frac{q q' }{ G^2}\left(g h u b+ G \partial_y\left(\frac{u}{G}\right)\right)dx\wedge d\rho\bigg],
\end{align}
where $b$ is an arbitrary function of $(\rho,x,y)$ and we employ the notation
\beq
\tilde{\partial}_{x}= \partial_x- \frac{\lambda}{G}\partial_{y}.
\eeq
The PDE that gets imposed is the following
\beq
2G(\partial_{y}\lambda-\partial_{x}G)+ 2g^2 h u q q'=0\label{eq:susyPDEgauged}
\eeq
At this point what remains of \eqref{eq:gbps0}-\eqref{eq:gbps8} merely fixes the various $d=4$ fluxes that appear in the $d=10$ flux $F_+$. We find that
\begin{align}
f_+&= \frac{\partial_{\rho}h}{ u}+ D\phi\wedge \bigg[\frac{ h q}{q'} \partial_{\rho}\left(\frac{\lambda}{G}\right)d\rho+\left(\frac{qq'}{G}\partial_y\left(\frac{h}{G}\right)-\frac{g h^2 b q q'}{G^2}\right) dx\nn\\[2mm]
&~~~~~~~~~~~~~~~~~~~~~~~~- \frac{q}{q'}\left(\frac{g^2 u h^2 q q'}{G^2}+ G \tilde{\partial}_{x}\left(\frac{h}{G}\right)\right)Dy\bigg],\nn\\[2mm]
e^{2A}g_+&=-\frac{h b}{u}+D\phi\wedge \bigg[\frac{2g^2 e^k u h q^2}{G^2}d\rho+\frac{ h G q}{g  q'}\partial_{\rho}\left(\frac{\lambda}{G}\right)Dy+ \frac{h qq'}{g}\partial_{\rho}\left(\frac{1}{G}\right)dx\bigg],\nn\\[2mm]
e^{3A}(1+\star_4\lambda)g_-&=-2 d\rho-2\frac{u h q q'}{G} D\phi\wedge dx\wedge dy,
\end{align}
Now we have dealt with all of the supersymmetry constraints, most of which have been solved by locally fixing the  form of the embedding - what remains to be solved is the PDE \eqref{eq:susyPDEgauged}.

Unfortunately, once the Bianchi identities of the fluxes in \eqref{eq:gneq0bianchis} are considered this class become rather hard to tame, no doubt in part because the metric in the coordinates we have chosen is non-diagonal. Classes exist, but the only ones that he have found that are governed by sensible PDEs are sub-classes of what we derive in type IIB modulo T-duality. It would be interesting to return to this class after working out how to diagonalise it,  a 3-manifold can always be made diagonal locally, but for now we will move on. 

\subsection{Type IIB embeddings for \texorpdfstring{$g\neq 0$}{g=/=0}}\label{sec:gaugedIIB}
In type IIB it is possible to extract the following general conditions that imply the parts of \eqref{eq:gbps0}-\eqref{eq:gbps8} that are independent of the RR fluxes at constant values of $\varphi$
\begin{subequations}
\begin{align}
&d(e^{4A-\Phi}a_2)+ g e^{2A-\Phi-\varphi}(a_1  H_1-2a_2 g e^{C} \tilde{W}),~~~~~ d(e^{2A-\Phi} a_1)+ \frac{1}{2}e^{3A-\varphi} P d\varphi,\label{eq:bpsgneqIIB1}\\[2mm]
&d(e^{4A+C-\Phi} U)= \left(e^{-C}\tilde{W}- i V\right)\wedge(e^{4A+C-\Phi} U),~~~~\partial_{\varphi}A=0,\label{eq:bpsgneqIIB2}\\[2mm]
&d(e^{2A+C-\Phi}a_2\tilde{W})+e^{2A-\Phi}a_1\left(\iota_{\partial_{\phi}}(H_3+d\varphi\wedge H_2)-\frac{1}{2}e^{A+\Phi-\varphi}\iota_{\partial_{\phi}}{\cal X}_2\right)=0,\label{eq:bpsgneqIIB3}\\[2mm]
&d(e^{4A+C-\Phi}a_1\tilde{W})-e^{2A-\Phi}a_2\bigg(e^{2A}\iota_{\partial_{\phi}}(H_3+d\varphi\wedge H_2)\nn\\[2mm]
&~~~~~~~~~~~~~~~~~~~~~~~~~~~~~~~~~~~~+ge^{-\varphi}(e^{C}H_1^{(3)}\wedge \tilde{W}-\iota_{\partial_{\phi}}H_1\text{vol}_2)\bigg)=0,\label{eq:bpsgneqIIB4}\\[2mm]
&(d+e^C dV\wedge )(e^{2A-\Phi}a_2\text{vol}_2)+\frac{1}{2}e^{3A-\varphi}d\varphi\wedge {\cal X}_2^{(3)}-e^{2A-\Phi}a_1 (H_3^{(3)}+d\varphi\wedge H^{(3)}_2)=0,\label{eq:bpsgneqIIB5}\\[2mm]
&(d+e^C dV\wedge )(e^{4A-\Phi}a_1 \text{vol}_2)-g e^{2A-\Phi-\varphi}(2a_1 g e^{C} \tilde{W}+a_2 H_2^{(3)})\wedge \text{vol}_2\nn\\[2mm]
&~~~~~~~~~~~~~~~~~~~~~~~~~~~~~~~~~~~+e^{4A-\Phi}a_2 (H_3^{(3)}+d\varphi \wedge H^{(3)}_2)=0,\label{eq:bpsgneqIIB6}\\[2mm]
& (e^{C}H_2^{(3)}+i (\iota_{\partial_{\phi}}H_2)\wedge \tilde{W})\wedge U=0,~~~~a_1\left(e^{C}H_2^{(3)}\wedge \tilde{W}-(\iota_{\partial_{\phi}}H_2)\wedge \text{vol}_2\right)=0,\label{eq:bpsgneqIIB7}
\end{align}
\end{subequations}
where we use the shorthand  $\text{vol}_2=\text{Re}U\wedge \text{Im}U$, we have decomposed
\beq
(1-\star_4\lambda)g_{+}= P\left(1- \text{vol}(\text{M}_4)\right)+ {\cal X}_2
\eeq
for $(P,{\cal X}_2)$ a real function and primitive (1,1)-form on M$_4$ without loss of generality, and the superscript 3 refers to the general decomposition of a k-form on M$_4$ as
\beq
C_k=  C_k^{(3)}+ D\phi\wedge C_{k-1}.
\eeq
Note that \eqref{eq:bpsgneqIIB7} only contains  non-trivial content when the tensor multiplet is non-trivial while in general \eqref{eq:bpsgneqIIB1} contains the term
\beq
a_1 \iota_{\partial_{\phi}}H_1=0.
\eeq
It is possible to take combinations of \eqref{eq:bpsgneqIIB1}, \eqref{eq:bpsgneqIIB3}, \eqref{eq:bpsgneqIIB5}  and their exterior derivatives to establish that the Bianchi identities of the NS flux are implied when the tensor multiplet is trivial (or more specifically when $d\varphi=0$) and $a_1 \neq 0$. In terms of the original SU(2)-structure forms $(j,~\omega)$ the above conditions imply
\begin{align}
&d(e^{4A-\Phi}\omega)=0,\nn\\[2mm]
&d(e^{4A-\Phi} a_1^{-1} j)- \frac{2g^2e^{C-\varphi}}{e^{2A}}\tilde{W}\wedge(e^{4A-\Phi} a_1^{-1} j)\bigg\lvert_{d\varphi\to 0}=0,~~~~\text{if}~ a_1\neq 0,\nn\\[2mm]
&d(e^{2A-\Phi} a_2^{-1} j)- \frac{g e^{-\varphi}a_1}{e^{2A}a_2}H^{(3)}_1\wedge (e^{2A -\Phi}a_2^{-1}
j)\bigg\lvert_{d\varphi\to 0}=0,~~~~\text{if}~ a_2\neq 0
\end{align}
which make it clear that with $g\neq 0$ there is no conformal CY$_2$ class like that which exists for $g=0$, however M$_4$ is conformally a Kahler manifold at constant values of $\varphi$  whenever $a_2 \neq 0$ and either $a_1 H_1=0$ or $\frac{g e^{-\varphi}a_1}{e^{2A}a_2^2}H^{(3)}_1$ is a total derivative for $\varphi$ constant.

\subsubsection{A class with a tensor multiplet governed by a Toda-like equation}\label{eq:todaclass}
In this section we derive the general class of solutions with the phase of the SU(2)-structure on M$_4$ tuned as $a=a_1+i a_2=1$, unlike the majority of the cases we have encountered in this work, this class is entirely insensitive to whether or not the tensor multiplet is non-trivial.\\
~\\
Upon fixing $(a_1=1,~a_2=0)$ one has that \eqref{eq:bpsgneqIIB2} and \eqref{eq:bpsgneqIIB4} give rise to
\begin{subequations}
\begin{align}
&d(e^{4A+C-\Phi}\tilde{W})=0,\label{eq:cond1}\\[2mm]
&d(e^{4A+C-\Phi} U)= \left(e^{-C}\tilde{W}- i V\right)\wedge(e^{4A+C-\Phi} U).\label{eq:cond2}
\end{align}
\end{subequations} 
It follows from \eqref{eq:cond1} that we have an integrable almost product structure, which means  that if we solve it as
\beq
e^{4A+C-\Phi}\tilde{W}=\lambda_0 d\rho,
\eeq
for $\rho$ a local coordinate and $\lambda_0$ a constant we include for later convenience, then coordinates exist on M$_3$ such that
\beq
ds^2(\text{M}_3)=  e^{-8A-2C+2\Phi}\lambda^2_0d\rho^2+ g^{(2)}_{ij}(\varphi,\textbf{y},\rho) dy_i dy_j,
\eeq
with respect to which $V$ only has legs in $y_i$. We then have from \eqref{eq:cond2} that we can choose $y_i$ such that
\beq
e^{4A+C-\Phi} U=\lambda_0 e^{\Delta}d(y_1+i y_2),
\eeq
with $\Delta$ a function of $(\varphi,\rho, y_i)$, at least a priori. Substituting the above definition of $(U,~\tilde{W})$ into  \eqref{eq:bpsgneqIIB1}-\eqref{eq:bpsgneqIIB7} we find that they fix the following
\begin{align}
e^{2A-\Phi}&= f,~~~~ e^{4A+2C-\Phi}\partial_{\rho}\Delta= \lambda_0,~~~~H_3=H_1=0,\nn\\[2mm]
 H_2&= \frac{e^{3A}}{2 e^{\varphi} f }{\cal X}_2,~~~~ e^{3A}P=-2 e^{\varphi}f',~~~~ V= \epsilon_{ij}\partial_{y_i}\Delta dy_j
\end{align}
where $f=f(\varphi)$, $(\Delta,~A)$ are independent of $\varphi$ and  ${\cal X}_2$, beyond being primitive, is unconstrained. From these conditions it follows that $\partial_{\rho}\Delta=0$ is impermissible when $g\neq 0$. In addition to this we find a single PDE
\beq
f e^{\varphi}e^{4A}\left(2\partial_{y_i}^2\Delta+ \partial_{\rho}^2e^{2\Delta}\right)= 4g^2 \lambda_0 e^{2\Delta}\partial_{\rho}\Delta.
\eeq
 Differentiating this with respect to $\varphi$ implies that we must have either
\beq
\partial_{\phi}(fe^{\varphi})=0~~~~\text{or}~~~~2\partial_{y_i}^2\Delta+ \partial_{\rho}^2e^{2\Delta}=0,\label{eq:choise}
\eeq
where the second choice is a 3d Toda equation. If we solve the above in terms of the Toda equation we must fix $g=0$, which also turns off the the ${\cal A}$ term in $D\phi$. This ultimately leads to M$_4$ being conformally any CY$_2$ containing a U(1) isometry under which $\omega$ is charged, which are indeed defined in terms of solutions to the Toda equation \cite{Lima:2022hji}. As we are interested in embeddings of the gauged 6d theory we should instead take the first option in \eqref{eq:choise} which can be solved without loss of generality as
\beq
f= e^{-\varphi}~~~~\Rightarrow~~~e^{4A}\left(2\partial_{y_i}^2\Delta+ \partial_{\rho}^2e^{2\Delta}\right)= 4g^2 \lambda_0 e^{2\Delta}\partial_{\rho}\Delta.\label{eq:PDEnice}
\eeq 
What remains of \eqref{eq:gbps0}-\eqref{eq:gbps8}  can then be shown to fix the terms in the decomposition of $F_-$ as 
\begin{align}
f_-&=- \lambda_0\left(e^{2\Delta}(2g^2 \lambda_0e^{-8A}+\partial_{\rho}(e^{-4A}))dy_1\wedge dy_2+ \epsilon_{ij}\partial_{x_i}(e^{-4A})dy_j\wedge d\rho\right)\wedge D\phi,\\[2mm]
e^{2A}g_-&=-\frac{2g\lambda_0}{e^{4A}\partial_{\rho}\Delta}D\phi,~~~~ e^{3A}g_+=2+H_2.
\end{align}
At this point the conditions for supersymmetry are reduced to finding a solution to the PDE in \eqref{eq:PDEnice}, but we still need to impose the Bianchi identities of the fluxes. Solving that of $e^{2A}g_{-}$ leads to
\beq
{\cal X}_2=0,~~~~e^{4A}= 2\frac{g^2\lambda_0}{\partial_{\rho}\Delta},
\eeq
and when these hold all of \eqref{eq:gneq0bianchis} are implied by \eqref{eq:PDEnice}.\\
~\\
In summary we find a class of embeddings of the form
\begin{align}
ds^2&= \frac{1}{\sqrt{\partial_{\rho}\Delta}}\bigg[2g^2g^{(6)}_{\mu\nu}dx^{\mu}dx^{\nu}+e^{\varphi}\bigg(D\phi^2+(\partial_{\rho}\Delta)^2\left(d\rho^2+e^{2\Delta}(dy_i)^2\right)\bigg)\bigg],\nn\\[2mm]
e^{-\Phi}&=\frac{\sqrt{\partial_{\rho}\Delta}}{2g^2} e^{-\varphi},~~~~ D\phi=d\phi+2g {\cal A}+V\nn\\[2mm]
F_3&=\frac{1}{2g^2}\left(dV-2g{\cal F}\right)\wedge D\phi+2 {\cal G},~~~~V=\epsilon_{ij}\partial_{y_i} \Delta dy_j,\label{eq:gaguedtensorembedding1}
\end{align} 
where we have fixed $\lambda_0= 2g^2$ without loss of generality and $\Delta=\Delta(\rho,y_i)$. Embeddings are defined by solutions to the Toda like equations
\beq
2\partial_{y_i}^2\Delta+ \partial_{\rho}^2e^{2\Delta}=2 (\partial_{\rho}e^{\Delta})^2,\label{eq:easygneqopde}
\eeq 
which is a deformation, in terms of the term on the right hand side, of the defining PDE that CY$_2$ manifolds that contain a charged U(1) isometry are defined in terms of. We note as a possible point of interest that ${\cal N}=2$ AdS$_5$ solutions of $d=11$ supergravity  are likewise defined in terms of a Toda equation \cite{Lin:2004nb}. Each solution to \eqref{eq:easygneqopde} defines a consistent truncation to  full $d=6$ Einstein-Maxwell gauged supergravity. The internal space in this case is not conformally a CY$_2$ manifold, neither for that matter is it it conformally Kahlar. This is actually the only class of embeddings that exists for $(a_1,~a_2)=(1,~0)$ and $g\neq 0$, which is to say that if one turns off the tensor multiplet and runs the analysis of this section again one finds \eqref{eq:gaguedtensorembedding1}, only with $\varphi=0$ and ${\cal G}$ appropriately constrained - the embedding does not become more general in the absence of the tensor multiplet. 

We note that the class of \eqref{eq:easygneqopde} has some similarities to the uplift in \cite{Cvetic:2003xr}, at least after S-dualising such that this yields a type IIB solution with non-trivial $F_3$ flux only. Specifically the $d=6$ dilaton appears in the same fashion in both metrics and $d=10$ dilatons and the only non-trivial flux is the RR 3-form. There are two main differences: First ${\cal A}$ appears explicitly in \eqref{eq:gaguedtensorembedding1} as the connection of a  circle fibration over the external space, where as in \cite{Cvetic:2003xr} it is a $\mathbb{T}^2$ fibration over the external space. Second the internal space in \cite{Cvetic:2003xr} is uniquely fixed and explicitly non-bounded, where as \eqref{eq:easygneqopde} is defined in terms of the solutions of a PDE. This raises the hope that it may contain examples for which the internal space is bounded -  we will explore this possibility later in section \ref{sec:towardscompacted}, were we do indeed find a bounded embedding, albeit one containing singularities we do not recognise as physical.

\subsubsection{A second class with a tensor multiplet and recovering \cite{Cvetic:2003xr}} \label{sec:secondfullembedding}
It is possible to establish that the only other possibility for realising a type IIB embedding for the whole of $d=6$ Einstein-Maxwell gauged supergravity is when one tunes the phase of the SU(2)-structure as $a=a_1+i a_2=i$ without loss of generality. Proving this explicitly is rather lengthy as it is possible to solve all the supersymmetry constraints under the weaker assumption that $a_2\neq 0$. It is not until one considers the Bianchi identities of the fluxes, specifically that of $e^{2A}g_-$, that either $d\varphi=0$ or $a_1=0$ gets forced on the class. For this reason we will start our derivation from $a_1=0$, we assure the reader that we have confirmed this is indeed required for $d\varphi\neq 0$ under our assumptions.\\
~\\
Upon fixing $(a_1,a_2)=(0,1)$ it is possible to extract a similar system of one form constraints to the previous class from \eqref{eq:bpsgneqIIB1}-\eqref{eq:bpsgneqIIB7}, namely
\begin{subequations}
\begin{align}
&d(e^{6A+C-\Phi}\tilde{W})=\frac{1}{2}e^{7A-\Phi-\varphi}d\varphi\wedge \left({\cal X}^{(3)}+ e^{C} P \tilde{W}\right) ,\label{eq:cond1s}\\[2mm]
&d(e^{4A+C-\Phi} U)= \left(e^{-C}\tilde{W}- i V\right)\wedge(e^{4A+C-\Phi} U).\label{eq:cond2s}
\end{align}
\end{subequations} 
These informs us that, for similar reasons to the previous section, coordinates exist such that
\beq
e^{6A+C-\Phi}\tilde{W}= e^{k}f(\varphi)d\rho,~~~~~e^{4A+C-\Phi} U= e^{\Delta}d(y_1+i y_2),
\eeq
where $f=f(\varphi)$, $k=k(\rho)$ and $e^{\Delta}$ is independent of $\varphi$ but otherwise free a priori. Inserting this definition of the vielbein on M$_3$ into \eqref{eq:bpsgneqIIB1} -\eqref{eq:bpsgneqIIB7} then fixes the fields in our ansatz as
\begin{align}
e^{4A}&= \frac{e^{2\Delta} \partial_{\rho}\Delta}{\tilde{G}q'},~~~e^{2(A+C)}=\frac{e^{\varphi}e^{2\Delta}}{2g^2\tilde{G} q},~~~ e^{4A-\Phi}=q,\nn\\[2mm]
f&=e^{\varphi},~~~e^{k}=\frac{q q'}{2g^2}~~~P=0,~~~\iota_{\phi}H_1=0,~~~H_2=0,~~~V=\epsilon_{ij}\partial_{y_i}\Delta dy_j\nn\\[2mm]
e^{3A}{\cal X}_2&=2 \left(\frac{2 g^2}{\tilde{G}}dy_1\wedge dy_2+\frac{q'e^{2\varphi}}{g^2}d\rho\wedge D\phi\right),~~~~e^{3A}H_3=-e^{-\varphi} g H_1\wedge j,
\end{align}
where $q=q(\rho)$, $\tilde{G}=\tilde{G}(y_i)$, and furnishes us with the PDE
\beq
\partial_{y_i}^2\Delta=0.\label{eq:class2gneq0genpde}
\eeq
The remaining terms in \eqref{eq:gbps0}-\eqref{eq:gbps8} then fix the components of the RR flux as
\begin{align}
f_-&=-\epsilon_{ij}\partial_{y_i}\left(\frac{\tilde{G}qq'}{\partial_{\rho}\Delta}\right)dy_j+ \frac{1}{\partial_{\rho}\Delta}\partial_{\rho}\left(\frac{\tilde{G}qq'}{\partial_{\rho}\Delta}\right)D\phi+ g qe^{-6A}\star_4 H_1,\nn\\[2mm]
e^{2A}g_-&=-q e^{-4A}\star_4(H_1\wedge j_2)-\frac{2g \tilde{G}q'}{q\partial_{\rho}}dy_1\wedge dy_2\wedge D\phi,\nn\\[2mm]
e^{3A}g_+&= 4 g^2 \tilde{G}dy_1\wedge dy_2.
\end{align}
At this point conditions for supersymmetry have been reduced to the PDE in \eqref{eq:class2gneq0genpde}, but we still need to impose the Bianchi identities. These fix
\beq
H_1= \frac{1}{q} h_idy_i,~~~~e^{2\Delta}= e^{2\Delta^{(2)}}(q^2-\rho_0^2),~~~~~\tilde{G}= e^{2\Delta^{(2)}}G,
\eeq
with $(h_i,~\Delta^{(2)},~G)$ depending on $y_i$ only and $\rho_0$ a constant, and  require that
\begin{align}
&\epsilon_{ij}\partial_{y_i}h_j=0,~~~~\partial_{y_i}(G^2h_i)=0,~~~~ (h_i)^2= 4 e^{2\Delta^{(2)}}g^2 \rho_0^2,~~~~ \partial_{y_i}^2 G=0.\label{eq:bianchiscase2eneq0}
\end{align}
This exhausts the conditions that need to be imposed to define an embedding, however we note that \eqref{eq:class2gneq0genpde} has become $\partial_{y_i}^2\Delta^{(2)}=0$ and $\Delta^{(2)}$ only appears in the combinations 
\beq
d\tilde{s}^2_2=e^{2\Delta^{(2)}}(dy_i)^2,~~~~\tilde{\text{vol}}_2=e^{2\Delta^{(2)}}dy_1\wedge dy_2.
\eeq
It is a simple matter to confirm that this implies that the metric on $d\tilde{s}^2_2$ is flat, so we can fix $\Delta^{(2)}=0$ without loss of generality.
\\
~\\
In summary we find a class of embeddings of  the form
\begin{align}
ds^2&= \sqrt{\frac{\rho}{G}}\bigg[g^{(6)}_{\mu\nu}dx^{\mu}dx^{\nu}+\frac{e^{\varphi}}{2g^2}\left(\frac{d\rho^2}{\rho^2-\rho_0^2}+ \frac{\rho^2-\rho_0^2}{\rho^2}D\phi^2\right)\bigg]+  2g^2e^{-\varphi}\sqrt{\frac{G}{\rho}}(dy_i)^2,\nn\\[2mm]
e^{-\Phi}&=G,~~~~D\phi=d\phi+2g {\cal A},~~~~H=\frac{1}{2g \rho^2}h_i dy_i\wedge\left(2\rho g {\cal F}- d\rho\wedge D\phi\right),\nn\\[2mm]
F_1&=-\epsilon_{ij}\partial_{y_i}G dy_j,~~~~F_5=(1+ \star)\bigg(-\frac{e^{\varphi}}{g}d\rho\wedge \star_6 {\cal F}+ 4 g^2 G dy_1\wedge dy_2\wedge {\cal G}\bigg),\nn\\[2mm]
F_3&=\frac{G}{2g \rho^2}\epsilon_{ij}h_idy_j\wedge\left(2g\rho {\cal  F} -d\rho\wedge D\phi\right).\label{eq:genclass2}
\end{align}
where we have chosen to use diffeomorphism invariance to fix $q=\rho$, and $(h_i,~G)$ have support on $y_i$. One has a consistent truncation whenever the Bianchi identities, \textit{i.e.} \eqref{eq:bianchiscase2eneq0}, are satisfied with $\Delta^{(2)}=0$. We note that the internal space becomes conformally Kahlar when $h_i= 0$, but this also requires $\rho_0=0$.

It is interesting to find another class of embeddings that is compatible with the whole of Einstein-Maxwell gauged supergravity. Clearly in this case $G$ plays the role of a D7 brane warp factor as it appears in the correct places in the metric and $\Phi$ and is harmonic on $y_i$. However the interval spanned by $\rho$ in this case is quite clearly unbounded. When $\rho_0\neq 0$ the $\rho$ interval is bounded below at $\rho=\rho_0$ where the $(\rho, D\phi)$ directions vanish regularly with the rest of the warping constant in $\rho$. If $\rho_0=0$ the interval terminates at $\rho=0$ where there is a curvature singularity we do not recognise as corresponding to a physical object. In either case however $\rho$ is not bounded from above and $\rho\to \infty$ is at infinite proper distance\footnote{One way to see this is to compute the $d=6$ Newton constant $G_6$ as in \eqref{eq:6dnewton}, for the case at hand $G_6\to 0$.}. Thus at best this class of embeddings is on an equal footing to the non-compact consistent truncation of \cite{Cvetic:2003xr}. In fact the class of this section contains \cite{Cvetic:2003xr} modulo T-duality as we will now show.\\
~\\
\textbf{Recovering \cite{Cvetic:2003xr}}\\
~\\
If we impose that $\partial_{y_i}$ are both isometries then $(G,h_i)$ must be constant and the only constraint that needs to be imposed to have an embedding is $(h_i)^2= 4 g^2 \rho_0^2$, which we choose to solve as
\beq
h_1=-2 g \rho_0,~~~~h_2=0,~~~~dG=0.\label{eq:cvetictuning}
\eeq
This results in the NS flux in \eqref{eq:genclass2} being expressible as
\beq
H=dB,~~~~B= -\frac{\rho_0}{\rho}D\phi\wedge dy_1,
\eeq
while $F_1=0$, $F_3$ is strictly parallel to $dy_2$ and $F_5$ decomposes into terms  strictly parallel to $dy_1\wedge dy_2$ and strictly orthogonal to it.  Thus if we T-dualise on both of $\partial_{y_i}$ we will be mapped to a solution in type IIB with the RR 3-form the only non-trivial flux, as the bi-linears are not charged under $\partial_{y_i}$ this will preserve supersymmetry. Following the T-duality rules in \cite{Kelekci:2014ima} and assuming that $y_i\sim y_i +2\pi$ we are mapped to the following
\begin{align}
ds^2&=\sqrt{\frac{\rho}{G}}\bigg[g^{(6)}_{\mu\nu}dx^{\mu}dx^{\nu}+\frac{e^{\varphi}}{2g^2}\left(\frac{d\rho^2}{\rho^2-\rho_0^2}+ \frac{\rho^2-\rho_0^2}{\rho^2}D\phi^2+\left(dy_1-\frac{\rho_0}{\rho}D\phi\right)^2+dy_2^2\right)\bigg],\nn\\[2mm]
e^{-\Phi}&= \frac{2 g^2G^{\frac{3}{2}}}{\sqrt{\rho}}e^{-\varphi},\label{eq:cvsol1}\\[2mm]
F_3&=\frac{G \rho_0}{\rho^2}\left(dy_1- \frac{\rho_0}{\rho}D\phi\right)\wedge (2g \rho {\cal F}-d\rho \wedge D\phi)+G\left(4g^2  {\cal G}-2 \frac{2g}{\rho^2}(\rho^2-\rho_0^2)D\phi\wedge {\cal F}\right),\nn
\end{align} 
with the other fluxes all trivial. When $\rho_0 \neq 0$ this is in-fact the embedding of \cite{Cvetic:2003xr} in disguise, or rather its S-dual. To make this fact more clear one can first fix
\beq
\rho=\rho_0 \cosh(2\rho),~~~~g=2 \bar{g},~~~G=\rho_0= \frac{1}{8 \bar{g}^2}, ~~~y_1=\alpha+\beta,~~~y_2=2 \sqrt{2}\bar{g} z,~~~\psi=-\alpha+\beta,
\eeq
then S-dualise as
\beq
ds^2\to e^{-\Phi}ds^2,~~~~F_3\to H,~~~~\Phi\to -\Phi.
\eeq
This maps \eqref{eq:cvsol1} to
\begin{align}
ds^2&=e^{-\varphi}g^{(6)}_{\mu\nu}dx^{\mu}dx^{\nu}+dz^2+\frac{1}{2\bar{g}^2}\bigg(d\rho^2+\frac{\cosh^2\rho}{\cosh^2(2\rho)}(d\alpha-2 \bar{g}{\cal A})^2+\frac{\sinh^2\rho}{\cosh^2(2\rho)}(d\beta+2 \bar{g}{\cal A})^2\bigg),\nn\\[2mm]
H&=\frac{\sinh\rho\cosh\rho}{2\bar{g}^2\cosh^2(2\rho)}(d\alpha-2 \bar{g}{\cal A})\wedge (d\beta+2 \bar{g}{\cal A})+2 {\cal G}\\[2mm]
&+\frac{1}{2\bar{g}}\bigg(\frac{\cosh^2\rho}{\cosh(2\rho)}(d\alpha-2 \bar{g}{\cal A})-\frac{\sinh^2\rho}{\cosh(2\rho)}(d\beta+2 \bar{g}{\cal A})\bigg)\wedge(2{\cal F}),~~~e^{-\Phi}= \sqrt{\cosh(2\rho)}e^{\varphi}.\nn
\end{align}
Comparing this to the d=10 flux and dilaton in (47) of \cite{Cvetic:2003xr}, and string frame metric of (49) therein, it is not hard to see that the embeddings are the same -- one should identify
\beq
2\varphi=\phi,~~~ 2{\cal G}= H_{(3)},~~~~2{\cal A}=A,~~~~2{\cal F}= F_{(2)},~~~g^{(6)}_{\mu\nu}dx^{\mu}dx^{\nu}=ds^2_6,
\eeq
where the quantities on the right hand side of these equalities are the $d=6$ fields in \cite{Cvetic:2003xr}. 

Thus the class of embeddings in this section are a generalisation of the embedding in \cite{Cvetic:2003xr}, however as  they are like-wise manifestly un-bounded we will not comment of them further. 
\subsubsection{A class without the tensor multiplet and \texorpdfstring{$a_1=0$}{a1=0}}

We will now consider embeddings of minimal $d=6$ supergravity coupled to the vector  multiplet only and with $g\neq 0$. Unlike the type IIB classes with $g=0$, it is not possible to solve \eqref{eq:bpsgneqIIB1}-\eqref{eq:bpsgneqIIB7} in this limit, namely
\beq
\varphi=0,~~~\star_6{\cal G}=-{\cal G}
\eeq
without deciding whether $a_1=0$ or not. As such we shall begin by fixing $(a_1,~a_2)=(0,~1)$ in this section and then consider the case of $a_1\neq 0$ in the next section.\\
~\\
Upon turning off the tensor multiplet and fixing $a_1=0$ we have that \eqref{eq:bpsgneqIIB1}, \eqref{eq:bpsgneqIIB2} and \eqref{eq:bpsgneqIIB4} imply that we can fix the vielbien in terms of local coordinates $(\rho,y_1,y_2)$ as
\beq
e^{2A+C-\Phi}\tilde{W}= \frac{d\rho}{2g^2q},~~~~e^{4A+C-\Phi}U= e^{\Delta}d(y_i+i y_2),~~~e^{4A-\Phi}=q,
\eeq
for $q=q(\rho)$ and $\Delta=\Delta(\rho,y_i)$. The remaining conditions in \eqref{eq:bpsgneqIIB1}-\eqref{eq:bpsgneqIIB7} then imply that $\partial_{\rho}\Delta\neq 0$, and fix
\begin{align}
e^{2A+2C-\Phi}&=\frac{q'}{2g^2\partial_{\rho}\Delta},~~~~V=\epsilon_{ij}\partial_{y_i}\Delta dy_j~~~~ e^{2A}H_3=-g H_1\wedge j_2,
\end{align}
where $H_1$ is an arbitrary $\partial_{\phi}$ respecting 1-form on M$_4$, and impose the following PDE
\beq
\partial_{y_i}^2\Delta+ \frac{2g^4}{q'}\partial_{\rho}\left(\frac{1}{e^{4A}q'}\partial_{\rho}\left(e^{2\Delta}\right)\right)=0.\label{eq:notensorgneq0pde1}
\eeq
As before, the remaining non-trivial parts of  \eqref{eq:gbps0}-\eqref{eq:gbps8} now just fix the RR fluxes, this time as
\begin{align}
f_-&=-\epsilon_{ij}\partial_{y_i}(e^{-4A}q)dy_j+\frac{1}{\partial_{\rho}\Delta}\partial_{\rho}(e^{-4A}q)D\phi +g q e^{-6A} \star_4 H_1,\nn\\[2mm]
e^{2A}g_-&=\frac{q}{g e^{2A}}\star_4 H_3- \frac{2g e^{2\Delta}}{e^{4A}q}D\phi\wedge dy_1\wedge dy_2,\nn\\[2mm]
e^{3A}(1+\star_4\lambda)g_+&=\frac{q'}{g^2}D\phi\wedge d\rho+\frac{4g^2\partial_{\rho}\Delta}{e^{4A}q'}dy_1\wedge dy_2,
\end{align}
at which point the conditions for symmetry have been reduced to finding a solution of \eqref{eq:notensorgneq0pde1}. As previously, we need to solve the Bianchi identities of the fluxes to have a solution, only now we have no tensor multiplet, so there are less of them. It is possible to establish that these fix
\beq
\iota_{\partial_{\phi}}H_1=0,~~~~~e^{A}= \frac{\sqrt{2}g}{h^{\frac{1}{4}}},~~~~H_1=  \frac{1}{q}h_i dx_i+b_0\frac{\partial_{\rho}\Delta}{h}d\rho,
\eeq
where $h=h(y_i),~ h_i=h_i(\rho,y_i)$ and $b_0$ is a constant, and impose the PDEs
\begin{align}
&\epsilon_{ij}\partial_{y_i}h_j=0,~~~~\partial_{\rho}h_i=b_0 \partial_{y_i}\left(\frac{q \partial_{\rho}\Delta}{h}\right),~~~~(h_i)^2+\frac{b_0^2 qe^{2\Delta}}{(q')^2h}(\partial_{\rho}\Delta)^2= \frac{2 g^2q^3}{q'}\partial_{\rho}\left(\frac{1}{q^2}e^{2\Delta}\right)\nn\\[2mm]
&\partial_{y_i}(h^2h_i)+b_0 \frac{q^2h^2}{2q'}\partial_{\rho}\left(\frac{1}{q^2q'}\partial_{\rho}\left(e^{2\Delta}\right)\right)=0,~~~~\partial_{y_i}^2h=0,\label{eq:class1notensorcondsgneq0}
\end{align}
which exhausts the embedding equations.\\
~\\
To summarise we find a class of embeddings  with NS sector of the form
\begin{align}
ds^2&= 2g^2\sqrt{\frac{q}{h}}g^{(6)}_{\mu\nu}dx^{\mu}dx^{\nu}+\frac{\partial_{\rho}\Delta}{\sqrt{q}}\bigg[\frac{q'}{\sqrt{ h }}\left(d\rho^2+\frac{1}{\partial_{\rho}\Delta}D\phi^2\right)+\frac{\sqrt{h}}{q q'}e^{2\Delta}(dy_i)^2\bigg],~~~e^{-\Phi}=\frac{h}{4g^2},
\nn\\[2mm]
H&=\frac{1}{2g}d(D\phi\wedge H_1)-\frac{1}{2g}H_1\wedge dV+\frac{b_0}{2g q^2 q'}(\partial_{\rho}e^{\Delta})^2dy_1\wedge dy_2\wedge d\rho,\nn\\[2mm]
H_1&=\frac{1}{q}h_i dy_i+b_0\frac{\partial_{\rho}\Delta}{h}d\rho,~~~~~D\phi=d\phi+V+2g {\cal A},~~~~V=\epsilon_{ij}\partial_{y_i}\Delta dy_j,
\end{align}
where $h$ has support on $y_i$, $\Delta=\Delta(\rho,y_i)$ and $q$, which can be fixed with a coordinate transformation, on $\rho$. The background also supports the following non-trivial RR fluxes
\begin{align}
F_1&=-\frac{1}{4g^4}\epsilon_{ij}\partial_{y_i}h dy_j,~~~F_5=(1+\star)\left(\frac{q'}{g^2}D\phi\wedge d\rho\wedge {\cal G}-\frac{q'}{g^2}d\rho\wedge \star_6{\cal F}\right),\nn\\[2mm]
F_3&=-\frac{h}{8 g^5 q^2}\left(q' \epsilon_{ij}h_i dy_j\wedge d\rho+ \frac{b_0 \partial_{\rho}\Delta}{q'}e^{2\Delta}dy_1\wedge dy_2\right)+ \left(\frac{h}{4 g^2 q}\epsilon_{ij} h_i dy_j-\frac{b_0}{4g^2}D\phi \right)\wedge{\cal F},
\end{align}
which is to say everything possible. Supersymmetry requires that the PDE
\beq
2\partial_{y_i}^2 \Delta+ \frac{h}{q'}\partial_{\rho}\left(\frac{1}{q'q}\partial_{\rho}\left(e^{2\Delta}\right)\right)=0,
\eeq
holds, while we have a consistent truncation whenever \eqref{eq:class1notensorcondsgneq0} are also satisfied.

This class of embeddings is not obviously unbounded like that of the previous section, though we do not currently know whether it contains anything bounded. We do know that a simple separation of variables ansatz like that of section \ref{sec:towardscompacted} leaves the $\rho$ interval semi infinite. We note that $h$ appears where one would expect a D7 warp factor to appear and obeys the correct PDE for such an interpretation. While there is no tensor multiplet turned on, this class is still compatible with non-trivial $d=6$ solutions such as the Mink$_4\times$S$^2$ solution of Salam-Sezgin. As such the class has promise and it would be interesting to study it in more detail.

\subsubsection{A class without the tensor multiplet and \texorpdfstring{$a_1\neq 0$}{a=/=0}}
The final class we will consider also has no tensor multiplet, but we now assume that $a_1 \neq 0$, meaning that we can divide by it. It is possible to show that
\beq
d(e^{4A+C-\Phi}a_1^{-1}\tilde{W})=0,
\eeq
must hold under this assumption which together with \eqref{eq:bpsgneqIIB2} allows us to define the vielbein in terms of local coordinates $(\rho,y_i)$ for $i=1,2$ as
\beq
e^{4A+C-\Phi}a_1^{-1}\tilde{W}= \lambda_0 d\rho,~~~~e^{4A+C-\Phi}U= \lambda_0 e^{\Delta}d(y_1+i y_2),
\eeq
where $\lambda_0$ is a constant that can be chosen to any convenient value and $\Delta$ has support on $(\rho,y_i)$. With this definition \eqref{eq:bpsgneqIIB1}-\eqref{eq:bpsgneqIIB7} reduce to
\begin{align}
e^{2A-\Phi}a_1&=1,~~~~e^{2(A+C)}=\frac{\lambda_0 a_1^2}{\partial_{\rho}\Delta},~~~~~V= \epsilon_{ij}\partial_{y_i}\Delta dy_j,\nn\\[2mm]
H_3&= -\frac{1}{2g}D\phi\wedge dH_1+\lambda_0 d(\frac{e^{2\Delta}a_2 \partial_{\rho}\Delta}{e^{2A}a_1} dx_1\wedge dx_2)+ \lambda_0 \frac{a_1 a_2}{e^{2A}}dV\wedge d\rho,\nn\\[2mm]
H_1&=2g \lambda_0 e^{-2A}a_2 a_2 d\rho-\frac{1}{g}d\left(\frac{e^{2A}a_2}{a_1}\right)
\end{align}
where as before $\partial_{\rho}\Delta=0$ is not possible. We also find the PDE
\beq
e^{4A}2\partial_{y_i}^2\Delta+ \partial_{\rho}\left(a_1^{-2}\partial_{\rho}\left(e^{2\Delta}\right)\right)=4 \lambda_0 g^2 e^{2\Delta}\partial_{\rho}\Delta,\label{eq:finalpde}
\eeq
which is a generalisation of what we found in section \ref{eq:todaclass}. What remains of \eqref{eq:gbps0}-\eqref{eq:gbps8} that is not implied by the above fixes the RR flux components as
\begin{align}
f_-&=-\epsilon_{ij}\partial_{y_i}\left(\frac{a_2}{e^{2A}a_1}\right)dy_j-\frac{1}{\partial_{\rho}\Delta}\partial_{\rho}\left(\frac{a_2}{e^{2A}a_1}\right)D\phi- \lambda_0 \bigg[\frac{(a_1^2+1)^2}{a_1^2}\epsilon_{ij}\partial_{y_i}\left(\frac{a_1^2}{e^{4A}(a_1^2+1)}\right)dy_j\wedge d\rho\nn\\[2mm]
&+e^{2\Delta}\left(\frac{2g^2\lambda_0}{e^{8A}}+\frac{(a_1^2+1)^2}{2a_1^4}\partial_{\rho}\left(\frac{a_1^2}{e^{4A}(a_1^1+1)}\right)\right)dx_1\wedge dx_2\bigg]\wedge D\phi,\nn\\[2mm]
e^{2A}g_-&=-\frac{2 g \lambda_0}{e^{4A}\partial_{\rho}\Delta}D\phi-\frac{1}{g e^{2A}a_1}\left(\epsilon_{ij}\partial_{y_i}\left(\frac{e^{2A}a_2}{a_1}\right)-\frac{1}{\partial_{\rho}\Delta}\partial_{\rho}\left(\frac{e^{2A}a_2}{a_1}\right)D\phi\right)\wedge(a_2+a_1 j),\nn\\[2mm]
e^{3A}(1&+\star_4\lambda)g_+=2\left(1- \frac{a_2}{a_1}j+\frac{1}{2}\text{vol}(\text{M}_4)\right).
\end{align}
Again we have reduced the conditions for the embedding manifold to preserve supersymmetry to a single PDE, this time \eqref{eq:finalpde}. We can now derive the conditions that must hold for the fluxes to obey their Bianchi identities - that of the NS 3-form is implied by the supersymmetry constraints for this class, so it is only the second two constraints in \eqref{eq:gneq0bianchis} that need to be imposed. To express these we find it convenient to introduce $h=h(\rho,y_i)$ such that
\beq
e^{-4A}=h.
\eeq
 We then find that the Bianchi identities of $(f_-,~e^{2A}g_-)$ fix
\beq
a_2=\frac{a_1 G}{\sqrt{h}},~~~~ d\left(\frac{G^2\partial_{\rho}h+2 g^2 h^3}{h^2 \partial_{\rho}\Delta}\right)=0, \label{eq:finalcond1}
\eeq 
where $G$ has support on $y_i$. Notice that when $G=0$, which fixes $a_2=0$, the second of these can be used to fix $h$, but this appears more difficult to solve when $G\neq 0$. In addition we also find that the following PDEs
\begin{align}
&\partial_{y_i}^2G=0, \label{eq:finalcond2}\\[2mm]
&\partial_{y_i}^2h+ \partial_{\rho}\left(e^{2\Delta}\partial_{\rho}\left(h\right)\right)+ 4 g^2 \lambda_0 e^{\Delta}h\partial_{\rho}(e^{\Delta}h)+\frac{G^2}{2h^2}\left(\partial_{\rho}\left(e^{2\Delta}\partial_{\rho}\left(h^2\right)\right)-4 e^{2\Delta}(\partial_{\rho}h)^2\right)=0,\nn\\[2mm]
&2 G h\partial_{y_i} G \partial_{y_i}h-h^2 (\partial_{y_i} G)^2-G^2 (\partial_{y_i} h)^2= e^{2\Delta}G^2(\partial_{\rho}h)^2 \left(1+ \frac{G^2}{h}\right)\nn\\[2mm]
&+ 2 g^2 \lambda_0e^{2\Delta}  h^2\left(2g^2 \lambda_0 h^3+2 G^2 \partial_{\rho}h- h^2 \partial_{\rho}\Delta\right)\nn
\end{align}
are required to hold. When all these constraints are satisfied one has a consistent truncation.\\
~\\
In summary we have found a class of embeddings with NS sector of the form
\begin{align}
ds^2&= \frac{1}{\sqrt{h}} g^{(6)}_{\mu\nu}dx^{\mu}dx^{\nu}+\lambda_0\frac{\sqrt{h}}{\partial_{\rho}\Delta}\bigg[\frac{1}{\Xi}\left((\partial_{\rho}\Delta)^2d\rho^2+D\phi^2\right)+ e^{2\Delta}(\partial_{\rho}\Delta)^2(dy_i)^2\bigg],~~~e^{-\Phi}=\sqrt{\Xi h},\nn\\[2mm]
H&=-\frac{1}{2g} D\phi\wedge H_1+\lambda_0d(e^{2\Delta}G \partial_{\rho}\Delta dy_1\wedge dy_2)+ \frac{\lambda_0 G}{\Xi}dV\wedge d\rho+H_1\wedge {\cal F},\\[2mm]
H_1&=  \frac{2g \lambda_0 G }{\Xi}d\rho-\frac{1}{g}d\left(\frac{G}{h}\right),~~~~\Xi=1+ \frac{G^2}{h},~~~~D\phi=d\phi+V + 2g{\cal A},~~~~ V=\epsilon_{ij}\partial_{y_i}\Delta dy_j,  \nn
\end{align}
where $\lambda_0$ is a constant, $G$ depends on $y_i$ and $(\Delta, h)$ on $(\rho,y_i)$. In addition to this the background supports the following RR fluxes
\begin{align}
F_1&= -\epsilon_{ij} \partial_{y_i}G dy_j,~~~~F_5=(1+\star)\left(2\frac{G}{\Xi}D\phi\wedge d\rho\wedge {\cal G}+\frac{G\partial_{\rho}h}{g h^2}d\rho\wedge\star_6{\cal F}\right),\nn\\[2mm]
F_3&=2{\cal G}-\left(\frac{G}{g}\epsilon_{ij}\partial_{y_i}\left(\frac{G}{h}\right)dy_j- \frac{2g^2 \lambda_0 h^3+G^2 \partial_{\rho}h}{g h^2 \partial_{\rho}\Delta}D\phi\right)\wedge {\cal F}\\[2mm]
&-\lambda_0\bigg[\frac{(\Xi+1)^2}{2\Xi}\epsilon_{ij}\partial_{y_i}\left(\frac{g}{\Xi}\right)dy_j\wedge d\rho+e^{2\Delta}\left(2g^2 \lambda_0 h^2+ \frac{(\Xi+1)^2}{2}\partial_{\rho}\left(\frac{g}{\Xi}\right)\right)dy_1\wedge dy_2\bigg]\wedge D\phi.\nn
\end{align}
Supersymmetry requires that 
\beq
2\partial_{y_i}^2\Delta+ \partial_{\rho}\left(\Xi\partial_{\rho}\left(e^{2\Delta}\right)\right)=4 \lambda_0g^2 e^{2\Delta}h \partial_{\rho}\Delta.
\eeq
When this holds one has a consistent truncation to the gauged $d=6$ theory without a tensor multiplet whenever the second of \eqref{eq:finalcond1} and all of \eqref{eq:finalcond2} hold.

This class is a little complicated, it is not currently clear to us whether it contains bounded embeddings beyond the $G=0$ limit where it reduces to the class of embeddings in section \ref{eq:todaclass}. None the less, as this class is consistent with several interesting $d=6$ solutions it would be a worth while endeavour to explore it in more detail - but that lies beyond the scope of this already very long work.

\section{Towards bounded embeddings of Einstein-Maxwell gauged supergravity}\label{sec:towardscompacted}
In this section we explore the possibility of embedding full Einstein-Maxwell gauged supergravity into type IIB in terms of a bounded internal space. As such our focus will be on the class of embeddings derived in section \ref{eq:todaclass}, though the conditions we derive to identify when an internal space is bounded apply to all of our embeddings.  We will show that at least one bounded class does indeed exist, which follows from essentially the most simple solution to the toda-like equation that defines the class. However this embedding is plagued by singularities that are probably not physical - we at least have not been able to identify them as coming from well defined objects in string theory. As such, while this embedding solves the type II equations of motion almost everywhere, it likely does not solve them at the loci of the singularities, at least for sources with an understood string origin such as D-branes or O-planes. As such the results of this section should be viewed as a promising, but very preliminary, step towards the construction of a truly well defined embedding of the full $d=6$ theory.\\
~\\
One can determine whether an embedding is bounded by computing the effective 6 dimensional Newtons constant $G_6$, in particular this should be finite. The Einstein frame action of type IIB supergravity contains the Einstein-Hilbert term
\beq
\frac{1}{2\kappa_{10}}\int d^{10}x\sqrt{-\det g^{(10)}_E} R_E,\label{Eq:ehef}
\eeq
where the $E$ subscripts indicate Einstein frame. Given how the $d=6$ metric is embedded into type II supergravity \eqref{eq: 10dmet}, one can extract $G_6$ from the term proportional to the $d=6$ Einstein-Hilbert term
\beq
\int d^{6}x \sqrt{-\det g^{(6)}} R^{(6)},
\eeq
contained in \eqref{Eq:ehef}. A short computation leads to the string frame expression
\beq
\frac{1}{G_6}\propto  \int_{\text{M}_4}e^{4A-2\Phi}\text{vol}(\text{M}_4),\label{eq:6dnewton}
\eeq
where we note that the integrand is guaranteed to be independent of $\varphi$ through the internal supersymmetry condition \eqref{eq:gbpspairing}. We conclude that a given embedding is bounded if \eqref{eq:6dnewton} is finite.

To find a concrete embedding for the class of section \ref{eq:todaclass} we need to find a solution to the Toda like equation \eqref{eq:easygneqopde} - we proceed with the separation of variables ansatz
\beq
e^{\Delta}= e^{\mu(y_1,y_2)} q(\rho)~~~\Rightarrow~~~ \partial^2_{y_i}\mu+ e^{2\mu}q q''=0.
\eeq
This implies that the Riemann surface spanned by $y_i$ has constant curvature such that its Ricci scalar is $R^{(2)}=2 \kappa$ when
\beq
\partial_{y_i}^2\mu+ \kappa e^{2\mu}=0,
\eeq
which yields a 2-sphere, torus or hyperboloid when $\kappa>0$, $\kappa=0$ or $\kappa<0$ respectively. This means that the function $q$ obeys
\beq
q q''=\kappa.
\eeq
When $\kappa=0$ this makes $q$ a linear function that cannot bound $\rho$ interval at both ends, while for $\kappa \neq 0$ it is solved as
\beq
q= \sqrt{\frac{2}{\pi }}\rho_0\text{exp}\left(-\text{efr}^{-2}\left(\frac{\sqrt{-\kappa}\rho}{\rho_0}\right)\right),
\eeq
where $\text{efr}$ is  the error function
\beq
\text{erf}(x)=\frac{2}{\sqrt{\pi} }\int_0^{x} e^{-y^2}dy,
\eeq
$\text{efr}^{-2}(x)$ its inverse squared and $\rho_0$ is a constant. We only have that $q$ is real when $\kappa<0$, and when this is so we can set $\kappa=-1$ without loss of generality. This makes $q$ a semi-circular contour starting at $\rho=-\rho_0$ and ending at $\rho=\rho_0$, with a maximum at $\rho=0$, when $\rho_0>0$, about which $q$ is symmetric. We then have that the metric and dilaton are given by
\beq
ds^2= \sqrt{\frac{q}{q'}}\bigg[2g^2g^{(6)}_{\mu\nu}dx^{\mu}dx^{\nu}+e^{\varphi}\bigg(D\phi^2+q'\left(\frac{d\rho^2}{q}+ds^2(\mathbb{H}^2)\right)\bigg)\bigg],~~~
e^{-\Phi}=\frac{\sqrt{\frac{q'}{q}}}{2g^2} e^{-\varphi}.
\eeq
We can make $\mathbb{H}^2$ compact by taking a discrete quotient of it such that $\mathbb{H}^2\to \mathbb{H}^2/\Gamma$, so the question is whether $(q,q')$ bound the $\rho$ interval to some finite range where the metric remains positive, which requires 
\beq
q'>0,~~~~q>0,
\eeq
on the interior of the interval. The inequalities follow when $\rho_0>0$ for $\rho$ in the range $ -\rho_0 <\rho<  0$ while at the boundaries of this interval we find
\begin{align}
&(q,~q')\sim \sqrt{2}\sqrt{|\log(\rho+\rho_0)|}\left(\rho+\rho_0,~1\right),~~~~\rho\to -\rho_0^+,\nn\\[2mm]
&(q,~q')\sim \left(\sqrt{\frac{2}{\pi}}\rho_0,-\sqrt{\frac{\pi}{2}}\frac{\rho}{\rho_0}\right),~~~~\rho\to 0^-.
\end{align}
From this it follows that the the interval is bounded from below at $\rho=-\rho_0$ and above at $\rho=0$ where the space is singular. Close to $\rho=0$ the warp factor of the external directions and $d=10$ dilaton behave like a D5 brane in flat space, but the internal direction do not obviously conform to such an interpretation. Conversely as $\rho\to -\rho_0$ the dilaton and external warp factors have $\rho$ dependence consistent with an O5 plane, but again the internal directions don't appear to be consistent with this. Thus we cannot say with confidence that the singularities that bound this embedding are physical, indeed they likely are not.   Putting aside the probable un-physicality of the singularities let us establish whether $G_{6}^{-1}$ is actually bounded in this case. We find that
\beq
e^{4A-2\Phi}\text{vol}(\text{M}_4)= \frac{\text{erf}^{-2}\left(\frac{\rho}{\rho_0}\right)}{g^2}\text{vol}(\mathbb{H}^2/\Gamma)\wedge d\rho\wedge d\phi,
\eeq
which if we substitute into \eqref{eq:6dnewton} and integrate $\rho\in [-\rho_0,0]$ and $\phi\in [0,2\pi)$ yields
\beq
\frac{1}{G_6}\propto \frac{2\pi\rho_0}{2g^2}\text{Vol}(\mathbb{H}^2/\Gamma),
\eeq
which being constant, clearly is non-divergent. We conclude that the simple embedding we have derived is indeed bounded.

We have thus found a prototype consistent truncation about a bounded embedding manifold, unfortunately though it has some spurious singularities bounding the $\rho$ interval which means it likely does not solve the matter coupled $d=10$ equations of motion for sensible sting sources. Notice though that the separation of variable ansatz we have made is essentially the simplest way to solve the defining PDE of the class of embeddings in section \ref{eq:todaclass}, it is extremely probable that further embeddings exist, and perhaps some do not suffer the pathology of this prototype. It would be worth while exploring these possibilities and whether any embedding is at least bounded by obviously physical singularities, that is however beyond the scope of this work.

\section{Conclusions}\label{eq:section 8}
Continuing in the spirit of \cite{Gauntlett:2007ma}, in this work we have further developed the application of G-structure and bi-spinor techniques to construct consistent truncations to minimal (gauged) supergravities coupled to matter multiplets. This is interesting because many novel solutions lie in such theories. One could of course employ exceptional field theory methods to achieve this goal following \cite{Cassani:2019vcl}, but we find that they are not necessary in this context, and it can be useful to have several different approaches to reach a common aim.

Our focus here has been on embedding $d=6$ Einstein-Maxwell (gauged) supergravity into type II supergravity. We reviewed the 6d theory and provided geometric conditions for its supersymmetry preservation in section \ref{sec: the 6dtheory}. We also present some interesting known solutions the theory contains in section \ref{sec:6dsoltions}. In section \ref{sec:two} we provided general conditions on the internal $d=4$ embedding manifold that ensure that when $d=6$ supersymmetry holds $d=10$ supersymmetry is implied. We further prove under this assumption that the EOM of type II are implied by those of the $d=6$ theory for these embeddings, and argue that they should also hold in the absence of external supersymmetry. We assume that when the gauge field ${\cal A}$ appears in the metric, as it must in the presence of R-symmetry gauging, it appears housed in a single U(1) isometry. Upon this foundation we construct classes of  embeddings for the un-gauged limit of the theory in sections \ref{eq:Mink6vac} - \ref{eq:section 6}. We then focus on classes of gauged embeddings in section \ref{eq:section 7}. Among the many classes two  stand out as the most promising candidates for constructing concrete embeddings: First is the class of section  \ref{sec:universal} that provides a universal embedding of minimal $d=6$ supergravity coupled to a vector multiplet and no R-symmetry gauging. The second is an embedding of the full gauged theory that is governed by a Toda-like equation in  section \ref{eq:todaclass}. We show in section \ref{sec:towardscompacted} that this contains at least one prototype  bounded embedding. However, possessing singularities that we have been unable to associate to well defined objects in string theory, this likely does not solve the matter coupled type II equations of motion at the loci of the singularities. However it should be appreciated that the prototype embedding descends from the simplest solution to the Toda-like equation, there will be many more solutions, and it is quite possible that some are fully well defined.

At this point we should acknowledging an elephant in the room, namely that $d=6$ Einstein-Maxwell supergravity is not an anomaly free theory. Indeed anomaly cancellation for minimal $d=6$ supergravity coupled to $n_T$ tensor, $n_V$ vector and $n_H$ hyper multiplets requires, among other things, that  \cite{Bonetti:2011mw,Guo:2025mlb}
\beq
29n_T+n_H -n_V=273,\label{eq:anomolycancel}
\eeq  
whether or not the R-symmetry is gauged. In either case additional conditions must also be imposes and in the presence of R-symmetry gauging these are quite restrictive which makes fully anomaly free $d=6$ gauged theories quite rare - see \cite{Becker:2025xgy} for the state of the art. However one thing is clear from \eqref{eq:anomolycancel} alone - one needs many more fields turned on than Einstein-Maxwell supergravity possess to realise anomaly cancellation in $d=6$. Given that type II supergravity is well known to be anomaly free this leads to an apparent contradiction for the uplifts we have constructed. Our view is that uplifts should exist which generalise the classes that appear in this work, and for which anomaly cancellation is realised. Given the large number of $d=6$ fields that would require housing in $d=10$, constructing these generalised embeddings seems daunting, but one needs to start somewhere and we have provided the first steps in this direction.  
~\\

Some future directions:\\

Having performed a detailed classification of the possible embeddings of $d=6$ Einstein-Maxwell (gauge) supergravity, the next step is to systematically explore the possible embeddings in the most promising classes - namely those of section \ref{sec:universal} and  \ref{eq:todaclass}. We hope to report on this in the future. 

We made two assumptions in this work which it might be beneficial to relax. The first is that the spinors the internal manifold supports have equal norm - \textit{i.e.} we consistently imposed $\cos\beta=0$ after section \ref{sec:universal}. This is well motivated for Mink$_6$ vacua and so also for uplifts of the un-gauged $d=6$ theory with $g=0$, but beyond tractability, we do not have an argument for imposing this when $g\neq 0$. The second assumption was that we sought only embeddings for the full 6d theory and its consistent sub-sectors - \textit{i.e.} we sought consistent truncations. However, when the tensor multiplet is turned on it is constructing an embedding consistent with  ${\cal F}\wedge {\cal F} \neq 0$ and $d\varphi\neq 0$ that is often so restrictive. Solutions, such as the scale separated AdS$_3$ solutions of \cite{Proust:2025vmv}, exist for which both these terms are zero and it may be possible to construct more general embeddings (that are not true consistent truncations) for them. It is notable that supersymmetry is blind to the flux component $(1-\star_4\lambda)g_{\pm}$ when $d\varphi=0$, and perhaps there are choices that can be made for it that more easily solve the Bianchi identities in this limit.

Another possible generalisation of this work  it could be interesting to pursue is  embeddings for the SU(2) gauged version of the $d=6$ theory considered here.  Solutions in this theory were classified in \cite{Cariglia:2004kk} and include a Mink$_3\times$S$^3$ solution that yields an Einstein static universe like background upon reduction to $d=4$ - this could have interesting consequences for string cosmology. It would also be interesting to see whether our methods can be used to construct embeddings of de Sitter solutions, such as those of \cite{Burgess:2024jkx,Guo:2025mlb}.  

Relating to the application of G-structures and bi-spinors to consistent truncations more  broadly, and along more of a holography vein: It would be very interesting to construct embeddings for $d=5$ minimal gauged supergravity coupled to Abelian vector multiplets with U(1) R-symmetry gauging \cite{Gunaydin:1984ak}. Such solutions where classified in  \cite{Gauntlett:2003fk,Gutowski:2004yv,Gutowski:2005id}, with the inclusion of hyper multiplets considered in \cite{Bellorin:2007yp}. Finding embeddings for this theory would have interesting application for asymptotically AdS$_5$ black holes and the AdS$_5$/CFT$_4$ correspondence. Along similar lines, embeddings for $d=4$ ${\cal N}=2$ minimal gauged supergravity coupled to vector multiplets would also be valuable. Such solutions where classified in \cite{Caldarelli:2003pb,Cacciatori:2008ek,Klemm:2009uw} and are know to contain interesting  asymptotically AdS$_4$ black hole solutions \cite{Cacciatori:2009iz}. 

\section*{Acknowledgements}
The work of the authors is supported by the Ram\'on y Cajal fellowship RYC2021-033794-I, and by grants from the Spanish government MCIU-22-PID2021-123021NB-I00 and principality of Asturias SV-PA-21-AYUD/2021/52177. The work of NM was supported in part by Simons Foundation award number 1023171-RC.

\appendix

\section{Conventions}\label{eq:convensions}
In this section we briefly clarify the conventions we use in this work. They are in fact identical to appendix A of \cite{Legramandi:2023fjr} where more details are given.\\
~\\
We use the following notation for the contraction of forms: A if $C_k$ is a $k$-form then
\beq
C_k^2= \frac{1}{k!}C_{A_1...A_k}C^{A_1...A_k},~~~~(C_k)^2_{MN}= \frac{1}{(k-1)!}C_{M A_1...A_{k-1}}C_N^{~A_1...A_{k-1}}.
\eeq
In terms of a vielbein $e^{\underline{M}}$ the Hodge dual in $d$ dimensions is defined as
\beq
\star(e^{\underline{M}_1}\wedge ...\wedge e^{\underline{M}_k})=\frac{1}{(d-k)!}\epsilon^{\underline{M}_1...\underline{M}_k}_{~~~~~~~\underline{N}_{k+1}....\underline{M}_{d}}e^{\underline{N}_{k+1}}\wedge ...\wedge e^{\underline{N}_d},
\eeq
which in particular, we would like to stress, are not the conventions coded into the EDCRGTCcode.m Mathematica package! Indeed they differ by a sign when the Hodge dual is applied to an odd form in even dimensions, in other instances there is no difference. We define the Dirac slash to be
\beq
\slashed{C}_k= \frac{1}{k!}C_{A_1...A_k}\Gamma^{A_1...A_k},
\eeq
though for the most part we we will leave the slash implicit, \textit{i.e} terms like  $C_k\zeta$ for $\zeta$ a spinor should be understood as $\slashed{C}_k\zeta$. Likewise if two forms $(X,Y)$ appear like  $X Y$, without a wedge, this should be read as $\slashed{X}\slashed{Y}$.

\section{A canonical frame for Lorentzian bi-linears in \texorpdfstring{$d=6$}{d=6}}\label{canonicalframe}
For a single negative chirality Lorentzian spinor $\zeta_-$ there always exists a canonical frame where
\beq
\hat\gamma^{(6)} \zeta_-=-\zeta_-,~~~~\gamma^{(6)}_{01}\zeta_-=\zeta_-,~~~~\gamma^{(6)}_{23}\zeta_-=i \zeta_-,~~~~\gamma^{(6)}_{24}\zeta_-=\zeta^c_-.
\eeq 
In such a frame it is not hard to establish that
\begin{align}
k&=f (- e^0+e^1),~~~~v=\frac{1}{2f}(e^0+e^1),\nn\\[2mm]
J&=e^{23}+e^{45},~~~~\Omega=(e^2+i e^3)\wedge (e^4+i e^5),
\end{align}
where $f=\zeta_-^{\dag}\zeta_-$ and so that indeed
\beq
g^{(6)}_{\mu\nu}dx^{\mu}dx^{\nu}=2 k v+ds^2(\text{M}_{\text{SU}(2)})= -(e^{0})^2+(e^1)^2+\sum_{i=2}^{5} (e^i)^2.
\eeq
With respect to this frame it is trivial to establish the following identities
\begin{align}
\star_6k&=-\frac{1}{2}k\wedge J\wedge J,~~~\star_6(k\wedge J)=-k\wedge J,~~~\frac{1}{2}\star_6 (k\wedge J\wedge J)=-k,\nn\\[2mm]
\star_6v&=\frac{1}{2}v\wedge J\wedge J,~~~\star_6(v\wedge J)=v\wedge J,~~~\frac{1}{2}\star_6 (v\wedge J\wedge J)=v,\nn\\[2mm]
\star_6 1&=-\frac{1}{2}k\wedge v\wedge J\wedge J,~~~\star_6 J=-k\wedge v\wedge J,~~~\frac{1}{2}\star_6(J\wedge J)=-k\wedge v,\nn\\[2mm]
\star_6(k\wedge v)&=\frac{1}{2}J\wedge J,~~~\star_6(k\wedge v\wedge J)=J,~~~\frac{1}{2}\star_6(k\wedge v\wedge J\wedge J)=1,
\end{align}
which hold in any frame preserving the orientation 
\beq
\frac{1}{2}k\wedge v\wedge J\wedge J=- \text{vol}_6= -e^{0...5}
\eeq

\section{Deriving the missing constraints in \texorpdfstring{$d=6$}{d=6}}\label{eq:6dpairing}
We claim in the main text that \eqref{eq:BPS3} gives rise to 48 constraints while \eqref{eq:geometric3} only 45, this can be seen as follows: A negative chirality Lorentian spinor in 6 dimensions can depend on 8 independent real functions.  One can decompose any negative chirality Lorentian spinor in 6 dimensions in terms of a particular spinor $\zeta_-$ through the basis 
\beq
(\gamma^{(6)})_{\alpha\beta}\zeta_-.
\eeq
Naively this basis appears to be 15 dimensional while a negative chirality spinor can only depend on 8 independent real functions. However not all components of $(\gamma^{(6)})_{\alpha\beta}\zeta_-$ are non-zero and independent, indeed since $\zeta_-$ supports an SU(2)$\ltimes  \mathbb{R}^4$-structure, there are 7 independent $X^{\alpha\beta}(\gamma^{(6)})_{\alpha\beta}$ combinations, for $X^{\alpha\beta}$ a real antisymmetric matrix, that annihilate $\zeta_-$ - hence the basis is indeed 8 dimensional. This means that we can expand the spin covariant derivative as
\beq
\nabla_{\mu}\zeta_- = Q_{\mu}^{~\alpha\beta}(\gamma^{(6)})_{\alpha\beta}\zeta_-,\label{eq:genspinexpansion}
\eeq
where $Q_{\mu}^{~\alpha\beta}$, which parameterises the torsion classes of the SU(2)$\ltimes  \mathbb{R}^4$-structure,  is real and anti symmetric in $\alpha\beta$. One can further assume that each spatial direction of $Q_{\mu}^{~\alpha\beta}$ only contains 8 independent components, yielding a total of $6\times 8=48$. The condition \eqref{eq:BPS3} clearly fixes all 48 independent components of $Q_{\mu}^{~\alpha\beta}$, the question is how many components does \eqref{eq:geometric3} fix? One can easily establish this through the identities
\begin{align}
&-\nabla_{\mu}k_{\nu}= \overline{\nabla_{\mu}\zeta}_-\gamma_{\nu}\zeta_-+\overline{\zeta}_-\gamma_{\nu}\nabla_{\mu}\zeta_-,\nn\\[2mm]
&2d\slashed{\psi}^{(6)}_-=\nabla\zeta\otimes \overline{\zeta}_-+\zeta\otimes \overline{\nabla\zeta}_-+(\gamma^{(6)})^{\mu}\zeta_-\otimes \overline{\nabla_{\mu}\zeta_-}-\nabla_{\mu}\zeta_-\otimes\overline{\zeta}_-(\gamma^{(6)})^{\mu},\label{eq:derivativeidentities}
\end{align}
and similarly for $d\tilde{\psi}^{(6)}_-$. These allow us to relate the expansion \eqref{eq:genspinexpansion} to the derivative terms in \eqref{eq:geometric3}, for instance
\beq
\nabla_{(\mu}k_{\nu)}=0~~~~\Rightarrow~~~~k^{\beta}Q_{(\mu \nu) \beta}=0. 
\eeq
In this way we establish that 45 components of $Q_{\mu}^{~\alpha\beta}$ get fixed by \eqref{eq:geometric3}. What remains undetermined requires us to define a null 1-form $v$ such that
\beq
v.k=1,
\eeq
in terms of which we can decompose
\beq
g^{(6)}_{\mu\nu}dx^{\mu}dx^{\nu}= 2 k v+ ds^2(\text{SU(2)}).
\eeq
The components that do not enter \eqref{eq:geometric3} are
\beq
v^{\mu}Q_{\mu}^{~\underline{\alpha}_4\underline{\beta}_4},
\eeq
where the 4 subscript indicates vielbein directions along the $ds^2(\text{SU(2)})$ directions. Only 3 of these are independent and non-zero and can be parameterised as
\beq
v^{\mu}Q_{\mu}^{~\alpha\beta}J_{\alpha\beta},~~~~v^{\mu}Q_{\mu}^{~\alpha\beta}\Omega_{\alpha\beta}\label{eq:missingcomponents}.
\eeq

To derive some conditions that do give us these components we follow a method from \cite{Tomasiello:2011eb}. First we observe that
\beq
\overline{\zeta_-} \slashed{v}(\gamma^{(6)})^{\mu}\slashed{v}\zeta_-=-2v^{\mu},
\eeq
which allows one to realise $v$ as a bi-linear. Some useful identities going forward are
\beq
\overline{\zeta_-} \slashed{v}\zeta_-=-1,~~~~~\slashed{k}\slashed{v}\zeta_-=2\zeta_-,~~~~\slashed{k}\slashed{k}=\slashed{v}\slashed{v}=0
\eeq
Next we observe that
\beq
\{\nabla,\slashed{v}\}\zeta_-= 2v^{\nu}\nabla_{\nu}\zeta_-+ d\slashed{v}\zeta_-+ (\nabla.v)\zeta_-,\label{eq:tobereturnedto}
\eeq
and that 
\begin{align}
-2 \nabla.v&= \nabla_{\mu}(\overline{\zeta_-} v)(\gamma^{(6)})^{\mu}v\zeta_-+\overline{\zeta_-} v(\gamma^{(6)})^{\mu}\nabla_{\mu}(v\zeta)\nn\\[2mm]
&=-2 \nabla.v+ 2v^{\mu}\big(\overline{\nabla_{\mu}\zeta_-}v\zeta_-+\overline{\zeta_-}v\nabla_{\mu}\zeta_-\big)+\overline{\zeta}_-[v,dv]\zeta_-
\end{align}
where we have left the Dirac slash implicit. This informs us that
\beq
\text{Re}\bigg[ 2v^{\mu}\overline{\zeta}_-v\nabla_{\mu}\zeta_-+\overline{\zeta}_-vdv\zeta_-\bigg]=0,
\eeq
this does inform us that 
\beq
\iota_k\iota_vdv=0,
\eeq
but does not fix any additional torsion classes - indeed given that $k^{\mu}\partial_{\mu}$ is a null Killing vector and $v$ is a vielbein direction the stronger condition $\iota_vdv=0$ must hold. If we return to \eqref{eq:tobereturnedto} we can now establish that
\begin{align}
\overline{\zeta}_-v\nabla(v\zeta_-)&=-\nabla.v+i \text{Im}\bigg[ 2v^{\mu}\overline{\zeta}_-v\nabla_{\mu}\zeta_-+\overline{\zeta}_-vdv\zeta_-\bigg],\nn\\[2mm]
&=-\nabla.v- \frac{i}{2} J^{\alpha\beta}(dv)_{\alpha\beta}-2iv^{\mu}Q_{\mu}^{~\alpha\beta}J_{\alpha\beta}.
\end{align}
We would now like to give a pairing constraint that defines the remaining torsion classes, which is to say a condition that holds inside the bracket
\beq
(X,Y)_6= X\wedge \lambda(Y)\bigg\lvert_{6}.
\eeq
The reason this object is useful is the following identities
\begin{align}
&\frac{1}{2^{[\frac{D}{2}]}}\text{Tr}(\star X  Y)=(-1)^{\deg{X}}(X,Y),\nn\\[2mm]
&(X\Psi_{\pm}Y,C)=\mp\frac{1}{32}\overline{\epsilon}_1X C Y \epsilon_2\text{vol}_{D},\nn\\[2mm]
&\overline{\epsilon}_1X C Y\epsilon_2=\mp  \text{Tr}(Y\lambda(\Psi_{\pm})XC),\label{eq:pairingproperties}
\end{align}
where $D$ is the dimension of the space in question, $(\epsilon_1,\epsilon_2)$ are spinors on that space, and when that space is even dimensional
\beq
\slashed{\Psi}_{\pm}=\epsilon_1\otimes \overline{\epsilon}_2,
\eeq
we will not need to consider odd $D$ in this work.

We will first aim for a condition involving $(v\psi_-^{(6)}v,d(v\psi^{(6)}_-)\gamma_{\alpha\beta})_6$. Now, under the Dirac slash
\begin{align}
2\overline{\zeta}_-v d(v\psi^{(6)}_-)\gamma^{(6)}_{\alpha\beta}v\zeta_-&=\overline{\zeta}_-v \{\nabla_{\mu}(v\zeta_-\otimes \overline{\zeta}_-),(\gamma^{(6)})^{\mu}\}\gamma^{(6)}_{\alpha\beta}v\zeta_-\nn\\[2mm]
&=\overline{\zeta}_-v\bigg[\nabla(v\zeta_-)\otimes\overline{\zeta}_-+2v^{\mu}\zeta_-\otimes \nabla_{\mu}\overline{\zeta}_-\bigg]\gamma^{(6)}_{\alpha\beta}v\zeta_-\label{eq:vpsiminusexp}\\[2mm]
&=\bigg[\overline{\zeta}_-v\nabla(v\zeta_-)\bigg]\overline{\zeta}_-\gamma^{(6)}_{\alpha\beta}v\zeta_--2v^{\mu}\nabla_{\mu}\overline{\zeta}_-\gamma^{(6)}_{\alpha\beta}v\zeta_-\nn\\[2mm]
&=(\nabla.v+ \frac{i}{2} J^{\mu\nu}(dv)_{\mu\nu}+2iv^{\rho}Q_{\rho}^{~\mu\nu}J_{\mu\nu})(i J_{\alpha\beta}+2 k_{[\alpha}v_{\beta]})+2 v^{\rho}Q_{\rho}^{~\mu\nu}P_{\mu\nu\alpha\beta},\nn
\end{align}
where
\beq
P_{\mu\nu\alpha\beta}=\overline{\zeta}_-\gamma^{(6)}_{\mu\nu}\gamma^{(6)}_{\alpha\beta}v\zeta_-.
\eeq
While $ v^{\rho}Q_{\rho}^{~\mu\nu}P_{\mu\nu\alpha\beta}$ does not give us  $v^{\rho}Q_{\rho}^{~\mu\nu}$ directly, it is a complex antisymmetric matrix containing the same information, so in particular the missing components \eqref{eq:missingcomponents}.  One can show that 
\beq
v^{\rho}Q_{\rho}^{~\mu\nu}P_{\mu\nu\alpha\beta}=-(2v^{\rho}Q_{\rho}^{~\mu\nu}k_{\mu}v_{\nu}+ i v^{\rho}Q_{\rho}^{~\mu\nu}J_{\mu\nu})(i J_{\alpha\beta}+2 k_{[\alpha}v_{\beta]})+v^{\rho}Q_{\rho}^{~\mu\nu}\overline{\Omega}_{\mu\nu}\Omega_{\alpha\beta}+v^{\rho}Q_{\rho}^{~\mu\nu}\tilde{P}_{\mu\nu\alpha\beta}\nn,
\eeq
where the final term contains the already determined parts of $v^{\rho}Q_{\rho}^{~\mu\nu}$ and is expanded in a basis of $(k_{[\alpha},E^1_{\alpha]},k_{[\alpha},E^2_{\alpha]})$ for $(E^1,E^2)$ canonical complex vielbein directions such that
\beq
J= \frac{i}{2}(E^1\wedge \overline{E}^1+E^2\wedge \overline{E}^2),~~~~\Omega= E^1\wedge E^2.
\eeq
In particular this means that the $v^{\rho}Q_{\rho}^{~\mu\nu}J_{\mu\nu}$ dependent terms in the final line of \eqref{eq:vpsiminusexp} cancel and we are left with only
\beq
2\overline{\zeta}_-v d(v\psi^{(6)}_-)\gamma^{(6)}_{\alpha\beta}v\zeta_-=2v^{\rho}Q_{\rho}^{~\mu\nu}\overline{\Omega}_{\mu\nu}\Omega_{\alpha\beta}+...,\label{eq:6dpaiinginter}
\eeq
where $...$ contains only the already determined torsion classes. However given that $\Omega_{\alpha\beta}\overline{\Omega}^{\alpha\beta}=8$, while the contraction of this with anything else on the RHS of \eqref{eq:6dpaiinginter} gives zero, we actually only need 
\beq
\overline{\zeta}_-v d(v\psi^{(6)}_-)\overline{\Omega} v\zeta_-=4v^{\rho}Q_{\rho}^{~\mu\nu}\overline{\Omega}_{\mu\nu}.
\eeq
If we instead use \eqref{eq:BPS3} in the third line of \eqref{eq:vpsiminusexp} we find that
\beq
\overline{\zeta}_-v d(v\psi^{(6)}_-)\overline{\Omega} v\zeta_-=2e^{\varphi}\star_6(v\wedge \overline{\Omega}\wedge {\cal G}).
\eeq
This is equivalent to
\beq
v\wedge \overline{\Omega}\wedge\bigg[ d(v \psi^{(6)}_-)\bigg\lvert_3+\frac{1}{4}e^{\varphi} {\cal G}\bigg]=0,
\eeq
where we only take the 3-form part of $d(v \psi^{(6)}_-)$,  but 
\beq
v \psi^{(6)}_-=-\frac{1}{8}(1-k\wedge v)\wedge e^{-i J}, 
\eeq
so we arrive at
\beq
v\wedge\Omega\wedge\bigg[  d(k\wedge v-i J)+ 2 e^{\varphi} {\cal G}\bigg]=0,
\eeq
which gives another 2 of the torsion classes.

It remains to find $v^{\mu}Q_{\mu}^{~\alpha\beta}J_{\alpha\beta}$, it seem logical to consider
\begin{align}
2\overline{\zeta}_-v d(v\tilde{\psi}^{(6)}_-)\gamma^{(6)}_{\alpha\beta}v\zeta_-&=\bigg[\overline{\zeta}_-v\nabla(v\zeta_-)\bigg]\overline{\zeta^c}_-\gamma^{(6)}_{\alpha\beta}v\zeta_--2v^{\mu}\nabla_{\mu}\overline{\zeta^c}_-\gamma^{(6)}_{\alpha\beta}v\zeta_-\label{eq:tildepsipairing}\\[2mm]
&=(\nabla.v+ \frac{i}{2} J^{\mu\nu}(dv)_{\mu\nu}+2iv^{\rho}Q_{\rho}^{~\mu\nu}J_{\mu\nu})\Omega_{\alpha\beta}+2 v^{\rho}Q_{\rho}^{~\mu\nu}\overline{\zeta^c}_-\gamma^{(6)}_{\mu\nu}\gamma^{(6)}_{\alpha\beta}v\zeta_-,\nn\
\end{align}
This time the final term can be expanded as
\beq
 v^{\rho}Q_{\rho}^{~\mu\nu}\overline{\zeta^c}_-\gamma^{(6)}_{~\mu\nu}\gamma^{(6)}_{\alpha\beta}v\zeta_-=(-2 v^{\rho}Q_{\rho}^{~\mu\nu}k_{\mu}v_{\nu}+i v^{\rho}Q_{\rho}^{\mu\nu}J_{\mu\nu})\Omega_{\alpha\beta}-i v^{\rho}Q_{\rho}^{~\mu\nu}\Omega_{\mu\nu}J_{\alpha\beta}+...
\eeq
where again $...$ contain the previously determined torsion classes and is orthogonal to what we write explicitly. This time the $ v^{\rho}Q_{\rho}^{\mu\nu}J_{\mu\nu}$ terms come with the same sign, as such we can extract what we lack by contracting \eqref{eq:tildepsipairing} with $\overline{\Omega}^{\alpha\beta}$. Making use of \eqref{eq:BPS3} in  \eqref{eq:tildepsipairing} we find
\beq
\overline{\zeta}_-v d(v\tilde{\psi}^{(6)}_-)\overline{\Omega} v\zeta_-=2i\star_6\bigg[k\wedge v\wedge J\wedge dv-2v\wedge(g{\cal A}\wedge J\wedge J-e^{\varphi}{\cal G}\wedge J)\bigg]+(\nabla.v)\star_6(k\wedge v\wedge J\wedge J)
\eeq
which is equivalent to 
\beq
v\wedge\bigg[8\overline{\Omega}\wedge d(v \tilde{\psi}^{(6)}_-)\bigg\lvert_3-(\nabla.v)k\wedge J\wedge J-2i k \wedge J\wedge dv-4i(g{\cal A}\wedge J\wedge J-e^{\varphi}{\cal G}\wedge J)\bigg]=0.
\eeq
However given that 
\beq
v \tilde{\psi}^{(6)}_-=\frac{1}{8}(1-k\wedge v)\wedge \Omega,
\eeq
 we have
\beq
v\wedge\bigg[d \Omega\wedge \overline{\Omega}-(\nabla.v)k\wedge J\wedge J-2i k \wedge J\wedge dv-4i(g{\cal A}\wedge J\wedge J-e^{\varphi}{\cal G}\wedge J)\bigg]=0
\eeq
We should stress though that  $v^{\mu}Q_{\mu}^{~\alpha\beta}J_{\alpha\beta}$ gets fixed by the complex part of this, the real part only contains $v^{\rho}Q_{\rho}^{~\mu\nu}k_{\mu}v_{\nu}$ which is already fixed, so it is only the imaginary part of this condition that is required to fix the final torsion class.
We now have geometric conditions that imply all 48 constraints in \eqref{eq:BPS3}.

\section{On the \texorpdfstring{$d=10$}{d=10} pairing constraints}\label{sec:missing10dconds}
Following \cite{Tomasiello:2011eb} we know that in order to have necessary and sufficient conditions for supersymmetry, in addition to solving 
\beq
\nabla_{(M}K_{N)}=0,~~~~~d\tilde{K}=\iota_{K}H,~~~~d_{H}(e^{-\Phi}\Psi_{\pm})=-(\iota_{K}+\tilde{K}\wedge)F_{\pm}\label{eq:diffcodnd10},
\eeq
we must also solve the $d=10$ pairing constraints. These are defined in terms of  additional two 1-forms $(V_1,V_2)$ which are such that 
\beq
V_1. K_1= V_2.K_2=\frac{1}{2}.
\eeq
The pairing constraints are
\begin{align}
&\bigg(V_1 \Psi V_2,~\Gamma^{MN}\bigg[\pm d_{H}(e^{-\Phi}\Psi_{\pm} V_2)+\frac{e^{\Phi}}{2}\star d(e^{-2\Phi} \star V_2)\Psi_{\pm}-F_{\pm}\bigg]\bigg)=0,\nn\\[2mm]
&\bigg(V_1 \Psi V_2,~\bigg[d_{H}(e^{-\Phi}V_1\Psi_{\pm} )-\frac{e^{\Phi}}{2}\star d(e^{-2\Phi} \star V_1)\Psi_{\pm}-F_{\pm}\bigg]\Gamma^{MN}\bigg)=0\label{eq:pairingeqs}.
\end{align}
where the bracket is defined as $(X,Y)= X\wedge \lambda(Y)\bigg\lvert_{10}$. We will now sketch how we extract the $d=4$ constraints that imply \eqref{eq:pairingeqs}.

In the case at hand we have 
\beq
K_1=-\frac{c}{8}e^{2A}\cos^2\left(\frac{\beta}{2}\right) k,~~~~K_2=-\frac{c}{8}e^{2A}\sin^2\left(\frac{\beta}{2}\right) k,
\eeq
so we can take
\beq
V_i= \mathfrak{f}_i v,~~~~\mathfrak{f}_1 =-\frac{4}{c \cos^2\left(\frac{\beta}{2}\right)}~~~~\mathfrak{f}_2 =-\frac{4}{c \sin^2\left(\frac{\beta}{2}\right)},
\eeq
where $v$ is the $d=6$ one form of section \ref{sec:6dsusy} which is such that $\iota_v k=1$ for $k^{\mu}\partial_{\mu}$ the $d=6$ Killing vector.
To proceed we observe that since $V_i$ lie along the external directions exclusively and ${\cal F}\wedge J\wedge J=0$ we have that
\beq
\star d(e^{-2\Phi}\star V_i)=e^{-2(A+\Phi)}\bigg(\mathfrak{f}_i(  \nabla^{(6)}.v)+ \partial_{\varphi}\log(e^{4A-2\Phi}\mathfrak{f}_i\sqrt{\det g^{(4)}})\iota^{(6)}_v d\varphi\bigg),
\eeq
where we add the $6$ superscript because these are computed on the unwarped external space and $g^{(4)}$ is the metric on the internal space (ignoring possible connection terms as they don't contribute). Likewise the first terms in the 2 pairing constraints can be manipulated to a more useful form, namely
\begin{align}
 d_{H}(e^{-\Phi}V_1\Psi_{\pm} )=d_H(e^{-\Phi}\iota_{V_1}\Psi_{\pm})-\iota_{V_1}(\tilde{K}+\iota_K)F_{\pm}+...,\nn\\[2mm]
\pm d_{H}(e^{-\Phi}\Psi_{\pm} V_2)=-d_H(e^{-\Phi}\iota_{V_2}\Psi_{\pm})+\iota_{V_2}(\tilde{K}+\iota_K)F_{\pm}+...\label{eq:usefulform}
\end{align}
where we have used the  final differential constraint in \eqref{eq:diffcodnd10} and $...$ vanishes inside the pairing constraint as either $V_1 V_1=0$  or $V_2V_2=0$. The reason \eqref{eq:usefulform} is useful is because the interior products with respect to $V_i$ only act non-trivially on the $k$ dependent terms in $\Psi_{\pm}$ and $(\tilde{K}+\iota_K)F_{\pm}$ as defined in \eqref{eq:10dbispinors} and \eqref{eq:fluxbit} respectively, for instance
\beq
\iota_{V_i}\Psi_{\pm}=\pm \frac{e^{-2A}\mathfrak{f}_i}{4}\bigg(e^{A} \text{Re}\psi_{\mp}+ e^{3A}J\wedge \text{Im}\psi_{\mp}-e^{3A}\text{Re}\big(\Omega\wedge\tilde{\psi}_{\mp}\big)-\frac{1}{2}e^{5A}J\wedge J\wedge \text{Re}\psi_{\mp}\bigg),
\eeq
where $(J,\Omega)$ span the SU(2)-structure in the external space. It is then possible to use \eqref{eq:bpscond6d1}-\eqref{eq:bpscond6d6} to take the derivatives with respect to the external directions in $d_H(e^{-\Phi}\iota_{V_i}\Psi_{\pm})$, and depending on the details of the embedding into $d=10$, the $d=4$ condition in either section \ref{sec:case1}, \ref{sec:case2} or \ref{sec:case3} to perform the internal derivatives - in this manner $d_H(e^{-\Phi}\iota_{V_i}\Psi_{\pm})$ can be expressed in a form where the only derivatives  that remain are $(\partial_{\varphi}A,~\partial_{\varphi}\Phi,~\partial_{\varphi} \mathfrak{f}_i,~d^{(4)}A,~ d^{(4)}\mathfrak{f}_i)$.  At this point one needs to insert definitions of the internal fluxes to deal with the $\iota_{V_i}(\tilde{K}+\iota_K)F_{\pm}$ and $F_{\pm}$ terms. For this purpose  \eqref{eq:sugbps4}, \eqref{eq:sugbps5}, \eqref{eq:sugbps6}, \eqref{eq:sugbps7}, \eqref{eq:sugbps8} or their analogues in sections \ref{sec:case2} or \ref{sec:case3} are sufficient to eliminate $(f_{\pm},~g_{\pm},~g_{\mp},h_{\mp})$ in favour of the NS fluxes and internal bi-linears - note that the only derivatives that need to be introduced in this process are $(\partial_{\varphi}A,~d^{(4)}A)$. At this point it is possible to compute all the components of \eqref{eq:pairingeqs}, which is a long and messy computation. We made use of Mathematica using the definition of the pairing as a trace in \eqref{eq:pairingproperties}.

Once the dust has settled we find in general that the pairing constants impose an additional 2 conditions that are not, at least obviously, contained in \eqref{eq:diffcodnd10} namely
\beq
\partial_{\varphi}(e^{2A}\sin\beta)=0,~~~~ \partial_{\varphi}(e^{4A-2\Phi}\sqrt{\det g^{(4)}})=0.
\eeq
Notice that these only have significance if the tensor multiplet is turned on. It does not really make a difference whether ${\cal A}$ appears in the metric or not, except with what is precisely meant by $\sqrt{\det{g^{(4)}}}$. When ${\cal A}$ appears in the metric more precisely
\beq
\sqrt{\det{g^{(4)}}}= e^{C}\sqrt{\det{g^{(3)}}}
\eeq
where $g^{(3)}$ is the metric on the 3 dimensional base space in the decomposition
\beq
ds^2(\text{M}_4)= ds^2(\text{M}_3)+ e^{2C} D\phi^2.
\eeq

\section{Integrability}\label{sec:Integrability}
In this section we sketch a proof that the uplifts we construct obey all the equations of motion of type II supergravity when the Bianchi identities of the NS and RR flux are satisfied (which we have been careful to ensure in the main text) and when the external equations of motion and  external supersymmetry holds. We argue in the main text that it should follow from this that the equations of motion in $d=10$ should hold even when external supersymmetry is absent, but we do not offer a proof of this.\\
~\\
Due to earlier works \cite{Lust:2008zd}, \cite{Giusto:2013rxa} we know that type II backgrounds that preserve supersymmetry, such that the $d=10$ Killing vector $K$ is null, have almost all of their EOM implied by 
\beq
dH=0,~~~~ d_HF_{\pm}=0.
\eeq 
What is not implied is a single component of Einstein's equations
\beq
{\cal E}_{MN}=0,~~~~{\cal E}_{MN}=R_{MN}+2 \nabla_{M}\nabla_M-\frac{1}{2}H_{MN}^2-\frac{e^{2\Phi}}{4}(F_{\pm})^2_{MN},
\eeq
specifically
\beq
K^M{\cal E}_{MN} K^N=0\label{eq:notimpliedEOM}.
\eeq
Our task is the to establish that this is implied for our type II uplifts when the EOM of Einstein-Maxwell gauged (or un-gauged) supergravity are assumed to hold. 

Generically when dealing with Einstein's equations the most complicated thing to deal with is the Ricci tensor. For us though this will be relatively trivial thanks to 2 useful identities: First for a $D+1$ dimensional U(1) fiber bundle
\beq
ds^2= ds^2(\text{B}_D)+ e^{2C}(d\phi+\tilde{{\cal A}})^2,~~~~\tilde{\cal F}=d\tilde{\cal A},
\eeq
the Ricci tensor  along the coordinates of $B_D$ is
\begin{align}
R_{MN}&= R^{(D)}_{MN}- 2\nabla^{(D)}_{(M}\nabla^{(D)}_{N)}C-e^{-C}\tilde{{\cal A}}_{(M}\nabla^{(D)}_P(e^{3C}\tilde{{\cal F}}^P_{~N)})-\frac{1}{2}e^{2C}\tilde{{\cal F}}^2_{MN}\nn\\[2mm]
&-e^{C}\tilde{{\cal A}}_M\tilde{{\cal A}}_N\left((\nabla^{(D)})^2(e^{C})-\frac{1}{2}e^{3C}\tilde{{\cal F}}^2\right).
\end{align}
When ${\cal A}$ appears in the metric of our uplift manifolds we have that
\beq
\tilde{{\cal A}}=p{\cal A}+ V,
\eeq
where $({\cal A},~V)$ only have components along the external and internal directions respectively. Notice that within \eqref{eq:notimpliedEOM} the only term  that can contribute is
\beq
R_{MN}= R^{(9)}_{MN} ...
\eeq
with $...$ vanishing inside \eqref{eq:notimpliedEOM} either because it is explicitly orthogonal to $K$ or, given that the metic can only dependent on the external coordinates through $\varphi$,  by the identities
\beq
{\cal L}_K \varphi=0,~~~~\iota_{K}{\cal F},~~~~\iota_{K}  {\cal A }=0, \label{eq:vanishesthero}
\eeq
which must hold because $K$ is parallel to $k$, the $d=6$ Killing vector.

The second useful identity is that  for  a $D$ dimensional metric and 
\beq
G_{MN}=  e^{2A} \hat G_{MN},
\eeq
we have
\beq
R_{MN}= \hat {R}_{MN}+ (D-2)\left(\hat{\nabla}_M\hat{\nabla}_NA-\hat{\nabla}_MA\hat{\nabla}_NA\right)-\left(\hat\nabla^2A+(D-2)(\hat\nabla A)^2\right)\hat G_{MN}.
\eeq 
Once more we see that only the first term can contribute to \eqref{eq:notimpliedEOM}. We thus conclude that whether or not ${\cal A}$ appears in the metric, the Ricci tensor for our uplifted backgrounds along the external directions is
\beq
R_{\mu\nu}= R^{(6)}_{\mu\nu}+...,
\eeq
with  ... irrelevant to the computation at hand. We then have through $d=6$ Einstein's equations that
\beq
(16)^2K^MR_{MN}K^N= 2e^{2\varphi}(\iota^{(6)}_k{\cal G})^2
\eeq
where on the RHS form contraction is performed with respect to $g^{(6)}_{\mu\nu}$ here and in the following expressions. 

Two more terms that appear in \eqref{eq:notimpliedEOM} are $\iota_KH$ and $\iota_K F_{\pm}$, the relevant terms for these are
\begin{align}
H&=H_0{\cal G}+e^{2\varphi}\star_6{\cal G}+...,\nn\\[2mm]
F_{\pm}&= e^{3A}{\cal G}\wedge g_{\mp}-e^{3A}\star_6{\cal G}\wedge \star_4\lambda(g_{\mp})-e^{4A}\star_6{\cal F}\wedge \star_4\lambda(g_{\pm})+e^{5A}\star_6 d\varphi\wedge h_{\mp}...
\end{align}
It is a relatively simple matter to establish that when supersymmetry holds
\beq
\frac{(16)^2}{2}(\iota_K H)^2= 2 c^2e^{2\varphi}\cos^2\beta (\iota^{(6)}_k{\cal G})^2.
\eeq
Then through a lengthier computation, making use of  \eqref{eq:sugbps4}, \eqref{eq:sugbps5}, \eqref{eq:sugbps6}, \eqref{eq:sugbps7}, \eqref{eq:sugbps8} or their equivalents in sections \ref{sec:case2} or \ref{sec:case3} as appropriate, it is also possible to establish that
\beq
e^{2\Phi}\frac{(16)^2}{4}(\iota_K F)^2= 2 c^2e^{2\varphi}\sin^2\beta (\iota^{(6)}_k{\cal G})^2.
\eeq
The only other term appearing in \eqref{eq:notimpliedEOM} is $K^M \nabla_M\nabla_N \Phi$ which is also zero through \eqref{eq:vanishesthero}. These results are independent of whether $g$ is non-trivial or ${\cal A}$ appears in the metric.

We have thus established that
\beq
K^M{\cal E}_{MN} K^N=(16)^2 2c^2e^{2\varphi} \left(1-\cos^2\beta-\sin^2\beta\right)\iota_{k}{\cal G}=0.
\eeq
So when supersymmetry holds and the Bianchi identities of $(H,~F_{\pm})$ are imposed then a solution in $d=6$ is lifted to a solution of type II supergravity for the uplifts we consider in this paper.     

\bibliographystyle{JHEP}

\bibliography{ref}

\end{document}